\documentclass[3p,times, review, authoryear, 12pt]{elsarticle}

\usepackage{amssymb}
\usepackage{amsfonts}
\usepackage{amsmath}
\usepackage{booktabs}
\usepackage{multirow}

\usepackage{algorithm} 
\usepackage{algorithmic}  
\usepackage[algo2e]{algorithm2e}

\DeclareMathOperator*{\essinf}{ess\,inf}

\newtheorem{assumption}{Assumption}
\usepackage[T1]{fontenc}
\usepackage{threeparttable}
\usepackage{caption}
\usepackage{subcaption}
\usepackage[marginal]{footmisc}
\usepackage{pifont}

\usepackage{longtable}
\usepackage{booktabs}
\usepackage{xr}
\usepackage{makecell}
 
\biboptions{comma,round}
\usepackage{hyperref}  
\makeatletter
\def\footnoterule{%
  \kern -3pt
  \hrule width 0.4\columnwidth
  \kern 2.6pt
}
\makeatother

\usepackage[figuresright]{rotating}
\usepackage{xcolor}

\newtheorem{definition}{Definition}
\newtheorem{proposition}{Proposition}
\newtheorem{theorem}{Theorem}
\newtheorem{proof}{Proof}
\newtheorem{lemma}{Lemma}

\newtheorem{convassumption}{Convergence Assumption}

\DeclareFontFamily{U}{matha}{\hyphenchar\font45}
\DeclareFontShape{U}{matha}{m}{n}{
	<5> <6> <7> <8> <9> <10> gen * matha
	<10.95> matha10 <12> <14.4> <17.28> <20.74> <24.88> matha12
}{}
\DeclareSymbolFont{matha}{U}{matha}{m}{n}
\DeclareFontSubstitution{U}{matha}{m}{n}

\DeclareFontFamily{U}{mathx}{\hyphenchar\font45}
\DeclareFontShape{U}{mathx}{m}{n}{
	<5> <6> <7> <8> <9> <10>
	<10.95> <12> <14.4> <17.28> <20.74> <24.88>
	mathx10
}{}
\DeclareSymbolFont{mathx}{U}{mathx}{m}{n}
\DeclareFontSubstitution{U}{mathx}{m}{n}
\DeclareMathDelimiter{\vvvert}{0}{matha}{"7E}{mathx}{"17}

\DeclareMathAlphabet{\dutchcal}{U}{dutchcal}{m}{n}
\SetMathAlphabet{\dutchcal}{bold}{U}{dutchcal}{b}{n}
\DeclareMathAlphabet{\dutchbcal} {U}{dutchcal}{b}{n}

\makeatletter
\def\ps@pprintTitle{%
 \let\@oddhead\@empty
 \let\@evenhead\@empty
 \let\@oddfoot\@empty
 \let\@evenfoot\@empty
}
\makeatother

\begin{document}

\begin{frontmatter}

\title{Distribution-valued Causal Machine Learning: Implications of Credit on Spending Patterns}

\author[label1]{Cheuk Hang LEUNG}
\address[label1]{Department of Data Science, City University of Hong Kong 
\vspace{0.5cm}
\begin{center} This version: August 18, 2025 \end{center}}
\author[label1]{Yijun LI}
\author[label1]{Qi WU}

\begin{abstract} 
    Fintech lending has become a central mechanism through which digital platforms stimulate consumption, offering dynamic, personalized credit limits that directly shape the purchasing power of consumers. Although prior research shows that higher limits increase average spending, scalar-based outcomes obscure the heterogeneous distributional nature of consumer responses. This paper addresses this gap by proposing a new causal inference framework that estimates how continuous changes in the credit limit affect the entire distribution of consumer spending. We formalize distributional causal effects within the Wasserstein space and introduce a robust Distributional Double Machine Learning estimator, supported by asymptotic theory to ensure consistency and validity. To implement this estimator, we design a deep learning architecture comprising two components: a Neural Functional Regression Net to capture complex, nonlinear relationships between treatments, covariates, and distributional outcomes, and a Conditional Normalizing Flow Net to estimate generalized propensity scores under continuous treatment. Numerical experiments demonstrate that the proposed estimator accurately recovers distributional effects in a range of data-generating scenarios. Applying our framework to transaction-level data from a major BigTech platform, we find that increased credit limits primarily shift consumers towards higher-value purchases rather than uniformly increasing spending, offering new insights for personalized marketing strategies and digital consumer finance.
\end{abstract}

\begin{keyword}
consumer credit, spending distributions, causal inference, double machine learning, deep learning.
\end{keyword}

\end{frontmatter}

\section{Introduction}
In recent years, fintech credit has emerged as a core component of leading digital retail ecosystems such as Amazon, Alibaba, and JD Digit. These BigTech firms weave short-term, revolving credit products directly into the checkout process, providing consumers with immediate access to liquidity that traditionally required engagement with external financial institutions. By collapsing the boundary between payment and borrowing, embedded credit effectively expands the effective budgets of consumers while offering retailers a powerful tool to simulate real-time demand \citep{li2021effects}. As of 2020, BigTech lenders had extended more than \$700 billion in credit worldwide \citep{cornelli2023fintech}, underscoring their increasing influence over both retail consumption and global credit markets.

At the core of credit services is the assignment of credit limits - a mechanism that directly controls the liquidity available to consumers at purchase. These limits are dynamically personalized using proprietary scoring algorithms that leverage demographic, historical transactional and financial data collected across the platform. Rather than serving merely as a passive financing constraint, the credit limit functions as an active instrument that shapes consumption behavior at the point of decision and directly influences not only the likelihood of purchase, but also the magnitude and composition of spending \citep{Li_Leung_Sun_Wang_Huang_Yan_Wu_Wang_Huang_2024}. 
    
Understanding how variations in assigned credit limits influence consumer spending behaviors is fundamentally important, as platforms can identify optimal credit levels that stimulate consumption without inducing excessive risk. This insight enables BigTech firms to strategically allocate credit to maximize transaction volume and revenue. However, determining these optimal credit levels involves inherently counterfactual scenarios, as the outcome of an alternative credit limit for a given consumer is always unobservable. This challenge is further compounded by the presence of confounding factors that simultaneously influence credit assignment decisions and consumer spending behaviors, thus obscuring the underlying causal relationship and complicating efforts to isolate the true effect of credit variation \citep{spirtes2010introduction}.

A growing body of empirical research has established that increases in credit availability—whether through higher card limits, relaxed lending terms, or digital financing options—tend to elevate aggregate consumer spending \citep{gross2002liquidity, soman2002effect, agarwal2007reaction, wilcox2011leave, aydin2022consumption}. These studies provide solid evidence that many consumers are limited by liquidity and respond to expanded credit access with elevated consumption. However, these studies have largely focused only on average treatment effects (ATE), such as total or mean expenditures, and offer limited insight into how consumers reallocate their spending and reshape their behaviors in response to incremental credit. The fundamental reason lies in the fact that these studies rely on classical causal inference frameworks that treat the outcome variable as a scalar quantity. In this paradigm, spending is typically aggregated by summing or averaging the monetary value of all transaction orders, thereby reducing complex behavioral profiles to single summary statistics. Although this approach simplifies identification and estimation, it inherently obscures heterogeneity in how credit is allocated across transactions. 
    
\begin{figure}
        \centering
        \includegraphics[width=0.6\columnwidth]{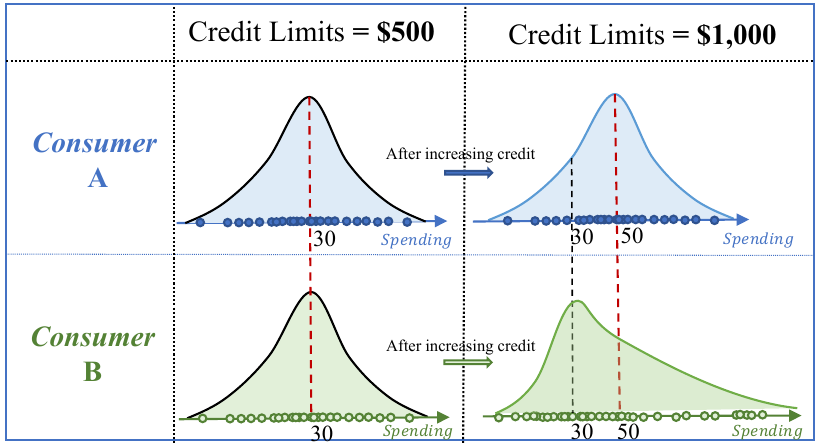}
        \caption{An illustration example of the shift of spending distribution due to the treatment effect. Each point represents the expenditure of an individual order, and all the spending collectively constitutes a spending distribution for a consumer.}
        \label{fig:example}
    \end{figure}

To illustrate, consider two consumers, A and B in Figure \ref{fig:example}. Suppose that both start with identical credit limits of \$500 and exhibit similar average spending levels around \$30. Consumer A makes moderately priced purchases—\$33, \$28, \$29—and upon receiving a higher credit limit of \$1,000, they uniformly increase spending to \$53, \$48, and \$49. In contrast, Consumer B initially spends \$33, \$28, \$29, and after the limit increase, allocates the additional credit almost entirely to high-end items, spending \$33, \$28, \$89. Although both groups exhibit the same post-treatment average of \$50, their behavioral responses are markedly different: the spending distribution of consumer A shows a shift to the right, while the spending distribution of consumer B exhibits a heavier right tail and an increase in skewness. Scalar-based approaches fail to capture such distributional dynamics, thus limiting their capacity to inform strategic credit design and behavioral targeting in practice.

To address this methodological gap, this paper introduces a distribution-valued causal inference framework for settings where the outcome of interest is a distribution rather than a scalar. Specifically, unlike scalar-based causal inference that operates in Euclidean space, we estimate the distributional average treatment effects in the Wasserstein space \citep{panaretos2020invitation}, which enables robust aggregation and comparison of distributions taking into account the geometric structure \citep{verdinelli2019hybrid, panaretos2019statistical}. Within this framework, we define two causal quantities: the Distributional Average Potential Outcome (Dist-APO) and the Distributional Average Treatment Effect (Dist-ATE), which serve as distributional analogs of the classical potential outcome and the average treatment effect. To estimate the Dist-APO, we develop a Distributional Double Machine Learning (Dist-DML) estimator grounded in \citep{chernozhukov2018double}. To implement our estimators and estimate key nuisance parameters, we design a unified deep learning architecture comprising two core components. The Neural Functional Regression Net (NFR Net) generalizes classical functional regression to capture nonlinear mappings from covariates and treatment levels to outcome distributions. In parallel, the Conditional Normalizing Flow Net (CNF Net) extends normalizing flow models to estimate generalized propensity scores under a continuous treatment regime. 
        
We validate our methodology through extensive simulation studies and a real-world application using proprietary data from a major E-commerce platform. In simulations, the Dist-DML estimator consistently outperforms benchmark methods, including Distributional Direct Regression (Dist-DR) and Distributional Inverse Propensity Weighting (Dist-IPW), by achieving lower bias and variance in recovering the true Dist-APO. Empirically, we validate our approach using data from a major digital retail platform, exploring how incremental adjustments in fintech credit influence consumer spending distributions. We find that increases in credit limits not only raise total spending, but also significantly reshape the distribution of expenditures, especially at higher quantiles, confirming that consumers tend to allocate additional fintech credit towards more expensive, discretionary purchases. These insights have strategic implications for online retail platforms and regulators in optimizing credit offerings, improving risk management strategies, and designing marketing strategies.

This study makes three contributions that advance both the causal inference methodology and the credit management practice of a platform:
    \begin{itemize}
        \item We introduce a new formalization of causal effects where the outcome is extended from a scalar to a distributional quantity, and the treatment variable is generalized from discrete to continuous. To capture the effects of such treatment variation, we define Dist-APO and Dist-ATE under the Wasserstein metric, which preserves the underlying geometry of outcome distributions. We further develop a robust and consistent estimator, Dist-DML, and establish its large-sample properties, providing a rigorous foundation for counterfactual analysis.
        
        \item We develop an end-to-end implementation centered on the proposed estimator to address high-dimensional confounding and continuous treatment spaces. The architecture combines a NFR Net, which maps covariates and treatment levels to full outcome distributions, with a CNF Net to estimate the generalized propensity score. This integrated design enables flexible and consistent estimation of distributional treatment effects in complex data-generating processes.
        
        \item We apply the proposed framework to a real-world transaction-level dataset collected from a major digital platform to uncover how changes in credit limits causally alter the shape of consumer spending distributions. Our results reveal that expanded credit access not only increases overall spending but disproportionately affects upper expenditure quantiles, suggesting that additional liquidity primarily induces consumers to shift toward higher-value purchases. These findings provide new operational insights for personalized credit allocation, targeted promotions, and platform-level financial decision making.
    \end{itemize}

\section{Literature Review}\label{sec:literature review}

\subsection{Credit Availability and Consumer Spending}
Classical consumption theory, grounded in the life cycle and permanent income hypotheses, posits that rational consumers smooth consumption over time by allocating resources in accordance with expected lifetime income \citep{modigliani1954utility, hall1978stochastic}. Within this framework, credit services merely facilitate intertemporal reallocation, allowing consumers to borrow against future income during low-income periods and repay during high-income periods, without affecting aggregate lifetime consumption. Consequently, temporary changes in credit access should not influence total spending, unless they reflect changes in lifetime resources.

However, a substantial body of empirical evidence challenges this neutrality by demonstrating that consumption is often excessively sensitive to credit conditions \citep{bacchetta1997consumption, breza2021measuring}. These findings suggest that many consumers face binding liquidity constraints or behavioral deviations from complete rationality. For example, financial deregulation and expansions of the credit market have been linked to substantial consumption booms across countries \citep{jappelli1989consumption}, while credit contractions have been observed to suppress consumption even when income remains unchanged. These patterns indicate that many people rely on credit not only for intertemporal smoothing but also as a binding component of current spending capacity.

A key mechanism through which credit affects spending is the mode of payment. Traditional theory implies that, conditional on budget constraints, the mode of payment should not alter spending. However, behavioral economics has shown that credit cards tend to increase spending by attenuating the psychological salience of payment. The foundational experiments conducted by \citet{feinberg1986credit} and the supporting studies given in \citep{hirschman1979differences, prelec2001always, raghubir2008monopoly} demonstrate that consumers spend more when using credit cards instead of cash, since the intangible nature of card-based payment weakens the ``pain of paying''. This decoupling of payment from consumption reduces transaction aversion and inflates willingness-to-pay.

Beyond the payment medium, credit limits serve as another channel to influence consumer behavior. Empirical studies have discovered that increasing credit limits tends to increase spending, especially for consumers who were close to their borrowing restrictions \citep{gross2002liquidity}. Similarly, \cite{aydin2022consumption} provides experimental evidence that newly available credit leads to sharp and sustained increases in expenditure. One possible explanation, offered by \citet{soman2002effect}, is that assigned credit ceilings act as implicit endorsements of financial standing, serving as psychological signals that justify greater consumption. This effect is particularly salient for financially inexperienced individuals who may interpret a generous limit as a reflection of future income potential.

Underlying the empirical insights above is an emphasis on credible identification strategies to recover causal effects. Since credit assignment and usage are often endogenous, researchers have sought exogenous variation to isolate the impact of credit access. Field experiments and Randomized Controlled Trials (RCTs) are always the gold standard for identifying causal effects in this domain \citep{aydin2022consumption, banerjee2015six}. When experiments are infeasible, scholars have leveraged natural experiments, difference-in-differences designs \citep{breza2021measuring, gross2002liquidity}, and instrumental variables \citep{agarwal2020mood, li2021effects}. Although effective, these strategies often face constraints related to data access, implementation costs, and ethical considerations. As a result, there is growing interest in methods that enable causal inference using observational data, particularly in complex and high-dimensional treatment settings.

\subsection{Causal Machine Learning}
Credible causal inference from observational data is challenging because each subject reveals only the outcome under the treatment actually received. The potential outcomes of alternative treatments remain unobserved. In addition, treatment assignment is usually correlated with observed and unobserved covariates, generating a confounding that biases naive comparisons of outcomes between various treatment levels \citep{hernan2010causal}.
	
In response to the challenges inherent in deriving causal inferences from observational data, a variety of methodologies have been developed. One such approach involves constructing the estimators for the target causal quantities while harnessing the capabilities of advanced machine learning techniques to estimate the nuisance parameters within these estimators. The simplest method, called the Direct Regression (DR) approach, regresses outcomes with treatments and covariates, but inherits bias when treatment assignment is endogenous. The inverse propensity weighting (IPW) method corrects this bias by constructing a pseudo-population and re-weighting observations with inverse generalized propensity scores \citep{rosenbaum1983central, hirano2003efficient}. However, using estimated propensity scores, especially when they are extreme, can lead to estimates with high variances. Double machine learning (DML) estimators mitigate bias and variance by orthogonalizing the estimating equations with respect to nuisance parameter error and using sample splitting to prevent overfitting \citep{chernozhukov2018double, farrell2015robust}. Subsequent work has extended DML to discrete treatments \citep{Huang_Leung_Yan_Wu_Peng_Wang_Huang_2021}, continuous treatments \citep{su2019non}, dynamic treatments \citep{bodory2022evaluating}, and combined treatments \citep{ye2023deep}, making it a versatile tool for high-dimensional causal analysis.
    
Despite this progress, most existing approaches for causal inference typically concentrate on estimating the causal quantities, such as average treatment effect or quantile treatment effect. Their key assumption is that, given the treatment, the realization of the outcome variables for each individual is a scalar point drawing from the same potential outcome distribution. Recent works by \cite{kennedy2021semiparametric} and \cite{martinez2023counterfactual} have shifted the focus toward directly estimating potential outcome distributions, rather than solely concentrating on counterfactual scalar values like means or specific quantiles. However, their approaches are also based on the assumption that all individuals share an identical distribution of potential outcomes when subjected to the same treatment.

In many real-world applications, the outcome for each individual is not a single realization but a distribution formed from multiple observations, such as the distribution of transaction amounts for a given consumer. This naturally connects to ideas from functional data analysis \citep{cai2022robust, chen2016variable}, where the outcome is treated as a continuous object rather than a scalar. Although early approaches have attempted to model such distributional responses in Euclidean space \citep{ecker2023causal}, it is now understood that Euclidean geometry may distort probabilistic properties when applied to distributional data \citep{panaretos2019statistical, verdinelli2019hybrid}. Alternative formulations that embed outcomes in non-Euclidean spaces, such as the Wasserstein space, provide a more principled way to capture variability across distributions. However, much of the existing work in this area has focused on discrete or binary treatments, limiting its relevance to applications involving continuous policy variables, such as credit limits in fintech platforms.
    
In numerous real-world scenarios, the outcome of an individual can be observed multiple times, collectively forming a unique distribution, such as the distribution of transaction amounts for a given consumer. This naturally connects to ideas from functional data analysis, which delves into data that continuously vary in a domain \citep{cai2022robust, chen2016variable}. Based on this concept, \cite{ecker2023causal} proposed a causal framework to analyze the impact of treatment on functional outcomes. However, their approaches are grounded in Euclidean space, in which the random structure of the distributional outcome can be destroyed \citep{verdinelli2019hybrid, panaretos2019statistical, lin2023causal}.

\section{Preliminary Backgrounds} \label{Backgrounds}	
\subsection{Notations}
We denote the treatment variable by $A$, a deterministic scalar variable taking continuous values in a subset $\mathcal{A}$ of $\mathbb{R}$; the outcome variable by $\dutchcal{Y}$ such that the realization for each individual is a distribution function; and the confounding variable/confounder by $\mathbf{X}=[X^{1},\cdots, X^{d}]\in\mathcal{X}\subseteq\mathbb{R}^{d}$ that exerts influence on both treatment $A$ and outcome $\dutchcal{Y}$ simultaneously. We assume that there exist $N$ independent units $(\mathbf{X}_{i},A_{i},\dutchcal{Y}_{i})_{i=1}^{N}$. For each unit, the realizations of $\mathbf{X}$ and $A$, together with a collection of observed values that can be characterized as the distributional outcome under the realized treatment, are observed. We also denote $\dutchcal{Y}(a)$ as the potential outcome variable associated with the specific treatment level $a$. When a unit actually receives the treatment $a$, $\dutchcal{Y}$ equals $\dutchcal{Y}(a)$, and we call $\dutchcal{Y}(a)$ the factual outcome; otherwise, $\dutchcal{Y}(a)$ is termed the counterfactual outcome and remains unobserved. Finally, we adopt a hat symbol to denote estimators (e.g., $\hat{\gamma}$ represents an estimator of the quantity $\gamma$).

\subsection{Causal Assumptions}\label{sec:Causal Assumptions}
Rooted in the potential outcome framework \cite{rubin1978bayesian, rubin2005causal}, our study is based on four key assumptions to identify causal quantities from observed data.
\begin{assumption}[SUTVA]\label{ass:SUTVA}\ 
It contains two parts:
\begin{enumerate}
\item The potential outcome of a individual is not influenced by the treatment assignment to other individuals .\label{SUTVA1}
\item For each unit, there are no different forms of treatment levels that lead to different potential outcomes.\label{SUTVA2}
\end{enumerate}
\end{assumption}
\begin{assumption}[Consistency]\label{ass:Consistency}\ 
If $A=a$, then $\dutchcal{Y}=\dutchcal{Y}(a)$.
\end{assumption}
\begin{assumption}[Ignorability]\label{ass:Ignorability}\
$A\perp\!\!\!\!\perp \dutchcal{Y}(a)\mid \mathbf{X}$\; for any $a\in\mathbb{R}$.
\end{assumption}
\begin{assumption}[Overlap]\label{ass:Positivity}\
Denote $p(a|\mathbf{x})$ as the density of $A=a$ conditioning on $\mathbf{X}=\mathbf{x}$ and $p(a,\mathbf{x})$ as the joint density function of the variables $(A,\mathbf{X})$ at $(a,\mathbf{x})$. There exists $c>0$ such that $\underset{a}{\inf}\;\underset{\mathbf{x}}{\essinf}\;p(a|\mathbf{x})\geq c$. Furthermore, we assume that $p(a,\mathbf{x})$ is a three-times differentiable function w.r.t. $a$ with all three derivatives uniformly bounded over the sample space.
\end{assumption}
We further explain the essentialness of the four assumptions in Appendix \ref{appendix:causal assumptions}.

\subsection{Wasserstein Space}\label{sec:wasserstein distance}
\begin{figure}
        \centering
        \includegraphics[width=0.6\columnwidth]{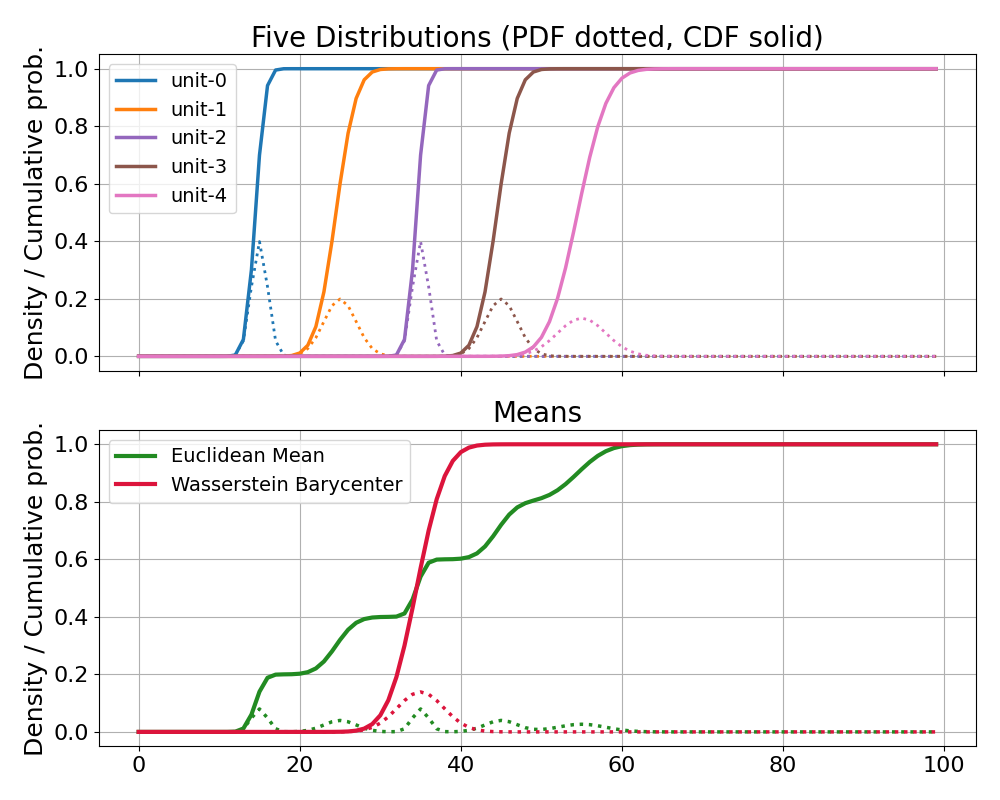}
        \caption{The Euclidean mean and Wasserstein mean (Barycenter) of 5 distributions. \label{fig:barycenter}}
    \end{figure}

We define the vector space $\mathcal{W}_{p}(\mathcal{I})\;(p\geq 1)$ that comprises cumulative distribution functions (CDFs) defined on $\mathcal{I}$ that satisfy the condition:
\begin{equation*}
	\begin{aligned}
		\mathcal{W}_{p}(\mathcal{I})=\bigg\{\textit{$\lambda$ is a CDF on $\mathcal{I}\subset\mathbb{R}$}\mid\int_{\mathcal{I}} t^{p}d\lambda(t)<\infty\bigg\}.
	\end{aligned}
\end{equation*}
To quantify the distance between two CDFs, a straightforward option for this purpose is the Euclidean $p$-measure. Under this measure, the distance between two CDFs $\lambda_{1}$ and $\lambda_{2}$ is calculated as the point-wise differences of the two CDFs in the domain $\mathcal{I}$. Mathematically, the Euclidean $p$-measure is defined as follows:
\begin{equation*}
	\bigg(\int_{\mathcal{I}}|\lambda_{1}(t)-\lambda_{2}(t)|^{p}dt\bigg)^{\frac{1}{p}}.
\end{equation*} 
However, the Euclidean $p$-measure, while simple, is not ideally suited to characterize the distance between two CDFs. One of its primary limitations is the manner in which it aggregates the values of various distributions. Using the Euclidean metric involves averaging the values of the distributions point by point. This process can potentially disrupt the structural properties of the resultant distribution, leading to a distortion or loss of its essential characteristics. To illustrate, consider the five normal distributions in the top figure of Figure \ref{fig:barycenter}. A Euclidean average of these curves produces the green density in the bottom figure of Figure \ref{fig:barycenter} where the result is multimodal and no longer Gaussian.  
    
To overcome the limitations of Euclidean $p$-measure, we turn to the $p$-Wasserstein metric \citep{villani2021topics, panaretos2019statistical, feyeux2018optimal}, which is formally defined as
\begin{definition}\label{def:Wasserstein metric}
Given two random variables $V_{1}$ and $V_{2}$, let the marginal CDFs of $V_{1}$ and $V_{2}$ be $\lambda_{1}$ and $\lambda_{2}$ that are defined in $\mathcal{I}$. In addition, let $\Lambda$ be the set that contains all the joint densities of $V_{1}$ and $V_{2}$. Suppose that the cost function $\gamma(\cdot,\cdot):\mathbb{R}\times\mathbb{R}\rightarrow\mathbb{R}$ adheres to the standard metric axioms: positivity, symmetry, and triangle inequality. The $p$-Wasserstein metric is given as $\mathbb{D}_{p}(\lambda_{1},\lambda_{2})$ such that
{\small
\begin{equation}
\begin{aligned}\label{eqt:Wasserstein metric}
\mathbb{D}_{p}(\lambda_{1},\lambda_{2})=\bigg\{\underset{\tilde{\lambda}\in\Lambda}{\inf}\int_{\mathcal{I}\times\mathcal{I}}\gamma(s,t)^{p}d\tilde{\lambda}(s,t)\bigg\}^{\frac{1}{p}}.
\end{aligned}
\end{equation}
}\noindent
\end{definition}
Here, $\gamma(\cdot,\cdot)$ represents the cost associated with transporting a point mass from position $s$ in the distribution $\lambda_{1}$ to position $t$ in the distribution $\lambda_{2}$. Thus, the integral $\int_{\mathcal{I}\times\mathcal{I}}\gamma(s,t)^{p}d\tilde{\lambda}(s,t)$ quantifies the total cost incurred in transporting the mass from $\lambda_{1}$ to $\lambda_{2}$. Consequently, $\mathbb{D}_{p}(\lambda_{1},\lambda_{2})$  is interpreted as the minimum total cost achievable among all possible joint distributions of $(\lambda_{1},\lambda_{2})$. We present a detailed illustration in the Appendix \ref{space comparison} to further distinguish the Wasserstein and Euclidean measures. 

The vector space {\small $\mathcal{W}_{p}(\mathcal{I})$} equipped with the metric {\small $\mathbb{D}_{p}(\cdot, \cdot)$} forms the $p$-Wasserstein space (formally represented as {\small $(\mathcal{W}_{p}(\mathcal{I}),\mathbb{D}_{p}(\cdot, \cdot))$}). Since the function $\gamma(s,t)$ in Definition \ref{eqt:Wasserstein metric} adheres to the metric axioms, the distance measure $\mathbb{D}_{p}(\cdot,\cdot)$ also satisfies the metric axioms, confirming that the $p$-Wasserstein space is a metric space. In the sequel, we specifically focus on the case where $p=2$ and $\gamma(s,t)=|s-t|$. This choice preserves the intrinsic geometry of the probability distributions and, therefore, produces a barycenter that retains the Gaussian shape of the original samples, as illustrated by the red curve in Figure \ref{fig:barycenter}.

\section{Distributional Outcome Causal Inference Framework}
\subsection{Dist-APO and Dist-ATE}\label{sec:causal quantities}

\begin{figure*}
\centering
\begin{subfigure}[b]{.9\textwidth}
\centering
\includegraphics[width=\textwidth]{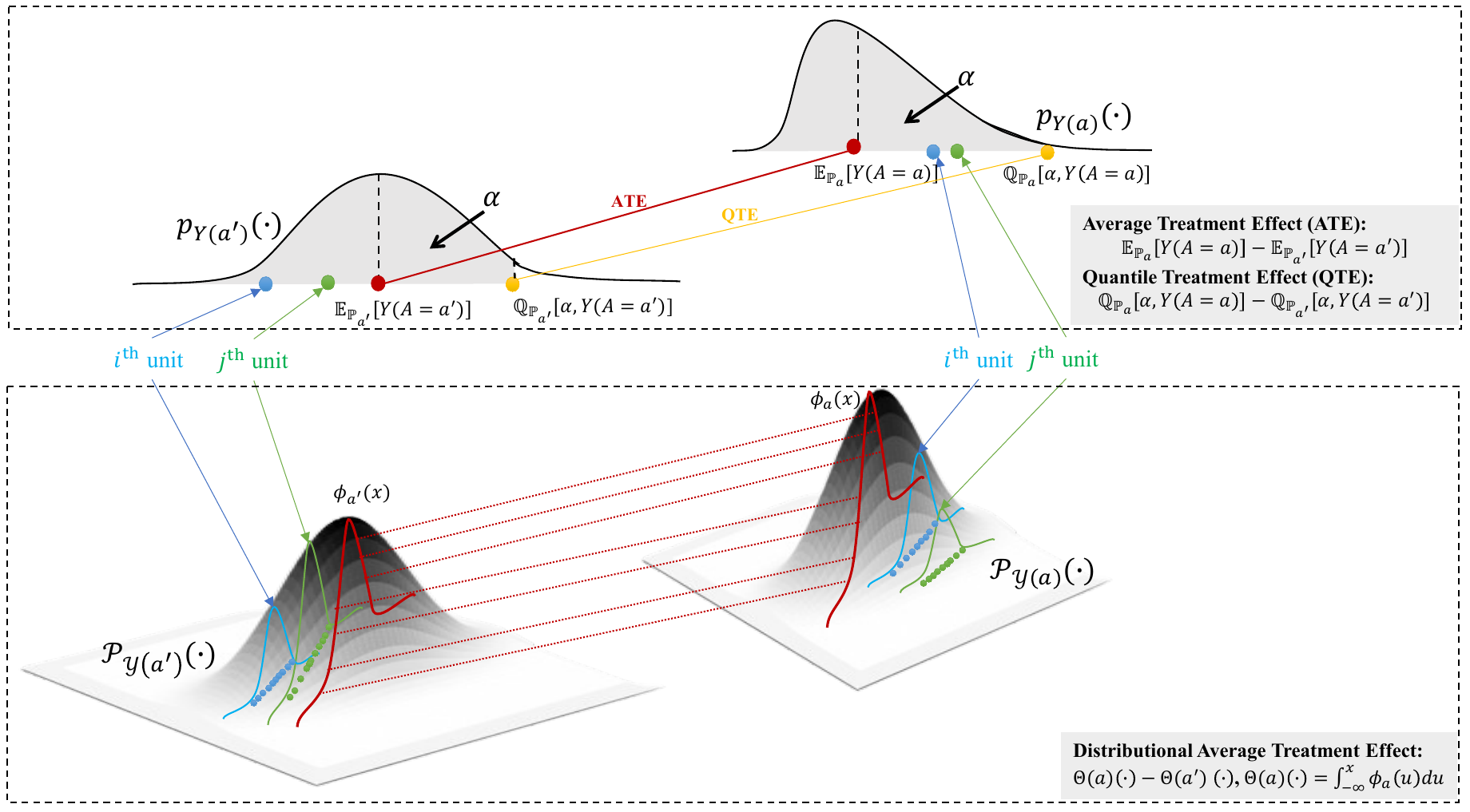}
\end{subfigure}
\caption{Comparisons of ATE, QTE, and the Dist-ATE.\label{fig:Comparisons}}
\end{figure*}

In scenarios where the outcome is a scalar, given the treatment $A = a$, the realization of the outcome variables for each individual is a scalar point drawn from the same potential outcome distribution. For example, as shown in the top figure of Figure \ref{fig:Comparisons}, the blue and green points represent the realizations of the $i^{th}$ ($j^{th}$) unit, respectively. Under this assumption, various causal quantities have been developed and explored. For example, the ATE between treatment $a$ and $a^{\prime}$ ($a \neq a^{\prime}$), denoted as $\theta(aa^\prime)$, measures the difference between the mean of the potential outcome $\mathrm{Y}(a)$ and the mean of the potential outcome $\mathrm{Y}(a^{\prime})$. Mathematically, \(\theta(aa^\prime)\) is defined with
\begin{equation}
\begin{aligned}
\theta(aa^\prime) = \mathbb{E}_{\mathbb{P}_{a}}[\mathrm{Y}(a)]- \mathbb{E}_{\mathbb{P}_{a^{\prime}}}[\mathrm{Y}(a^\prime)].
\end{aligned}
\end{equation}
Here, $\mathbb{E}_{\mathbb{P}_{a}}[\mathrm{Y}(a)]$ is the expectation of $\mathrm{Y}(a)$ in the probability measure $\mathbb{P}_{a}$, representing the average potential outcome when all individuals receive treatment $a$.
Similarly to ATE, but designed for the distributional outcomes $\dutchcal{Y}(a)$ and $\dutchcal{Y}(a^{\prime})$ of different treatments, we focus on a quantity termed Dist-ATE, which captures the causal effects across all quantiles of the distributional outcomes due to different treatments, providing a comprehensive understanding of the treatment-outcome relationships. We require the definition of the distributional average potential outcome (Dist-APO) given in Definition \ref{def:distributional-valued potential outcome}.
	
\begin{definition}\label{def:distributional-valued potential outcome}
The distributional average potential outcome is denoted as $\Theta(a)(\cdot)$, where $\Theta(a)(\cdot)=\bar{\dutchcal{Y}}^{-1}(a)(\cdot)$ and
\begin{equation}
\begin{aligned}\label{eqt:distributional-valued potential outcome}
\bar{\dutchcal{Y}}(a)(\cdot):=\underset{v\in\mathcal{W}_{2}(\mathcal{I})}{\arg\min}\;\mathbb{E}_{\mathcal{P}_{a}}\big[\mathbb{D}_{2}(\dutchcal{Y}(a),v)^{2}\big].
\end{aligned}
\end{equation}
\end{definition}

\begin{figure}
	\centering
	\includegraphics[scale=0.4]{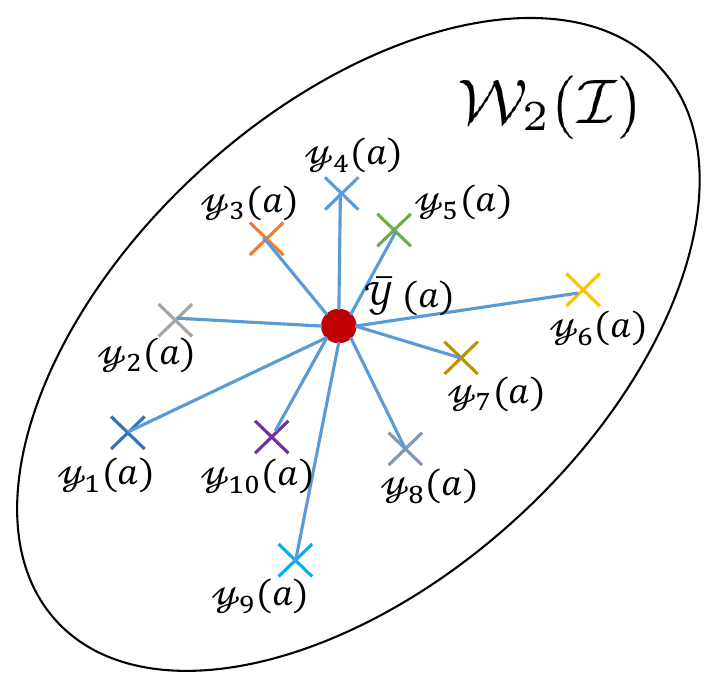}
	\caption{The illustration of Wasserstein mean in $\mathcal{W}_2(\mathcal{I})$ space. Each cross is a realization of $\dutchcal{Y}(a)$. $\bar{\dutchcal{Y}}(a)$ is the Wasserstein mean that minimizes the averaged squared distance to every realized distribution of $\dutchcal{Y}(a)$. }
	\label{fig: W2space}
\end{figure}

Unlike scalar outcomes, the realization of $\dutchcal{Y}(a)$ in the context of distributional outcomes consists of CDFs that are represented as points within the Wasserstein space $\mathcal{W}_{2}(\mathcal{I})$. This space (see Figure \ref{fig: W2space}) forms the basis of the probability space $(\mathcal{W}_{2}(\mathcal{I}), \mathcal{F}_{\mathcal{W}_{2}(\mathcal{I})}, \mathcal{P}_{a})$, where $\mathcal{W}_{2}(\mathcal{I})$ serves as the outcome space, $\mathcal{F}_{\mathcal{W}_{2}(\mathcal{I})}$  is the associated $\sigma$-algebra, and the probability measure $\mathcal{P}_{a}$ integrates to one over this space. The expectation $\mathbb{E}_{\mathcal{P}_{a}}[\cdot]$ is taken over the distributions rather than over the standard real-valued variables and calculates the averaged squared Wasserstein distance between every possible distribution of $\dutchcal{Y}(a)$ (e.g., $\dutchcal{y}_{1}(a), \cdots, \dutchcal{y}_{10}(a)$ in Figure \ref{fig: W2space}) and an arbitrary distribution and an arbitrary distribution $v$ in $\mathcal{W}_{2}(\mathcal{I})$. Consequently, $\bar{\dutchcal{Y}}(a)(\cdot)$ is specifically a CDF located in a position within $\mathcal{W}_{2}(\mathcal{I})$ that minimizes this average squared distance. $\bar{\dutchcal{Y}}(a)(\cdot)$ is also known as the Wasserstein mean or the Wasserstein barycenter, and its inverse, denoted $\bar{\dutchcal{Y}}^{-1}(a)(\cdot)$ (or $\Theta(a)(\cdot)$), serves as the quantile function of $\bar{\dutchcal{Y}}(a)(\cdot)$. In the sequel, we will omit $(\cdot)$ for simplicity, and thus $\bar{\dutchcal{Y}}(a)(\cdot), \Theta(a)(\cdot)$ will be $\bar{\dutchcal{Y}}(a), \Theta(a)$. 

To provide further clarity on the expected value $\mathbb{E}_{\mathcal{P}_{a}}[\mathbb{D}_{2}(\dutchcal{Y}(a),v)^{2}]$, consider a specific example in which the treatment level $a$ is set to $\frac{1}{2}$, and the random variable $\dutchcal{Y}(a)$ or $\dutchcal{Y}(\frac{1}{2})$ is defined such that each of its realizations is a normal distribution $\mathcal{N}(u,1)$, where the mean $u$ is drawn from a uniform distribution $\mathcal{U}([\frac{1}{2},\frac{3}{2}])$. In this setting, individual realizations of $\dutchcal{Y}(\frac{1}{2})$ might be, for example, $\mathcal{N}(\tilde{u}_{1}=\frac{3}{4},1)$ or $\mathcal{N}(\tilde{u}_{2}=\frac{6}{5},1)$, where $\tilde{u}_{1}$ and $\tilde{u}_{2}$ are numbers randomly chosen from $\mathcal{U}([\frac{1}{2},\frac{3}{2}])$. As a result, given $v\in\mathcal{W}_{2}(\mathcal{I})$, then
{\small
	\begin{equation*}
		\begin{aligned}
			\mathbb{E}_{\mathcal{P}_{a}}[\mathbb{D}_{2}(\dutchcal{Y}(a),v)^{2}]|_{a=\frac{1}{2}}=\int_{\frac{1}{2}}^{\frac{3}{2}}\mathbb{D}_{2}\big(\int_{-\infty}^{x}\frac{1}{\sqrt{2\pi}}e^{-\frac{(z-u)^{2}}{2}}dz,v(x)\big)^{2}du.
		\end{aligned}
	\end{equation*}
}\noindent
With the definition of Dist-APO, we can define the Dist-ATE between two treatments in Definition \ref{def:distributional-valued average treatment effect}.
	
\begin{definition}\label{def:distributional-valued average treatment effect}
The distributional average treatment effect between treatments $a$ and $a^\prime$, denoted $\Theta(aa^{\prime})$, is defined as the difference between $\Theta(a)$ and $\Theta(a^\prime)$. Mathematically, we have
\begin{equation}
\begin{aligned}\label{eqt:distributional-valued average treatment effect}
\Theta(aa^{\prime})\coloneqq\Theta(a)-\Theta(a^{\prime}).
\end{aligned}
\end{equation}
\end{definition}
To improve clarity, we summarize the comparison notations and definitions between the distributional outcome and the scalar outcome frameworks in Table \ref{table:Comparisons_main} and provide a detailed comparison in Appendix \ref{comparison}.

\begin{table*}[t]
	\caption{Comparisons between framework of distributional outcome and scalar outcome.\label{table:Comparisons_main}}
	\resizebox{\textwidth}{!}{
			\begin{tabular}{cccccccccccccccccccccc}
				\toprule
				& Distributional Outcome     &  Scalar Outcome \\
				\midrule
				Treatment/Covariates variable (with realization) & $A/\mathbf{X}$ ($a/\mathbf{x}$) & $A/\mathbf{X}$ ($a/\mathbf{x}$)  \\
				Outcome/Potential outcome variable (with realization) & $\dutchcal{Y}/\dutchcal{Y}(a)$ ($\dutchcal{y}/\dutchcal{y}(a)$) & $\mathrm{Y}/\mathrm{Y}(a)$ ($\mathrm{y}/\mathrm{y}(a)$) \\
				Ambient space of outcome variable ($\Omega$)  & $\mathcal{W}_{2}(\mathcal{I})$ & $\mathbb{R}$  \\
				Probability measure  & $\mathcal{P}(\omega)$, $\mathcal{P}_{a}(\omega)$, where $\omega\in\Omega$ & $\mathbb{P}(\omega)$, $\mathbb{P}_{a}(\omega)$, where $\omega\in\Omega$\\
				Metric & Wasserstein & Euclidean\\
				Realization of outcome variable  & distribution & scalar  \\
				Average Potential Outcome  & $\Theta(a)\in\mathcal{W}_{2}$  & $\theta(a)\in\mathbb{R}$ \\
				Average Treatment Effect  & $\Theta(aa^{\prime})=\Theta(a)-\Theta(a^\prime)\in\mathcal{W}_{2}$  & $\theta(aa^{\prime})=\theta(a)-\theta(a^{\prime})\in\mathbb{R}$ \\
				\bottomrule
			\end{tabular}
		}
	\end{table*}

\subsection{Dist-DML form}
As established in the previous section, the calculation of $\bar{\dutchcal{Y}}(a)$ poses a significant challenge, as it requires solving an optimization problem within the Wasserstein space, that is, $\bar{\dutchcal{Y}}(a)=\underset{v\in\mathcal{W}_{2}(\mathcal{I})}{\arg\min}\;\mathbb{E}_{\mathcal{P}_{a}}\big[\mathbb{D}_{2}(\dutchcal{Y}(a),v)^{2}\big]$. This process can be particularly demanding in terms of computational resources, especially when dealing with high-dimensional datasets or large sample sizes. To enhance the efficiency of the calculation process, it is imperative to circumvent the direct optimization step. This goal is achieved through Proposition \ref{prop:no optimization} that offers a methodological advancement by simplifying the computation of $\Theta(a)$.
\begin{proposition}\label{prop:no optimization}
The quantity $\Theta(a)$ can be reduced as $\mathbb{E}_{\mathcal{P}_{a}}\big[\dutchcal{Y}(a)^{-1}\big]$.
\end{proposition}
	
The detailed proof of Proposition \ref{prop:no optimization} is deferred to Appendix \ref{appendix:proofs of no optimization}.  This proposition elucidates that the Dist-APO $\Theta(a)$ can be conceptualized as the average of all quantile functions corresponding to the population entirely subjected to treatment $a$. However, directly estimating this quantity from the observed data poses another significant challenge. This difficulty arises because, for each individual unit, we can only observe and characterize the distribution of outcomes under a specific treatment that the individual actually received. It remains infeasible to directly characterize the distributions of the outcomes that would have occurred under alternative treatments. 

To overcome this limitation, we concentrate on exploring alternative forms of $\Theta(a)$ that facilitate the practical estimation of Dist-APO using the observed data. Consequently, we introduce the Dist-DML form for this purpose (apart from the Dist-DML form, there are two other forms: the Dist-DR form and the Dist-IPW form. These two forms are treated as the benchmark approaches and are deferred to the Appendix \ref{appendix:Dist-DR and Dist-IPW}). The Dist-DML form is developed from the Double Machine Learning Theorem as depicted in \cite{chernozhukov2018double}. The theorem provides a powerful framework that combines the benefits of both the Dist-DR form and the Dist-IPW form. The specific expression of $\mathbb{E}_{\mathcal{P}_{a}}[\dutchcal{Y}(a)^{-1}]$ based on the DML approach is summarized in Proposition \ref{prop:DML expectation form}.
\begin{proposition}\label{prop:DML expectation form}
Suppose Assumptions \ref{ass:SUTVA} - \ref{ass:Positivity} hold, we have
{\small
\begin{align}
\Theta(a)=\mathbb{E}_{(\mathbb{P}(A),\mathbb{P}(\mathbf{X}),\mathcal{P})}\bigg[m(a;\mathbf{X})
+\frac{\delta(A-a)}{p(a|\mathbf{X})}[\dutchcal{Y}^{-1}-m(a;\mathbf{X})]\bigg].\label{eqt:continuous DML expectation form}
\end{align}
}\noindent
\end{proposition}
The proof is elaborated in the Appendix \ref{appendix:proofs of DML expectation form}. Here, $m(a;\mathbf{X})=\mathbb{E}_{\mathcal{P}|\mathbb{P}(\mathbf{X})}[\dutchcal{Y}^{-1}|A=a,\mathbf{X}]$ can be obtained from the observed data using an appropriate regression model. $\delta(\cdot)$ is known as the Delta Dirac function such that (1) $\int_{\mathbb{R}}\delta(x)dx=1$; and (2) $\forall\;f\in\Omega$ with $0\in\Omega$, $\int_{\Omega}f(x)\delta(x)dx=f(0)$.
    
In both the Dist-DR form and the Dist-IPW form, the unbiased estimation of the Dist-APO $\Theta(a)$ is critically dependent on the accurate estimation of specific nuisance parameters. For the Dist-DR form, this parameter is functional regression $m(a;\mathbf{X})$ and for the Dist-IPW form, it is the generalized propensity score $p(a|\mathbf{X})$. Ideally, these estimations should align with the true nuisance parameters to ensure unbiased results. However, achieving such accuracy in real-world applications is often a significant challenge. However, the Dist-DML form offers a unique advantage in this context. It ensures the unbiasedness of $\Theta(a)$ even if either $m(a;\mathbf{X})$ or $p(a|\mathbf{X})$, but not both, are estimated with a certain inaccuracy. This doubly robust property provides a significant safeguard against potential modeling inaccuracies, ensuring that the estimation remains reliable as long as one of the two components is correctly specified.
	
\subsection{Dist-DML Estimator}\label{sec:Estimators}	
The construction of estimators based on the Dist-DML form, as described in Eqn. \eqref{eqt:continuous DML expectation form}, presents a unique challenge due to the inclusion of the Delta Dirac function $\delta(\cdot)$, which is a theoretical construction that cannot be implementable in practice. To overcome this problem, an approximation approach is utilized in which the Delta Dirac function is replaced with a sequence of kernel functions. The kernel sequence allows for the practical implementation of the concept embodied by the Delta Dirac function in statistical estimations. 
	\begin{definition}[Kernel function]\ \label{def:univariate kernel function}
		\begin{enumerate}
			\item Given that $K(\cdot):\mathbb{R} \rightarrow \mathbb{R}$ is a symmetric function (i.e., $K(v)=K(-v)$ $\forall v\in\mathbb{R}$). We say that $K(\cdot)$ is a \textit{kernel function} if it satisfies $\int_{\mathbb{R}}K(v)dv=1$.
			\item A kernel function $K(\cdot)$ is said to have order $\nu$ ($\nu$ is an even number) if $\int_{\mathbb{R}}v^{j}K(v)\;dv=0$ $\forall\;1\leq j\leq \nu-1$ and $\int_{\mathbb{R}}v^{\nu}K(v)\;dv<\infty$.
		\end{enumerate}
	\end{definition}
In this paper, we focus specifically on second-order kernel functions ($\nu=2$), which are frequently utilized in statistical estimations. A list of commonly used second-order kernel functions, along with their properties, can be found in Table \ref{tab:kernel functions} in Appendix \ref{appendix:tables and figures}. For any given kernel function $K(x)$, we define its scaled kernel with a bandwidth $h$, denoted as $K_{h}(x)$. The scaled kernel is defined as:
{\small
\begin{align*}
K_{h}(x):=\frac{1}{h}K\bigg(\frac{x}{h}\bigg)\quad\text{and}\quad \underset{h\rightarrow 0}{\lim}K_{h}(x)=\delta(x).
\end{align*}
}\noindent
Given that $\underset{h\rightarrow 0}{\lim}K_{h}(x)=\delta(x)$, we can replace $\delta(A-a)$ in Eqn. \eqref{eqt:continuous DML expectation form} with $K_{h}(A-a)$ for our estimation purposes. Consequently, the estimator for the Dist-APO using the Dist-DML form, denoted as $\hat{\Theta}^{DML}(a)$, is formulated using sample averaging:
{\small
\begin{equation}
\begin{aligned}\label{eqt:continuous DML estimator}
\hat{\Theta}^{DML}(a)=\frac{1}{N}\underset{i=1}{\overset{N}{\sum}} \bigg[m(a;\mathbf{X}_{i})+\frac{K_{h}(A_{i}-a)}{p(a|\mathbf{X}_{i})}(\dutchcal{Y}_{i}^{-1}-m(a;\mathbf{X}_{i}))\bigg].
\end{aligned}
\end{equation}
}\noindent
In practice, to avoid the overfitting problem that often occurs when Dist-DML estimators are used directly on the entire dataset, we implement the cross-fitting technique \citep{chernozhukov2018double}. Specifically, we first partition the total $N$ individuals into $\mathcal{K}$ disjoint groups. Each group, denoted as $\mathcal{D}_{k}$ $(k=\{1,\dots, \mathcal{K}\})$, contains $N_k$ individuals. The complementary data for each group, $\mathcal{D}_{-k}$, is formed by combining all other groups, i.e., $\mathcal{D}_{-k}={\cup}^{\mathcal{K}}_{r=1, r\neq k}\mathcal{D}_{r}$. Then, we use $\mathcal{D}_{-k}$ to learn the estimated functions $\hat{m}^{k}(a;\cdot)$ and the estimated generalized propensity score $\hat{p}^{k}(a|\cdot)$. Finally, we utilize $\mathcal{D}_{k}$ to compute $\hat{\Theta}^{DML,k}(a)$ using
{\small
\begin{align}\label{eqt:cross-fit estimator DML}
\begin{split}
\hat{\Theta}^{DML,k}(a)=\frac{1}{N_{k}}\underset{i\in\mathcal{D}_{k}}{\overset{}{\sum}} \bigg[\hat{m}^{k}(a;\mathbf{X}_{i})+\frac{K_{h}(A_{i}-a)}{\hat{p}^{k}(a|\mathbf{X}_{i})}(\dutchcal{Y}_{i}^{-1}-\hat{m}^{k}(a;\mathbf{X}_{i}))\bigg].
\end{split}
\end{align}
}\noindent
Consequently, we can obtain the cross-fitted estimators $\hat{\Theta}^{DML}(a)$ by averaging these individual estimates across all $\mathcal{K}$ groups:
\begin{align}\label{eqt:corss results}
\hat{\Theta}^{DML}(a)=\underset{k=1}{\overset{\mathcal{K}}{\sum}}\frac{N_k}{N}\hat{\Theta}^{DML,k}(a).
\end{align}

To end this section, we outline the above computation process in Algorithm \ref{Algorithm for cross-fitted estimators}.

	\begin{algorithm}[h] 
		\caption{Computations of $\hat{\Theta}^{DML}{(a)}$} \label{alg:estimators}
	\label{Algorithm for cross-fitted estimators}
		\begin{algorithmic}[1]
				\REQUIRE Realizations of $(A_{i},\mathbf{X}_{i},\dutchcal{Y}_{i})_{i=1}^{N}$. Determine the kernel function $K(\cdot)$.
				\STATE Estimate $\hat{\dutchcal{Y}}_i^{-1}$ for each unit $i\in \{1, \cdots, N\}$.
				\STATE Split $(A_{i},\mathbf{X}_{i},\hat{\dutchcal{Y}}_{i})_{i=1}^{N}$ to $\mathcal{K}$ disjoint units $\mathcal{D}_{k}$ where $k\in\{1,\cdots,\mathcal{K}\}$ and formulate $\mathcal{D}_{-k}$. The size of $\mathcal{D}_{k}$ is $N_{k}$.
				\FOR {$k \gets 1$ to $\mathcal{K}$}
				\STATE {Estimate $\hat{p}^{k}(a|\cdot)$ based on $\mathcal{D}_{-k}$.}\label{alg:probability estimation}
				\STATE {Estimate $\hat{m}^k(a; \cdot)$ based on $\mathcal{D}_{-k}$.}\label{alg:function estimation}
				\STATE {Compute $\hat{\Theta}^{DML, k}(a)$ based on $\mathcal{D}_{k}$ according to Eqns. \eqref{eqt:cross-fit estimator DML}.}
				\ENDFOR
				\STATE Compute $\hat{\Theta}^{DML}(a)$ according to Eqn. \eqref{eqt:corss results}.
			\end{algorithmic}
	\end{algorithm}

\section{Theory}\label{sec:theory}

We investigate the asymptotic properties of the proposed estimator $\hat{\Theta}^{DML}(a)$. To facilitate a clear and rigorous analysis, we begin by introducing several notations pertinent to our study. Consider $\mathbf{X}$ as a random variable with a distribution function denoted by $F_{\mathbf{X}}(\mathbf{x})$. In our analysis, we consider three types of spaces, namely (1) $\mathcal{L}^{2}(\mathcal{X};F_{\mathbf{X}})$, (2) $\mathcal{L}^{2}([0,1];\lambda)$ where $\lambda$ is the Lebesgue measure, and (3) $\mathcal{L}^{2}(\mathcal{X}\times[0,1];F_{\mathbf{X}}\times\lambda)$. Each space contains different forms of function:
		\begin{enumerate}
			\item $\mathcal{L}^{2}(\mathcal{X};F_{\mathbf{X}})$ contains $f$ such that $f:\mathcal{X}\rightarrow \mathbb{R}$;
			\item $\mathcal{L}^{2}([0,1];\lambda)$ contains $g$ such that $g:[0,1]\rightarrow \mathbb{R}$;
			\item $\mathcal{L}^{2}(\mathcal{X}\times[0,1];F_{\mathbf{X}}\times\lambda)$ contains $\Gamma$ such that $\Gamma:\mathcal{X}\times[0,1]\rightarrow \mathbb{R}$.
		\end{enumerate}
		Each of the defined spaces above is associated with the following norm:
		\begin{enumerate}
			\item {\small $\|f(\mathbf{X})\|_{2}^{2}=\int_{\mathcal{X}}  |f(\mathbf{x})|^{2}dF_{\mathbf{X}}(\mathbf{x})=\mathbb{E}_{\mathbb{P}(\mathbf{X})}[|f(\mathbf{X})|^{2}]$};
			\item {\small $\|g\|^{2}=\int_{[0,1]} g(t)^{2} dt$};
			\item $|\!|\!|\Gamma(\mathbf{X},t)|\!|\!|^{2}=\int_{\mathcal{X}\times[0,1]} \Gamma^{2}(\mathbf{x},t)\;dF_{\mathbf{X}}(\mathbf{x})dt=\int_{\mathcal{X}} \|\Gamma(\mathbf{x},t)\|^{2}dF_{\mathbf{X}}(\mathbf{x})$.
		\end{enumerate}
		In addition, we can define an inner product $\langle\cdot,\cdot\rangle$ for $\mathcal{L}^{2}([0,1];\lambda)$: Given $g,\;\tilde{g}\in\mathcal{L}^{2}([0,1];\lambda)$, we have
		\begin{align*}
			\langle g,\tilde{g}\rangle&=\int_{[0,1]} g(t)\tilde{g}(t) dt,\quad \text{where} \int_{[0,1]} |g(t)|^{2}dt, \; \int_{[0,1]} |\tilde{g}(t)|^{2}dt<\infty.
		\end{align*}
		
		Let $\mathbb{P}_{N}$ be the empirical average operator defined as $\mathbb{P}_{N}\mathcal{O}=\frac{1}{N}\sum_{s=1}^{N}\mathcal{O}_{s}$. We also denote the learned estimates of $m(a;\cdot)$ from dataset $\mathcal{D}_{-k}$ as $\tilde{m}^{k}(a;\cdot)$ and $\hat{m}^{k}(a;\cdot)$ for the true outcome distribution $\dutchcal{Y}$ and empirical outcome distribution $\mathcal{\hat{Y}}$, respectively. To quantify the estimation error, we define
{\small
\begin{align*}
\rho_{m}^{4}&=\underset{a\in\mathcal{A}}{\sup}\{\vvvert \tilde{m}^{k}(a)-m(a)\vvvert^{4}\}=\underset{a\in\mathcal{A}}{\sup}\{[\int_{\mathcal{X}} \|\tilde{m}^{k}(a;\mathbf{x})-m(a;\mathbf{x})\|^{2}dF_{\mathbf{X}}(\mathbf{x})]^{2}\}
\end{align*}
}\noindent
for $1\leq k\leq \mathcal{K}$. Similarly, we define
\begin{align*}
\rho_{p}^{4}=\underset{a\in\mathcal{A}}{\sup}\;\mathbb{E}_{\mathbb{P}(\mathbf{X})}[|\hat{p}^{k}(a|\mathbf{X})-p(a|\mathbf{X})|^{4}].
\end{align*}
\noindent		
With these notations and definitions in place, we proceed to present the convergence assumptions necessary to study the asymptotic properties of the proposed estimators. 
\begin{convassumption}\label{ass:assumption1}
$\hat{\dutchcal{Y}}_{1},\cdots,\hat{\dutchcal{Y}}_{N}$ are estimates of $\dutchcal{Y}_{1},\cdots,\dutchcal{Y}_{N}$ that are independent of each other under the probability measure $\hat{\mathcal{P}}$. Furthermore, there are two sequences of constants $\alpha_{N}=o(N^{-\frac{1}{2}})$ and $\nu_{N}=o(N^{-\frac{1}{2}})$ (which are thus $o(1)$ automatically) such that
{\small
\begin{align*}
&\underset{1\leq i\leq N}{\sup}\;\underset{v\in\mathcal{W}_{2}(\mathcal{I})}{\sup}\mathbb{E}_{\hat{\mathcal{P}}}[\mathbb{D}_{2}^{2}(\hat{\dutchcal{Y}}_{i},\dutchcal{Y}_{i})|\dutchcal{Y}_{i}=v]=O(\alpha_{N}^{2})\quad\textit{and}\quad\underset{1\leq i\leq N}{\sup}\;\underset{v\in\mathcal{W}_{2}(\mathcal{I})}{\sup}\mathbb{V}_{\hat{\mathcal{P}}}[\mathbb{D}_{2}^{2}(\hat{\dutchcal{Y}}_{i},\dutchcal{Y}_{i})|\dutchcal{Y}_{i}=v]=O(\nu_{N}^{4}).
\end{align*}
}\noindent
Here, $\mathbb{V}$ means the variance and $\mathbb{V}_{\hat{\mathcal{P}}}[\mathbb{D}_{2}^{2}(\hat{\dutchcal{Y}}_{i},\dutchcal{Y}_{i})|\dutchcal{Y}_{i}=v]$ is the variance of $\mathbb{D}_{2}^{2}(\hat{\dutchcal{Y}}_{i},\dutchcal{Y}_{i})$ conditioning on $\dutchcal{Y}_{i}=v$ where $v\in\mathcal{W}_{2}(\mathcal{I})$.

\end{convassumption}
		
\begin{convassumption}\label{ass:assumption3}$\forall\;a\in\mathcal{A}$ and $\forall\;1\leq k\leq \mathcal{K}$, we have 
\begin{enumerate}
\item $\underset{\mathbf{x}\in\mathcal{X}}{\sup}\|\tilde{m}^{k}(a;\mathbf{x})-m(a;\mathbf{x})\|=o_{P}(1)$;
\item  $\underset{\mathbf{x}\in\mathcal{X}}{\sup}\|\hat{p}^{k}(a|\mathbf{x})-p(a|\mathbf{x})\|=o_{P}(1)$.
\end{enumerate}
\end{convassumption}

\begin{convassumption}\label{ass:assumption5}
$\forall\;a\in\mathcal{A}$ and $1\leq k\leq \mathcal{K}$, we have 
\begin{equation*}
\begin{aligned}
\vvvert \hat{m}^{k}(a;\cdot)-\tilde{m}^{k}(a;\cdot)\vvvert=O_{P}(N^{-1}+\alpha_{N}^{2}+\nu_{N}^{2}).
\end{aligned}
\end{equation*}
\end{convassumption}
\begin{convassumption}\label{ass:assumption6}
There exist constants $c_{1}$ and $c_{2}$ such that $0<c_{1}\leq \frac{N_{k}}{N}\leq c_{2}<1$ for all $N$ and $1\leq k\leq \mathcal{K}$.
\end{convassumption}

The corresponding results for the asymptotic properties of $\hat{\Theta}^{DML}(a)$ are given in Theorem \ref{thm:asymptotic property}.

\begin{theorem}\label{thm:asymptotic property}
Let $h\rightarrow 0$, $Nh \rightarrow \infty$, and $Nh^{5} \rightarrow C\in[0,\infty)$. Suppose that $p(a|\mathbf{x})\in\mathcal{C}^{3}$ on $\mathcal{A}$ such that the derivatives (including the derivative of $0$ order) are uniformly bounded in the sample space for any $\mathbf{x}$. Furthermore, we assume that $\mathbb{E}_{\mathcal{P}|\mathbb{P}(\mathbf{X})}\big[\dutchcal{Y}^{-1}|A=a,\mathbf{X}\big]\in\mathcal{C}^{3}$ in $[0,1]\times\mathcal{A}$ and $\mathbb{E}_{\mathcal{P}|\mathbb{P}(\mathbf{X})}\big[\|\dutchcal{Y}^{-1}\||A=a,\mathbf{X}\big]\in\mathcal{C}^{3}$ in $\mathcal{A}$ are uniformly bounded in the sample spaces. Under the convergence assumptions, we have
{\small
\begin{equation}
\begin{aligned}\label{eqt:asymptotic result continuous}
&\sqrt{Nh}\big(\hat{\Theta}^{DML}(a)-\Theta(a)\big)=\sqrt{Nh}\bigg[\mathbb{P}_{N}\{\varphi(A,\mathbf{X},\dutchcal{Y})\}-\Theta(a)\bigg]+o_{P}(1),
\end{aligned}
\end{equation}
}\noindent
where $\varphi(A,\mathbf{X},\dutchcal{Y}):=\varphi(A,\mathbf{X},\dutchcal{Y})(t)=\frac{K_{h}(A-a)\{\dutchcal{Y}^{-1}(t)-m(a;\mathbf{X})(t)\}}{p(a|\mathbf{X})}+m(a;\mathbf{X})(t)$ and $\rho_{m}\rho_{p}=o(N^{-\frac{1}{2}})$, $\rho_{m}=o(1)$, $\rho_{p}=o(1)$. Additionally, we have
\begin{subequations}
\begin{align}\label{eqt:Gaussian IPW_continuous}
\sqrt{Nh}\{{\hat{\Theta}^{DML}(a)}-{\Theta(a)}-h^{2}B_{a}\}
\end{align}
converges weakly to a centred Gaussian process in $\mathcal{L}^{2}([0,1];\lambda)$ such that 
{\small 
\begin{equation*}
\begin{aligned}
B_{a}=&\frac{\int u^{2}K(u)du}{2}\times\bigg(\mathbb{E}_{\mathbb{P}(\mathbf{X})}\bigg[\partial_{aa}^{2}{m(a;\mathbf{X})}+\frac{2\partial_{a}{m(a;\mathbf{X})}\partial_{a}p(a|\mathbf{X})}{p(a|\mathbf{X})}\bigg]\bigg).
\end{aligned}
\end{equation*}
}\noindent
\end{subequations}
\end{theorem}
\noindent
The proofs for Theorem \ref{thm:asymptotic property} are provided in Appendix \ref{Proof of Theorem 1}. This theorem underscores a key advantage of the Dist-DML estimator. When estimators are constructed on the basis of the Dist-DML form, the requirement for accuracy in estimating nuisance parameters can be relaxed. Specifically, we only require the product of $\rho_{m}\rho_{p}$ equals $o((Nh)^{-\frac{1}{2}})$. This means, for instance, that both $\rho_{m}$ and $\rho_{p}$ could be $o((Nh)^{-\frac{1}{4}})$, which is less strict than what is needed for the Dist-DR or Dist-IPW estimators. In the case of the latter two estimators, both $\rho_{m}$ and $\rho_{p}$ must individually be $o((Nh)^{-\frac{1}{2}})$ to ensure accurate estimation.

We also give the covariance function of the central Gaussian process of Eqn. \eqref{eqt:Gaussian IPW_continuous}. 
{\small
\begin{equation*}
\begin{gathered}
\Psi(s)=\mathbb{E}_{(\mathbb{P}(A),\mathbb{P}(\mathbf{X}),\mathcal{P})}[\varphi(A,\mathbf{X},\dutchcal{Y})(s)],\quad\Psi(s,t)=\mathbb{E}_{(\mathbb{P}(A),\mathbb{P}(\mathbf{X}),\mathcal{P})}[\varphi(A,\mathbf{X},\dutchcal{Y})(s)\varphi(A,\mathbf{X},\dutchcal{Y})(t)].
\end{gathered}
\end{equation*}
}\noindent
The covariance function is $C(s,t)$ such that
\begin{align*}
C(s,t)
=&h\Psi(s,t)-h\Psi(s)\Psi(t).
\end{align*}
\noindent The leading term of $C(s,t)$ is given by 
{\small
\begin{equation*}
\begin{aligned}
C^{lea}(s,t)=\big(\int K^{2}(u)du\big)\mathbb{E}_{\mathbb{P}(\mathbf{X})}\bigg[\frac{\mathbb{CV}(s,t;a,\mathbf{X})}{p(a|\mathbf{X})}\bigg],
\end{aligned}
\end{equation*}
}\noindent
where $\mathbb{CV}(s,t;a,\mathbf{X})=\mathbb{E}_{\mathbb{P}(\mathbf{X})}[(\dutchcal{Y}^{-1}(s)-m(a;\mathbf{X})(s))(\dutchcal{Y}^{-1}(t)-m(a;\mathbf{X})(t))|A=a,\mathbf{X}]$.
\noindent
\noindent The asymptotic quantity
\begin{align*}
\mathbb{E}_{(\mathbb{P}(A),\mathbb{P}(\mathbf{X}),\mathcal{P})}[({\hat{\Theta}^{DML}(a)(s)}-{\Theta^{DML}(a)(s)})({\hat{\Theta}^{DML}(a)(t)}-{\Theta^{DML}(a)(t)})]
\end{align*}
equals $h^{4}B_{a}(s)B_{a}(t)+\frac{C^{lea}(s,t)}{Nh}$, and it allows us to choose a suitable $h$ for the estimators $\hat{\Theta}^{w}(a)$. For example, we can choose $h$ such that
{\small
\begin{align*}
\int_{[0,1]}\bigg[h^{4}B_{a}(s)B_{a}(t)+\frac{C^{lea}(s,t)}{Nh}\bigg]\bigg|_{s=t}dt
\end{align*}
}\noindent
is minimized. To compute the target quantity, we have to obtain $\hat{B}_{a}(t)$ and $\hat{C}(s,t)$ which are estimates of $B_{a}(t)$ and $C^{lea}(s,t)$. $\hat{B}_{a}(t)$ and $\hat{C}(s,t)$ are obtained as follows: denote $\hat{\Theta}^{DML;\beta}(a)(t)$ as the computation of $\hat{\Theta}^{DML}(a)(t)$ using the bandwidth $h=\beta$. Then $\hat{B}_{a}(t)$ is given as
\begin{equation*}
\begin{aligned}
\hat{B}_{a}(t)=\frac{{\hat{\Theta}^{DML;\beta}(a)(t)}-{\hat{\Theta}^{DML;\eta\beta}(a)(t)}}{\beta^{2}(1-\eta^{2})},\quad a\in(0,1)
\end{aligned}
\end{equation*}
followed by \cite{powell1996optimal}. In the sequel, we choose $\eta=0.5$ and $\beta=2h$. On the other hand, define
{\small 
\begin{equation*}
\begin{gathered}
\hat{\Psi}_{N_{k}}^{h}(s)=\frac{1}{N_{k}}\underset{i\in\mathcal{D}_{k}}{\overset{}{\sum}}\hat{\varphi}_{k}^{h}(A_{i},\mathbf{X}_{i},\dutchcal{Y}_{i})(s),\quad
\hat{\Psi}_{N_{k}}^{h}(s,t)=\frac{1}{N_{k}}\underset{i\in\mathcal{D}_{k}}{\overset{}{\sum}}\hat{\varphi}_{k}^{h}(A_{i},\mathbf{X}_{i},\dutchcal{Y}_{i})(s)\hat{\varphi}_{k}^{h}(A_{i},\mathbf{X}_{i},\dutchcal{Y}_{i})(t),
\end{gathered}
\end{equation*}
}\noindent
where $\hat{\varphi}_{k}^{h}(A_{i},\mathbf{X}_{i},\dutchcal{Y}_{i})(s)=\frac{K_{h}(A_{i}-a)(\hat{\dutchcal{Y}}_{i}^{-1}-\hat{m}^{k}(a;\mathbf{X}_{i}))(s)}{\hat{p}^{k}(a|\mathbf{X}_{i})}+\hat{m}^{k}(a;\mathbf{X}_{i})(s)$. Then $\hat{C}(s,t)$ is given as follows:
{\small
\begin{equation*}
\begin{aligned}
\hat{C}(s,t)=\frac{h}{\mathcal{K}}\underset{k=1}{\overset{\mathcal{K}}{\sum}}\{&\hat{\Psi}_{N_{k}}^{h}(s,t)-\hat{\Psi}_{N_{k}}^{h}(s)\hat{\Psi}_{N_{k}}^{h}(t)\},
\end{aligned}
\end{equation*}
}\noindent
As such, we may find $h^{\ast}$ such that $h^{\ast}=\underset{h}{\arg\min}\big\{\int_{[0,1]}\big[h^{4}\hat{B}_{a}(t)^{2}+\frac{\hat{C}(t,t)}{Nh}\big]dt\big\}$.

Finally, we can give an estimated range of values which includes the target quantity \(\hat{\Theta}^{DML}(a)(t)\) for each \(a\in\mathcal{A}\) and \(t\in[0,1]\). Recall that \(\hat{\Theta}^{DML}(a)=\hat{\Theta}^{DML}(a)(\cdot)\). The estimated range can be obtained based on the result given in Theorem \ref{thm:asymptotic property}. For example, given a fixed \(h\), if we want to have a range with confidence level \(1-\alpha\) for each \(a\in\mathcal{A}\) and \(t\in[0,1]\), then we have \(\Theta(a)\in \left[\hat{\Theta}^{DML}(a)-B_{a}h^{2}-\frac{q_{\frac{\alpha}{2}}}{\sqrt{Nh}},\hat{\Theta}^{DML}(a)-B_{a}h^{2}+\frac{q_{\frac{\alpha}{2}}}{\sqrt{Nh}}\right]\) where \(q_{\frac{\alpha}{2}}\) satisfies \(\mathbb{P}\{\underset{t\in[0,1]}{\sup}|\sqrt{Nh}\{\hat{\Theta}^{DML}(a)(t)-\Theta(a)(t)-B_{a}(t)h^{2}\}|\leq q_{\frac{\alpha}{2}}\}=1-\alpha\). To obtain an estimated range from the observed data, it remains to compute the quantities \(B_{a}\) and \(q_{\frac{\alpha}{2}}\) empirically. Previously, we demonstrated how to approximate \(B_{a}\) with \(\hat{B}_{a}\). We now discuss how to estimate \(q_{\frac{\alpha}{2}}\). To start, suppose that we draw \(\mathfrak{N}\) samples (\(G_{1},\cdots,G_{\mathfrak{N}}\)) from the centered Gaussian process with covariance \(\hat{C}(s,t)\) (see Appendix \ref{appendix:Gaussian Process Simulation}). For each \(G_{i}\), we compute \(g_{i}=\underset{t\in[0,1]}{\sup}\;|G_{i}(t)|\). We then obtain an estimate of \(q_{\frac{\alpha}{2}}\), denoted as \(\hat{q}_{\frac{\alpha}{2}}\), empirically by finding the quantile at the quantile level \(1-\frac{\alpha}{2}\) of \(g_{1},\cdots,g_{\mathfrak{N}}\). As a result, the estimated range is empirically equal to {\small \(\left[\hat{\Theta}^{DML}(a)-\hat{B}_{a}h^{2}-\frac{\hat{q}_{\frac{\alpha}{2}}}{\sqrt{Nh}},\hat{\Theta}^{DML}(a)-\hat{B}_{a}h^{2}+\frac{\hat{q}_{\frac{\alpha}{2}}}{\sqrt{Nh}}\right]\)}.
\noindent

\section{Models} \label{sec:models}
\begin{figure*}
\centering
\includegraphics[scale=0.53]{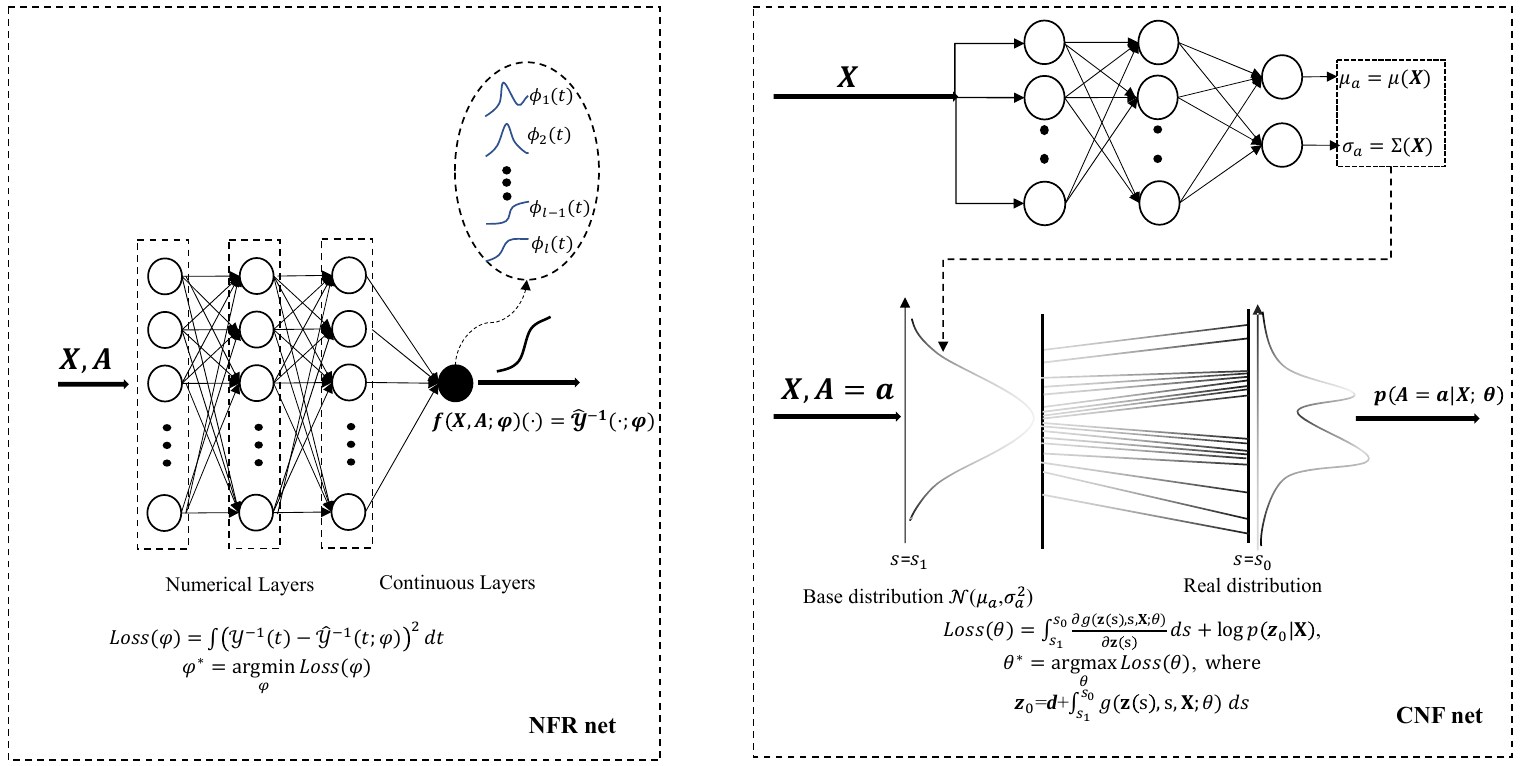}
\caption{A visualization of our proposed model. The L.H.S. is the NFR Net which is designed to learn the quantity $m(a;\cdot)$, while the R.H.S. is the CNF Net which aims to estimate the quantity $p(a|\mathbf{X})$ when $\mathbf{X}=\mathbf{x}$.\label{fig:proposed model}}
\end{figure*}

As Eqn. \eqref{eqt:cross-fit estimator DML}, it becomes essential to accurately estimate $\dutchcal{Y}^{-1}$, $p(a|\mathbf{X})$, and $m(a;\mathbf{X})$ based on the observed dataset. Estimation of $\dutchcal{Y}^{-1}$, denoted as $\dutchcal{\hat{Y}}^{-1}$, is relatively straightforward. We can estimate $\dutchcal{Y}$ empirically and then invert it to obtain the corresponding quantile function $\dutchcal{\hat{Y}}^{-1}$. Estimation of nuisance parameters $m(a;\cdot)$ and $p(a|\cdot)$ presents complex challenges due to the non-linear relationship between outcome distribution and covariates, as well as the high-order dependencies among covariates and treatment. To address the issue, we develop a comprehensive framework of deep learning. This framework consists of two distinct components: (1) NFR Net and (2) CNF Net. Each component is designed to effectively estimate different aspects of our model. The NFR Net is specifically designed to estimate $m(a;\mathbf{X})$, which aims to capture the functional relationship between the covariates $\mathbf{X}$, treatment $A$, and the outcome distribution $\dutchcal{Y}^{-1}$. The CNF Net focuses on estimating the propensity score $p(a|\mathbf{X})$, estimating the conditional density of receiving a specific treatment given covariates $\mathbf{X}$. A visual representation of our proposed model is provided in Figure \ref{fig:proposed model}. In this illustration, the NFR Net is shown on the left-hand side, and the CNF Net is depicted on the right-hand side.

%
		
\subsection{NFR Net}\label{sec:NFR net}
In cases where the outcome for each individual is scalar, neural networks, such as feed-forward neural networks, have demonstrated their ability to capture complex patterns between the outcome, treatment, and covariates. However, when the outcome for each individual is a distribution, as in our context, the application of these conventional models is not straightforward.
		
To address this challenge, we turn to functional-on-scalar regression, a method well suited for analyzing distributional outcomes \citep{ramsay2005fitting}. This approach utilizes a finite series of predetermined basis functions to approximate the regression equation. Mathematically, given a set of basis functions $\{\phi_{1},\cdots,\phi_{v}\}$ (e.g., B-spline basis), the linear functional regression model \citep{chen2016variable} can be expressed as follows:
{\small
\begin{align}
\dutchcal{\tilde{Y}}^{-1}(t)&=A\sum_{k=1}^{v} \alpha_{0 k} \phi_{k}(t)+\sum_{j=1}^{d} \beta_{j}X^{j}+\epsilon(t),\quad\beta_{j}=\sum_{k=1}^{v} \alpha_{j k} \phi_k(t).\label{eqt:literature basis expansion}
\end{align}
}\noindent
Here, $\dutchcal{\tilde{Y}}^{-1}(t)$ is the estimated outcome function, $(A, \mathbf{X})=[A, X^{1}, \cdots, X^{j}, \cdots, X^{d}]$ are predictors, $\alpha_{jk}$ ($0\leq j\leq d$ and $1\leq k\leq v$) are the regression parameters and $\epsilon(t)$ is the noise term. 
		
Eqn. \eqref{eqt:literature basis expansion} is a valuable approach that assumes an additive relationship between $\dutchcal{\tilde{Y}}^{-1}(t)$ and the predictors $(A,\mathbf{X})$. However, in many cases, this relationship is inherently non-linear and involves high-order dependencies. To address this complexity, we have designed the NFR Net, which is a deep learning architecture tailored to capture these intricate patterns. The NFR Net comprises two integral parts: (1) the numerical layers and (2) the continuous layer (see Figure \ref{fig:proposed model}). In our framework and settings, the numerical layers focus on learning a representation $\mathcal{F}(A, \mathbf{X}; \eta)$, which is a $u$-dimensional vector such that
{\small
\begin{align*}
\mathcal{F}(A, \mathbf{X}; \eta)=[\mathcal{F}_1(A, \mathbf{X}; \eta), \cdots, \mathcal{F}_u(A, \mathbf{X}; \eta)],
\end{align*}
}\noindent
where $\mathcal{F}_{i}(A, \mathbf{X}; \eta)$ represents the $i$-th linear component that contributes to the formation of the target distribution. $\mathcal{F}(A, \mathbf{X}; \eta)$ is then processed by a continuous layer to output the estimated function $\dutchcal{\tilde{Y}}^{-1}(t)$ with
{\small
\begin{align*}
\tilde{\dutchcal{Y}}^{-1}\left(t ; \eta,\alpha_{i j}\right)=\sum_{i=1}^u \mathcal{F}_{i}(A, \mathbf{X} ; \eta) \sum_{j=1}^v \alpha_{i j} \phi_j(t),
\end{align*}
}\noindent
where $\alpha_{i j}$ are the training parameters.
		
To train our model effectively, we define a loss metric $L$ (such as $\mathcal{L}_{1}$/$\mathcal{L}_{2}$ loss) that measures the difference between the empirical estimates $\dutchcal{\hat{Y}}^{-1}(t)$ and the estimates of the functional regression model $\dutchcal{\tilde{Y}}^{-1}(t)$, and focus on $\underset{\eta,\alpha_{ij}}{\min}\dutchcal{L}(\eta,\alpha_{i j})$, where
{\small
\begin{align*}
\dutchcal{L}(\eta,\alpha_{i j})\coloneqq\int_{0}^{1} L(\tilde{\dutchcal{Y}}^{-1}\left(t ; \eta,\alpha_{i j}\right), \hat{\dutchcal{Y}}^{-1}(t)) dt.
\end{align*}
}\noindent 
In practice, we can approximate the integral using the trapezoidal rule or Simpson’s rule by taking a number of discrete quantile points $t$.
		
\subsection{CNF Net}\label{sec:CNF net}
Estimating the density function from observed data is a pivotal task in various fields. Traditional approaches to this problem assumed specific forms for the target density \citep{varanasi1989parametric, efromovich2010orthogonal}, or employed kernel-based methods \citep{nadaraya1964estimating, watson1964smooth, silverman2018density}. However, each of these methods presents certain limitations. For example, assuming specific forms for the target density lacks prior knowledge about the true form of the target density, leading models not flexible enough to accurately capture the underlying distribution, especially in complex datasets. Furthermore, kernel-based methods, while more flexible, heavily depend on the choice of the appropriate bandwidth. 
		
Beyond traditional methods, we turn our focus to the use of normalizing flows \citep{dinh2014nice}, an advanced generative approach, to estimate density functions. Normalizing flows leverage the concept of transforming a base distribution $\mathbf{Z}$ (e.g., standard normal distribution) into a target distribution $\mathbf{Y}$ through a learnable, differentiable, and bijective function $G$, i.e., $\mathbf{y}=G(\mathbf{z})$. The relationship between the densities of the base distribution	$p_{\mathbf{Z}}(\mathbf{z})$ and the target distribution $p_{\mathbf{Y}}(\mathbf{y})$ is governed by the change of variables formula:
{\small
\begin{align*}
p_{\mathbf{Y}}(\mathbf{y})&=p_{\mathbf{Z}}\left(G^{-1}(\mathbf{y})\right)\left|\operatorname{det} \frac{\partial G(\mathbf{z})}{\partial \mathbf{z}}\right|^{-1}\\
\Rightarrow \log p_{\mathbf{Y}}(\mathbf{y})&=\log p_{\mathbf{Z}}\left(G^{-1}(\mathbf{y})\right)-\log \left|\operatorname{det} \frac{\partial G(\mathbf{z})}{\partial \mathbf{z}}\right|.
\end{align*}
}\noindent
Typically, the transformation function $G$ is parameterized using a sequence of neural networks, i.e., $G=G_{1} \circ \cdots \circ G_M$. A key consideration in designing these neural networks, particularly the weight matrices of $G_{j}$, for $j=1,\cdots,M$, is to ensure that they are triangular, which facilitates efficient computation of the determinant of the Jacobian \citep{dinh2016density, kingma2018glow}. However, a notable challenge with this design is the high computational cost, especially when dealing with large-scale data \citep{chen2018neural}. To mitigate this issue, we apply the continuous normalizing flow \citep{DBLP:conf/iclr/GrathwohlCBSD19} which converts the discrete transformation process into continuous dynamics, so that it achieves state-of-the-art results without the need for a triangular design. Such a continuous transformation is always governed by neural ordinary differential equations (Neural ODEs) that can be described by the following integral equation:
{\small
\begin{align}\label{eqt:ffjord_cnf}
\begin{bmatrix}
\mathbf{z}(\tau_0) \\
\log p_{\mathbf{Y}}(\mathbf{y}) - \log p_{\mathbf{Z}(\tau_{0})}(\mathbf{z}(\tau_{0}))
\end{bmatrix} = 
\begin{bmatrix}
\mathbf{y} \\
0
\end{bmatrix}+ \int_{\tau_1}^{\tau_0}
\begin{bmatrix}
g(\mathbf{z}(\tau), \tau) \\
\operatorname{Tr}\big(\frac{\partial g(\mathbf{z}(\tau), \tau)}{\partial \mathbf{z}(\tau)}\big)
\end{bmatrix}d\tau.
\end{align}
}\noindent
Here, $\tau_{0}$ and $\tau_{1}$ represent the initial and final flow times, respectively, with $\mathbf{z}(\tau_{0})=\mathbf{z}$ and $\mathbf{z}(\tau_{1})=\mathbf{y}$ being realizations of the base distribution $\mathbf{Z}$ and the target distribution $\mathbf{Y}$. The function $g$ is bijective, Lipschitz continuous in $\mathbf{z}$ and continuous in $\tau$, and $\operatorname{Tr}$ denotes the trace operator. This transformation process characterizes how a realized individual point of the base distribution `flows' through the ODEs to reach its counterpart in the target distribution.
		
Although continuous normalizing flows have been introduced primiarily to estimate unconditional density functions, we extend it to CNF Net (see Figure \ref{fig:proposed model}) to estimate the conditional density function, focusing on the propensity score $p(a|\mathbf{X})$. Specifically, we consider an augmented state $\mathbf{z}(\tau)=[z(\tau),\mathbf{X}(\tau)]^{\top}$ where $z(\tau)$ characterizes a flow from a base variable (initially at $z(\tau_{0})$ when $\tau=\tau_{0}$) to the treatment variable (ultimately at $z(\tau_{1})=a$ when $\tau=\tau_{1}$), while $\mathbf{X}(\tau)=\mathbf{X}$, $\forall\;\tau_{0}\leq \tau \leq \tau_{1}$. Consequently, the transformation form of the first equation in Eqn. \eqref{eqt:ffjord_cnf} becomes
		{\small
		\begin{align}\label{eqt:ffjord1_cond1_final}
			\begin{bmatrix}
				z(\tau_{0})\\
				\mathbf{X}(\tau_{0})
			\end{bmatrix}=\begin{bmatrix}
				a\\
				\mathbf{X}
			\end{bmatrix}+\int_{\tau_{1}}^{\tau_{0}}\begin{bmatrix}
				g(z(\tau),\mathbf{X}, \tau)\\
				\mathbf{0}
			\end{bmatrix}d\tau.
		\end{align}
	}
		Additionally, we establish a relationship between the logarithmic densities $\log p(z(\tau_{0}),\mathbf{X})$ and $\log p(a,\mathbf{X})$ based on the second equation of Eqn. \eqref{eqt:ffjord_cnf}, as formally stated in Proposition \ref{thm:ffjord conditional density}.
		
		\begin{proposition}\label{thm:ffjord conditional density}
			Let $\mathbf{z}(\tau)=[z(\tau),\mathbf{X}]^{\top}$ be a finite continuous random variable, and the probability density function of $\mathbf{z}(\tau)$ is $p(\mathbf{z}(\tau))=p(z(\tau),\mathbf{X})$ which depends on time $\tau$, where $\tau_{0}\leq \tau\leq \tau_{1}$. Given the governing equation of $\mathbf{z}(\tau)$ in Eqn. \eqref{eqt:ffjord1_cond1_final} and $g$ is Lipschitz continuous in $z$ and continuous in $\tau$ for any $\mathbf{X}$, we have
{\small
\begin{align}\label{eqt:ffjord cond density final}
\log p(a,\mathbf{X})=\log p(z(\tau_{0}),\mathbf{X}) + \int_{\tau_1}^{\tau_0}\left(\frac{\partial g(z(\tau), \mathbf{X}, \tau)}{\partial z(\tau)}\right)d\tau.
\end{align}
}\noindent
	
\end{proposition}
The formal proofs are deferred to the Appendix \ref{appendix:proofs of Theorem ffjord}. Then, by subtracting both sides of Eqn. \eqref{eqt:ffjord cond density final} by $\log p(\mathbf{X})$, we have
{\small
\begin{align}\label{eqt:ffjord cond density finalized}
\log p(a|\mathbf{X}) = \log p(z(\tau_{0})|\mathbf{X}) + \int_{\tau_1}^{\tau_0}\left(\frac{\partial g(z(\tau), \mathbf{X}, \tau)}{\partial z(\tau)}\right)d\tau.
\end{align}
}\noindent
This formulation indicates that the density $p(a|\mathbf{X})$ is dependent on the conditional base distribution $p(z(\tau_{0})|\mathbf{X})$. To model this base distribution, we assume that $z(\tau_{0})|\mathbf{X}$ follows a conditional normal distribution $\mathcal{N}(\mu(\mathbf{X}),\sigma^{2}(\mathbf{X}))$. Here, $\mu(\cdot)$ and $\sigma(\cdot)$ are two functions parametrized by feed-forward neural networks that represent the mean and standard deviation of the conditional normal distribution, respectively. In the final step of implementing our CNF Net framework, the training process revolves around maximizing the log-likelihood function, specifically $\log p(a|\mathbf{X})$.
\section{Numerical Experiment}\label{sec:experiment}
\subsection{Experiment Setting}
To verify our theories and models, we perform a numerical experiment in which the treatment takes continuous values, and the outcome for each sample is a specific distribution function. 
The data generation process (DGP) for our numerical experiment is designed to simulate the intricate dynamics often encountered in real-world datasets, aiming to assess the capability of our models in handling non-linear interactions and complex causal relationships. The DGP is formulated as follows:
{\small
\begin{align}
\begin{split}\label{eqt: DGP}
\dutchcal{Y}_{i}^{-1}\left(\cdot\right)&=c+(1-c)(\mathbb{E}[\gamma^{\top} \mathbf{X}_{i}]+\exp{(A_{i})})\times\underset{j=1}{\overset{\frac{n}{2}}{\sum}} w_{j}\mathbf{B}^{-1}\left(\alpha_{j}, \beta_{j}\right)+\epsilon_{i},\\
w_{j}&=\frac{\exp \left(\mathbf{X}_{i}^{2 j-1} \mathbf{X}_{i}^{2 j}\right)}{\underset{k=1}{\overset{\frac{n}{2}}{\sum}} \exp \left(\mathbf{X}_{i}^{2 k-1} \mathbf{X}_{i}^{2 k}\right)},\\
A_{i}  &\sim  \mathcal{N}(\gamma^{\top} \mathbf{X}_{i}, \log(1+\exp(\xi^{\top} \mathbf{X}_{i})),
\end{split}
\end{align}
}\noindent
where $\dutchcal{Y}_{i}^{-1}$ is the quantile function of the individual $i$, which is a complex function of the treatment $A_{i}$ and the covariates $\mathbf{X}_{i}$. $A_{i}$ follows a Gaussian distribution whose mean and variance are controlled by the covariates $\mathbf{X}_{i}$. $n$ is an even number that indicates the number of covariates. $\mathbf{B}^{-1}(\alpha, \beta)$ is the inverse CDF (quantile function) of the Beta distribution with the shapes' parameters $\alpha$ and $\beta$. We choose Beta distributions because they vary widely given different parameters. $c$ is the constant that controls the strength of the causal relationship between $A_{i}$ and $\dutchcal{Y}_{i}^{-1}$. $\epsilon_{i}$ is the noise that follows $\mathcal{N}(0,0.05)$. 
\begin{figure}
\centering
\includegraphics[scale=0.5]{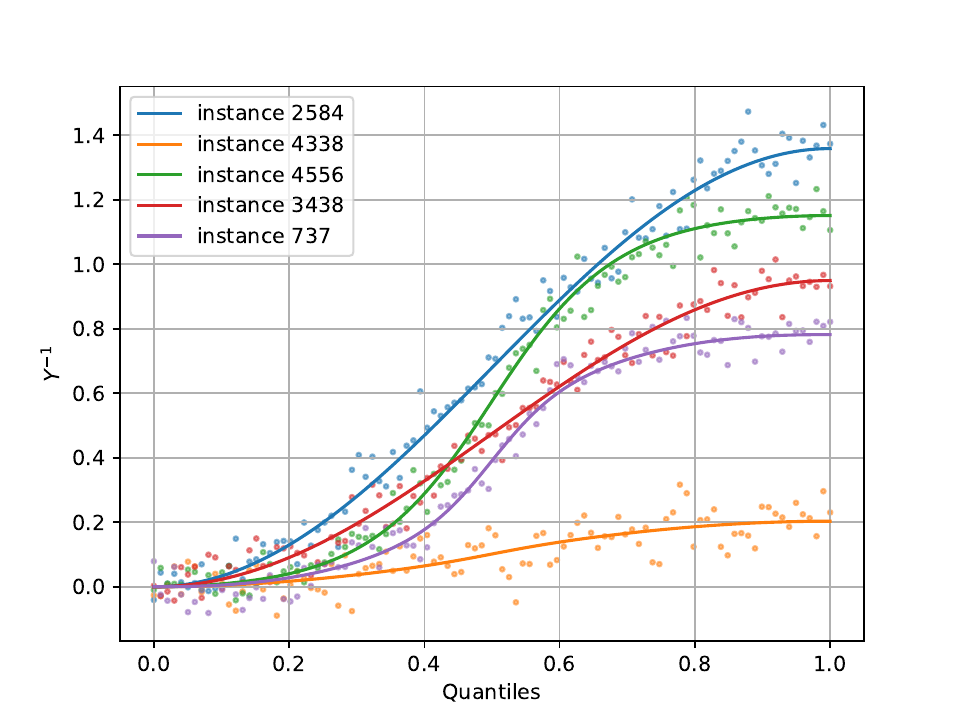}
\caption{The inverse CDF of 5 simulated samples}
\label{fig:samples}
\end{figure}
In the experiment, we configure the number of covariates $(n)$ to be 10, where $X^1, X^2 \sim \mathcal{N}(-2,1), X^3, X^4 \sim \mathcal{N}(-1,1), X^5, X^6 \sim\mathcal{N}(0,1)$, $X^7, X^8 \sim \mathcal{N}(1,1)$, and $X^9, X^{10} \sim \mathcal{N}(2,1)$. To add complexity to the outcome distributions, we utilize five inverse Beta CDFs, each set with distinct parameters. For each individual in our dataset, we generate 100 observations in accordance with our data generation process (Eqn. \eqref{eqt: DGP}) using the inverse transform sampling method. In total, 50,000 individuals are generated. Figure \ref{fig:samples} summarizes 5 simulated individuals, where the curve represents the true inverse CDF and the points indicate the corresponding observations for each unit. This visualization highlights the variability in the inverse CDFs between different treatments. The primary objective of the experiment is to estimate the potential outcome distributions for all individuals when treated with specific treatment values: $-0.5$, $0.0$, and $0.5$ (i.e., $A=-0.5$, $0.0$, $0.5$).
		
Two components, the NFR Net and CNF Net, are trained separately during an experiment to estimate the functional outcome $\dutchcal{Y}^{-1}$ and the conditional density $p(a|\mathbf{X})$. To optimize performance, the hyperparameters of NFR Net and CNF Net are tuned using the random search approach \citep{bergstra2012random}, and the finalized training parameters include a learning rate of 0.003, a batch size of 128, and a weight decay of 0.001. The Adam algorithm \citep{kingma2014adam} is set as the default optimizer. To ensure efficient convergence and prevent overfitting, an adaptive learning rate strategy is used, wherein if the validation loss does not decrease over 10 epochs, the learning rate would be reduced by half. The model that performs best during the training phase is preserved and subsequently used to compute counterfactual distributional outcomes. 
		
We implement a two-fold cross-fitting technique for training. Half of the individuals (25,000) are utilized for training purposes, while the remaining half are used to obtain the Dist-DML estimator and two benchmark Dist-DR and Dist-IPW estimators. To assess the performance of our estimators, we discretize both the ground truth outcome distribution $\Theta(a)$ and the estimated Dist-APO $\hat{\Theta}(a)$, comparing them across 9 quantiles, ranging from 0.1 to 0.9. The Mean Absolute Error (MAE) between these discretized outcomes serves as our primary metric of performance. To ensure the robustness of our results, the entire experiment is repeated 100 times. This repetition allows us to report both the mean and the standard deviation of the MAE, providing a comprehensive view of the performance and reliability of our estimators under varying conditions.
\begin{table*}[t]
\centering
\caption{The numerical experiment results of DR, IPW, and DML estimators on treatment $A=-0.5$, $0.0$, $0.5$\label{Tab: numerical res}}
 \resizebox{\textwidth}{!}{
\begin{tabular}{cllllllllll}
\toprule
& \multicolumn{1}{c}{\textbf{Q=0.1}} & \multicolumn{1}{c}{\textbf{Q=0.2}} & \multicolumn{1}{c}{\textbf{Q=0.3}} & \multicolumn{1}{c}{\textbf{Q=0.4}} & \multicolumn{1}{c}{\textbf{Q=0.5}} & \multicolumn{1}{c}{\textbf{Q=0.6}} & \multicolumn{1}{c}{\textbf{Q=0.7}} & \multicolumn{1}{c}{\textbf{Q=0.8}} & \multicolumn{1}{c}{\textbf{Q=0.9}} & \multicolumn{1}{c}{\textbf{Avg.}} \\ \midrule
\multicolumn{11}{c}{$A=-0.5$} \\ \midrule
\textbf{Ground} & 0.0068 & 0.0279 & 0.0654 & 0.1374 & 0.3053 & 0.4731 & 0.5452 & 0.5826 & 0.6038 &  \\
\textbf{Dist-DR} & \multicolumn{1}{c}{\begin{tabular}[c]{@{}c@{}}0.0007\\ (0.0022)\end{tabular}} & \multicolumn{1}{c}{\begin{tabular}[c]{@{}c@{}}0.0225\\ (0.0013)\end{tabular}} & \multicolumn{1}{c}{\begin{tabular}[c]{@{}c@{}}0.0928\\ (0.0018)\end{tabular}} & \multicolumn{1}{c}{\begin{tabular}[c]{@{}c@{}}0.1948\\ (0.0034)\end{tabular}} & \multicolumn{1}{c}{\begin{tabular}[c]{@{}c@{}}0.3118\\ (0.0055)\end{tabular}} & \multicolumn{1}{c}{\begin{tabular}[c]{@{}c@{}}0.4274\\ (0.0076)\end{tabular}} & \multicolumn{1}{c}{\begin{tabular}[c]{@{}c@{}}0.5265\\ (0.0093)\end{tabular}} & \multicolumn{1}{c}{\begin{tabular}[c]{@{}c@{}}0.5943\\ (0.0104)\end{tabular}} & \multicolumn{1}{c}{\begin{tabular}[c]{@{}c@{}}0.6160\\ (0.0106)\end{tabular}} &  \\
\textbf{Dist-DR (MAE)} & 0.0061 & 0.0054 & 0.0274 & 0.0573 & 0.0065 & 0.0457 & 0.0187 & 0.0117 & 0.0123 & 0.0212 \\
\textbf{Dist-IPW} & \multicolumn{1}{c}{\begin{tabular}[c]{@{}c@{}}-0.0028\\ (0.0001)\end{tabular}} & \multicolumn{1}{c}{\begin{tabular}[c]{@{}c@{}}0.0400\\ (0.0012)\end{tabular}} & \multicolumn{1}{c}{\begin{tabular}[c]{@{}c@{}}0.0889\\ (0.0026)\end{tabular}} & \multicolumn{1}{c}{\begin{tabular}[c]{@{}c@{}}0.1679\\ (0.0047)\end{tabular}} & \multicolumn{1}{c}{\begin{tabular}[c]{@{}c@{}}0.3171\\ (0.0086)\end{tabular}} & \multicolumn{1}{c}{\begin{tabular}[c]{@{}c@{}}0.4581\\ (0.0123)\end{tabular}} & \multicolumn{1}{c}{\begin{tabular}[c]{@{}c@{}}0.5321\\ (0.0143)\end{tabular}} & \multicolumn{1}{c}{\begin{tabular}[c]{@{}c@{}}0.5791\\ (0.0156)\end{tabular}} & \multicolumn{1}{c}{\begin{tabular}[c]{@{}c@{}}0.6220\\ (0.0168)\end{tabular}} & \multicolumn{1}{c}{} \\
\textbf{Dist-IPW (MAE)} & 0.0096 & 0.0120 & 0.0235 & 0.0305 & 0.0118 & 0.0150 & 0.0131 & 0.0035 & 0.0182 & 0.0153 \\
\textbf{Dist-DML} & \multicolumn{1}{c}{\begin{tabular}[c]{@{}c@{}}-0.0026\\ (0.0001)\end{tabular}} & \multicolumn{1}{c}{\begin{tabular}[c]{@{}c@{}}0.0405\\ (0.0005)\end{tabular}} & \multicolumn{1}{c}{\begin{tabular}[c]{@{}c@{}}0.0900\\ (0.0002)\end{tabular}} & \multicolumn{1}{c}{\begin{tabular}[c]{@{}c@{}}0.1697\\ (0.0007)\end{tabular}} & \multicolumn{1}{c}{\begin{tabular}[c]{@{}c@{}}0.3195\\ (0.0005)\end{tabular}} & \multicolumn{1}{c}{\begin{tabular}[c]{@{}c@{}}0.4612\\ (0.0010)\end{tabular}} & \multicolumn{1}{c}{\begin{tabular}[c]{@{}c@{}}0.5357\\ (0.0007)\end{tabular}} & \multicolumn{1}{c}{\begin{tabular}[c]{@{}c@{}}0.5833\\ (0.0009)\end{tabular}} & \multicolumn{1}{c}{\begin{tabular}[c]{@{}c@{}}0.6267\\ (0.0008)\end{tabular}} & \multicolumn{1}{c}{} \\
\textbf{Dist-DML (MAE)} & 0.0094 & 0.0126 & 0.0246 & 0.0322 & 0.0142 & 0.0120 & 0.0094 & 0.0007 & 0.0229 & 0.0153 \\ \midrule
\multicolumn{11}{c}{$A=0.0$} \\  \midrule
\textbf{Ground} & 0.0112 & 0.0459 & 0.1075 & 0.2260 & 0.5020 & 0.7780 & 0.8965 & 0.9581 & 0.9929 & \multicolumn{1}{c}{} \\
\textbf{Dist-DR} & \multicolumn{1}{c}{\begin{tabular}[c]{@{}c@{}}0.0103\\ (0.0034)\end{tabular}} & \multicolumn{1}{c}{\begin{tabular}[c]{@{}c@{}}0.0311\\ (0.0021)\end{tabular}} & \multicolumn{1}{c}{\begin{tabular}[c]{@{}c@{}}0.1406\\ (0.0033)\end{tabular}} & \multicolumn{1}{c}{\begin{tabular}[c]{@{}c@{}}0.3080\\ (0.0059)\end{tabular}} & \multicolumn{1}{c}{\begin{tabular}[c]{@{}c@{}}0.5028\\ (0.0092)\end{tabular}} & \multicolumn{1}{c}{\begin{tabular}[c]{@{}c@{}}0.6948\\ (0.0128)\end{tabular}} & \multicolumn{1}{c}{\begin{tabular}[c]{@{}c@{}}0.8560\\ (0.0160)\end{tabular}} & \multicolumn{1}{c}{\begin{tabular}[c]{@{}c@{}}0.9590\\ (0.0180)\end{tabular}} & \multicolumn{1}{c}{\begin{tabular}[c]{@{}c@{}}0.9762\\ (0.0185)\end{tabular}} & \multicolumn{1}{c}{} \\
\textbf{Dist-DR (MAE)} & 0.0009 & 0.0148 & 0.0330 & 0.0820 & 0.0008 & 0.0832 & 0.0405 & 0.0008 & 0.0167 & 0.0303 \\
\textbf{Dist-IPW} & \multicolumn{1}{c}{\begin{tabular}[c]{@{}c@{}}0.0078\\ (0.0003)\end{tabular}} & \multicolumn{1}{c}{\begin{tabular}[c]{@{}c@{}}0.0608\\ (0.0014)\end{tabular}} & \multicolumn{1}{c}{\begin{tabular}[c]{@{}c@{}}0.1321\\ (0.0031)\end{tabular}} & \multicolumn{1}{c}{\begin{tabular}[c]{@{}c@{}}0.2618\\ (0.0060)\end{tabular}} & \multicolumn{1}{c}{\begin{tabular}[c]{@{}c@{}}0.5108\\ (0.0115)\end{tabular}} & \multicolumn{1}{c}{\begin{tabular}[c]{@{}c@{}}0.7489\\ (0.0169)\end{tabular}} & \multicolumn{1}{c}{\begin{tabular}[c]{@{}c@{}}0.8693\\ (0.0197)\end{tabular}} & \multicolumn{1}{c}{\begin{tabular}[c]{@{}c@{}}0.9377\\ (0.0212)\end{tabular}} & \multicolumn{1}{c}{\begin{tabular}[c]{@{}c@{}}0.9900\\ (0.0224)\end{tabular}} & \multicolumn{1}{c}{} \\
\textbf{Dist-IPW (MAE)} & 0.0034 & 0.0149 & 0.0246 & 0.0358 & 0.0088 & 0.0291 & 0.0272 & 0.0204 & 0.0029 & 0.0185 \\
\textbf{Dist-DML} & \multicolumn{1}{c}{\begin{tabular}[c]{@{}c@{}}0.0080\\ (0.0002)\end{tabular}} & \multicolumn{1}{c}{\begin{tabular}[c]{@{}c@{}}0.0615\\ (0.0007)\end{tabular}} & \multicolumn{1}{c}{\begin{tabular}[c]{@{}c@{}}0.1346\\ (0.0003)\end{tabular}} & \multicolumn{1}{c}{\begin{tabular}[c]{@{}c@{}}0.2672\\ (0.0011)\end{tabular}} & \multicolumn{1}{c}{\begin{tabular}[c]{@{}c@{}}0.5195\\ (0.0005)\end{tabular}} & \multicolumn{1}{c}{\begin{tabular}[c]{@{}c@{}}0.7610\\ (0.0015)\end{tabular}} & \multicolumn{1}{c}{\begin{tabular}[c]{@{}c@{}}0.8841\\ (0.0008)\end{tabular}} & \multicolumn{1}{c}{\begin{tabular}[c]{@{}c@{}}0.9543\\ (0.0008)\end{tabular}} & \multicolumn{1}{c}{\begin{tabular}[c]{@{}c@{}}1.0070\\ (0.0009)\end{tabular}} & \multicolumn{1}{c}{} \\
\textbf{Dist-DML (MAE)} & 0.0031 & 0.0155 & 0.0271 & 0.0412 & 0.0175 & 0.0171 & 0.0124 & 0.0038 & 0.0141 & 0.0169 \\ \midrule
\multicolumn{11}{c}{$A=0.5$} \\  \midrule
\textbf{Dist-Ground} & 0.0184 & 0.0756 & 0.1770 & 0.3720 & 0.8264 & 1.2807 & 1.4758 & 1.5772 & 1.6344 & \multicolumn{1}{c}{} \\
\textbf{Dist-DR} & \multicolumn{1}{c}{\begin{tabular}[c]{@{}c@{}}0.0233\\ (0.0058)\end{tabular}} & \multicolumn{1}{c}{\begin{tabular}[c]{@{}c@{}}0.0443\\ (0.0034)\end{tabular}} & \multicolumn{1}{c}{\begin{tabular}[c]{@{}c@{}}0.2184\\ (0.0066)\end{tabular}} & \multicolumn{1}{c}{\begin{tabular}[c]{@{}c@{}}0.4924\\ (0.0130)\end{tabular}} & \multicolumn{1}{c}{\begin{tabular}[c]{@{}c@{}}0.8129\\ (0.0211)\end{tabular}} & \multicolumn{1}{c}{\begin{tabular}[c]{@{}c@{}}1.1276\\ (0.0297)\end{tabular}} & \multicolumn{1}{c}{\begin{tabular}[c]{@{}c@{}}1.3879\\ (0.0374)\end{tabular}} & \multicolumn{1}{c}{\begin{tabular}[c]{@{}c@{}}1.5461\\ (0.0428)\end{tabular}} & \multicolumn{1}{c}{\begin{tabular}[c]{@{}c@{}}1.5543\\ (0.0448)\end{tabular}} & \multicolumn{1}{c}{} \\
\textbf{Dist-DR (MAE)} & 0.0050 & 0.0313 & 0.0414 & 0.1203 & 0.0135 & 0.1531 & 0.0878 & 0.0311 & 0.0801 & 0.0626 \\
\textbf{Dist-IPW} & \multicolumn{1}{c}{\begin{tabular}[c]{@{}c@{}}0.0205\\ (0.0025)\end{tabular}} & \multicolumn{1}{c}{\begin{tabular}[c]{@{}c@{}}0.0929\\ (0.0115)\end{tabular}} & \multicolumn{1}{c}{\begin{tabular}[c]{@{}c@{}}0.2031\\ (0.0253)\end{tabular}} & \multicolumn{1}{c}{\begin{tabular}[c]{@{}c@{}}0.4190\\ (0.0528)\end{tabular}} & \multicolumn{1}{c}{\begin{tabular}[c]{@{}c@{}}0.8284\\ (0.1047)\end{tabular}} & \multicolumn{1}{c}{\begin{tabular}[c]{@{}c@{}}1.2173\\ (0.1547)\end{tabular}} & \multicolumn{1}{c}{\begin{tabular}[c]{@{}c@{}}1.4135\\ (0.1794)\end{tabular}} & \multicolumn{1}{c}{\begin{tabular}[c]{@{}c@{}}1.5183\\ (0.1925)\end{tabular}} & \multicolumn{1}{c}{\begin{tabular}[c]{@{}c@{}}1.5879\\ (0.2011)\end{tabular}} & \multicolumn{1}{c}{} \\
\textbf{Dist-IPW (MAE)} & 0.0021 & 0.0174 & 0.0261 & 0.0469 & 0.0020 & 0.0634 & 0.0622 & 0.0589 & 0.0465 & 0.0362 \\
\textbf{DML} & \multicolumn{1}{c}{\begin{tabular}[c]{@{}c@{}}0.0212\\ (0.0010)\end{tabular}} & \multicolumn{1}{c}{\begin{tabular}[c]{@{}c@{}}0.0949\\ (0.0065)\end{tabular}} & \multicolumn{1}{c}{\begin{tabular}[c]{@{}c@{}}0.2088\\ (0.0019)\end{tabular}} & \multicolumn{1}{c}{\begin{tabular}[c]{@{}c@{}}0.4302\\ (0.0083)\end{tabular}} & \multicolumn{1}{c}{\begin{tabular}[c]{@{}c@{}}0.8459\\ (0.0060)\end{tabular}} & \multicolumn{1}{c}{\begin{tabular}[c]{@{}c@{}}1.2411\\ (0.0168)\end{tabular}} & \multicolumn{1}{c}{\begin{tabular}[c]{@{}c@{}}1.4427\\ (0.0115)\end{tabular}} & \multicolumn{1}{c}{\begin{tabular}[c]{@{}c@{}}1.5511\\ (0.0091)\end{tabular}} & \multicolumn{1}{c}{\begin{tabular}[c]{@{}c@{}}1.6219\\ (0.0120)\end{tabular}} & \multicolumn{1}{c}{} \\
\textbf{Dist-DML (MAE)} & 0.0029 & 0.0193 & 0.0318 & 0.0582 & 0.0195 & 0.0396 & 0.0331 & 0.0260 & 0.0125 & 0.0270 \\ \bottomrule
\end{tabular}
}
\end{table*}
		
\begin{figure}
\centering
\includegraphics[width=\columnwidth]{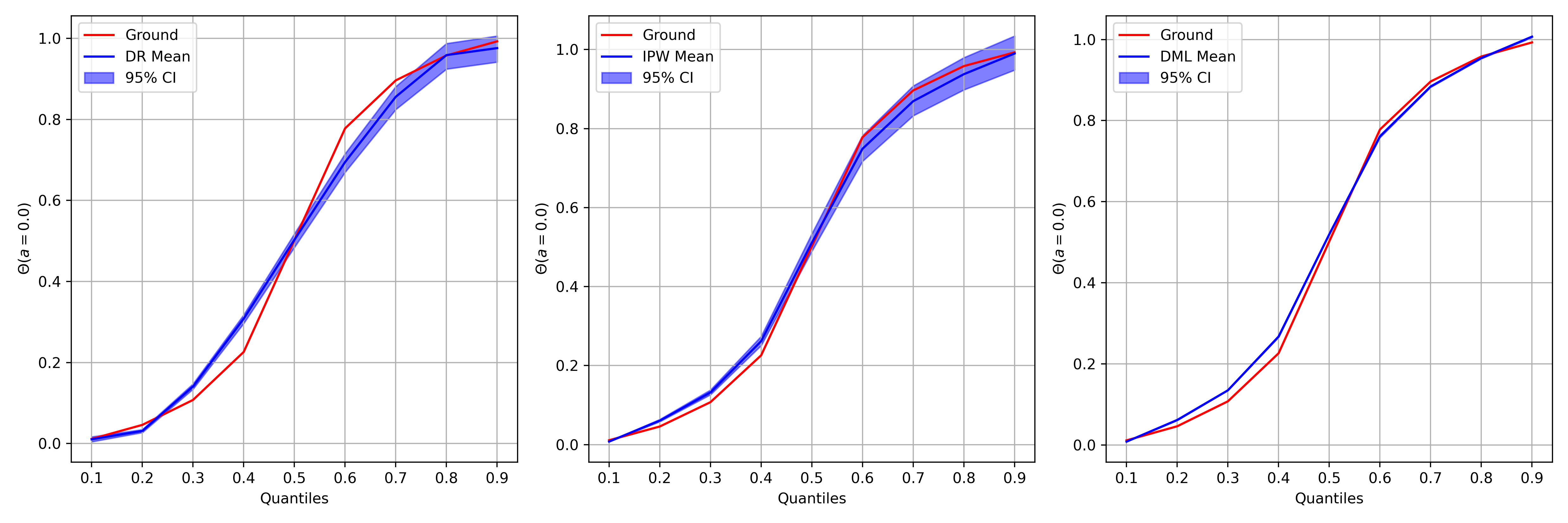}
\caption{The ground and estimated quantile function when A=0.0 based on Dist-DR (left), Dist-IPW (middle), and Dist-DML (right) Estimators\label{fig: numerical res}}
\end{figure}

\subsection{Comparison between Dist-DR, Dist-IPW, Dist-DML Estimators}
Table \ref{Tab: numerical res} presents the results of the numerical experiment, showing the efficacy of different estimators in recovering the distributions of potential outcomes at three treatment levels $A=-0.5, 0.0, 0.5$. The ground truth for the outcome distribution, derived from the DGP as specified in Eqn. \eqref{eqt: DGP}, is presented in the first row for each treatment level. The performance of the Dist-DR, Dist-IPW, and Dist-DML estimators in approximating this ground truth is then detailed, with the mean, standard deviation of the estimates, and the corresponding MAE provided. The results reveal that while all estimators demonstrate the ability to approximate the potential outcome distribution, the Dist-DML estimator stands out in terms of performance. Across all treatment levels, the Dist-DML estimator consistently achieves the lowest MAE, underscoring its robustness and precision. This superior performance aligns with theoretical expectations, as the Dist-DML estimator is designed to capitalize on the strengths of both the Dist-DR and Dist-IPW estimators, thereby enhancing its reliability and accuracy.
		
Figure \ref{fig: numerical res} complements this analysis by visually representing the ground truth, the estimated function, and the 95\% confidence interval, estimated over 100 experimental runs, when the treatment equals 0.0. This graphical depiction illustrates that the Dist-DR estimator tends to show a smaller variance but a larger bias, while the Dist-IPW estimator exhibits a larger variance but a smaller bias. The Dist-DML estimator, on the other hand, adeptly balances these aspects, manifesting both lower bias and lower variance. 
		
\subsection{Sensitivity Analysis}
		
In our theoretical exploration, it became apparent that the sample size and the bandwidth of the kernel function play a pivotal role in the convergence behavior of the Dist-DML estimator. In particular, an increase in the sample size tends to enhance the convergence speed of the estimator, suggesting that larger datasets can lead to more accurate and stable estimates. Meanwhile, a small bandwidth of the kernel function allows for a closer approximation to the Delta Dirac function, thereby potentially improving the precision of the estimator in capturing the true causal effect. Building on these theoretical insights, we further investigate the practical implications through simulation experiments.
\begin{figure}
\centering
\includegraphics[width=\columnwidth]{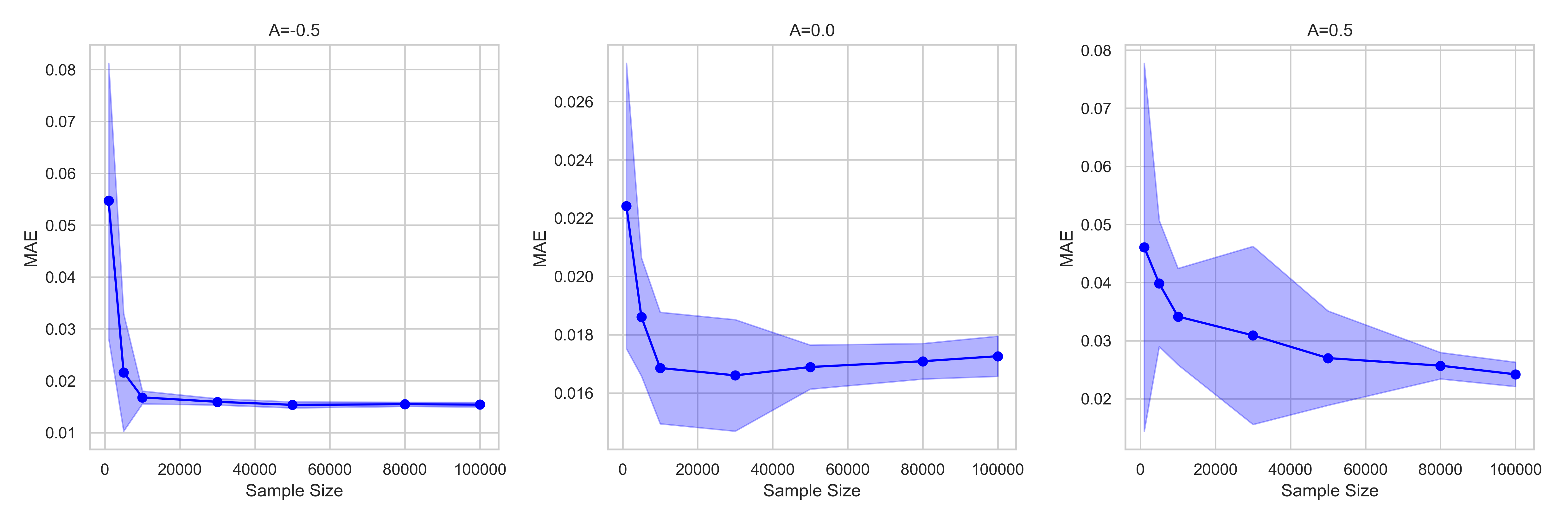}
\caption{The estimated MAE w.r.t various sample size for Dist-DML estimator.\label{fig: sample size}}
\end{figure}
	
\subsubsection{The Impact of Sample Size}
In this experiment, we fix the bandwidth to an optimal value and then adjust the sample size using the same data generation process, testing a range of 1000 to 100,000. Each experiment is conducted 100 times to ensure the reliability of the results. The results of these experiments are depicted in Figure \ref{fig: sample size}, which illustrates the estimated MAE with respect to different sample sizes for the Dist-DML estimator at three different treatment levels ($A=-0.5$, $A=0.0$, and $A=0.5$). Within each sub-plot, the blue line represents the mean MAE computed from the 100 repetitions, and the shaded area around the line indicates the standard deviation of the MAE across these repetitions. Each of the three plots corresponds to a specific treatment level and shows a blue line that represents the mean MAE derived from 100 repeated experiments, while the shaded area indicates the standard deviation of the MAE across these experiments.  The plots reveal that when the sample size is small (e.g., 1,000 or 5,000), the MAE is relatively larger, implying a less accurate estimation at all three treatment levels. However, as the sample size increases, there is a clear downward trend in the MAE, which suggests enhanced accuracy of the Dist-DML estimator. In addition, the reduction in the shaded area as the sample size grows indicates a decrease in the variance of the estimator, thus reflecting more robust and consistent results.

\subsubsection{The Impact of Bandwidth}
In this experiment, we maintain a constant sample size while varying the bandwidth of the kernel function. As stated in section \ref{sec:theory},  the optimal bandwidth, denoted as $h^\ast$, is determined for the Dist-DML estimator by minimizing $\int_{[0,1]}\big[h^{4}\hat{B}_{a}(t)^{2}+\frac{\hat{C}(t,t)}{Nh}\big]dt$. This selection process is detailed in the Appendix \ref{appendix:bandwidth selection}. We test a range of bandwidths that are multiples of $h^\ast$: specifically $\frac{h^\ast}{6}$, $\frac{h^\ast}{4}$, $\frac{h^\ast}{2}$, $h^\ast$, $2h^\ast$, $4h^\ast$, and $6h^\ast$.
		
\begin{figure}
\centering
\includegraphics[width=\columnwidth]{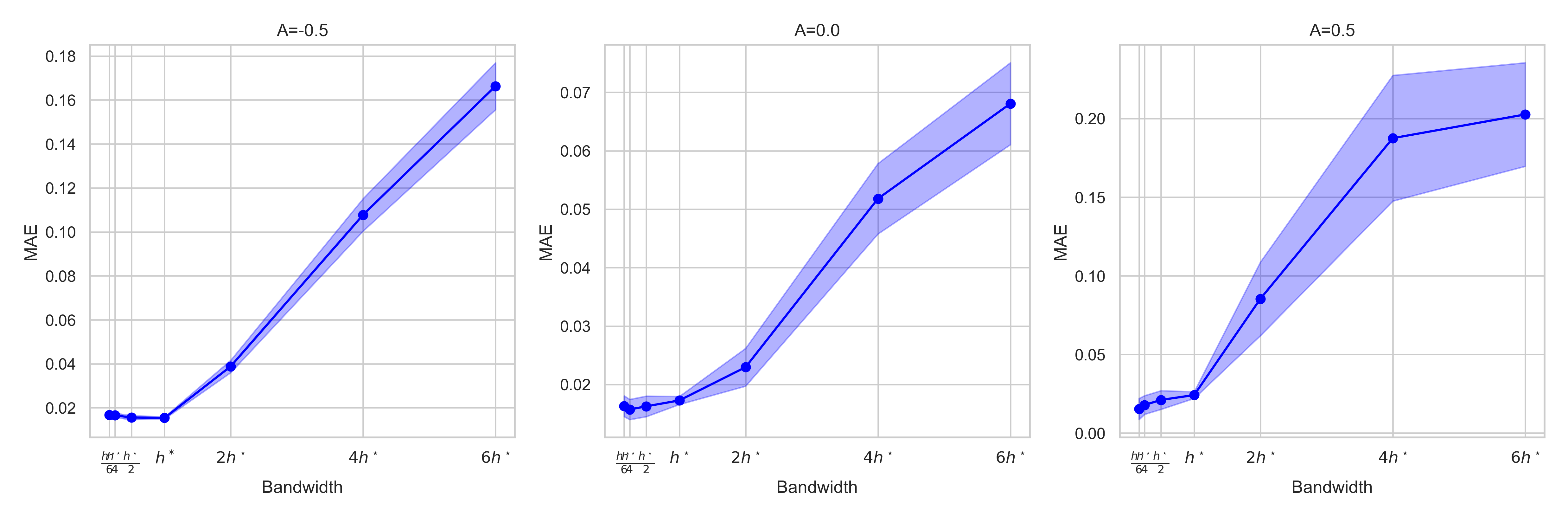}
\caption{The estimated MAE w.r.t the different bandwidth for Dist-DML estimator. }
\label{fig: bandwidth}
\end{figure}

The results, displayed in Figure \ref{fig: bandwidth}, illustrate the estimated MAE in relation to the varying bandwidths for the Dist-DML estimator. As the bandwidth narrows, the MAE decreases, suggesting increased precision in the estimations at the three treatment levels ($A=-0.5$, $A=0.0$ and $A=0.5$). This trend indicates that a more focused kernel function more accurately approximates the Delta Dirac function, leading to a more precise estimation of the Dist-ATE. The experiment also reveals that the smallest standard deviation in the estimated MAE occurs at the optimal bandwidth $h^\ast$, since it is selected when the covariance function is at its minimum.

\section{Empirical Application}
	
In recent times, the rapid advancement of financial technology (FinTech) has facilitated the proliferation of electronic platforms within the credit market, notably through the introduction of online consumer credit systems \citep{balyuk2023fintech}. These E-platforms, such as Taobao.com and JD.com, offer online marketplace services that enable consumers to make credit-based purchases without an immediate payment requirement. Simultaneously, by harnessing a comprehensive array of consumer data, including browsing, transaction, and credit records, these platforms leverage advanced machine learning algorithms to customize credit limits for individual users.
		
The capacity of E-commerce platforms to set differentiated credit limits for individual users raises a critical research question: how does adjustment in credit limits influence consumer spending behaviors? To investigate this problem, we employ our approach by using data collected from a leading and large E-commerce platform in China. The platform assigns unique credit limits to consumers based on a variety of factors, including their income, ages, genders, and historical behaviors such as shopping and credit records. The platform also offers a one-month interest-free loan for purchases, with the stipulation that the total loan amount must not exceed the user's credit limit. Our primary objective in this section is to analyze how alterations in credit limits influence the distribution of consumer spending, thereby contributing to a deeper understanding of consumer behavior in the context of the E-commerce platform. 
		
\begin{figure*}[t!]
    \centering
    \begin{subfigure}[t]{0.47\textwidth}
        \centering
        \includegraphics[scale=0.3]{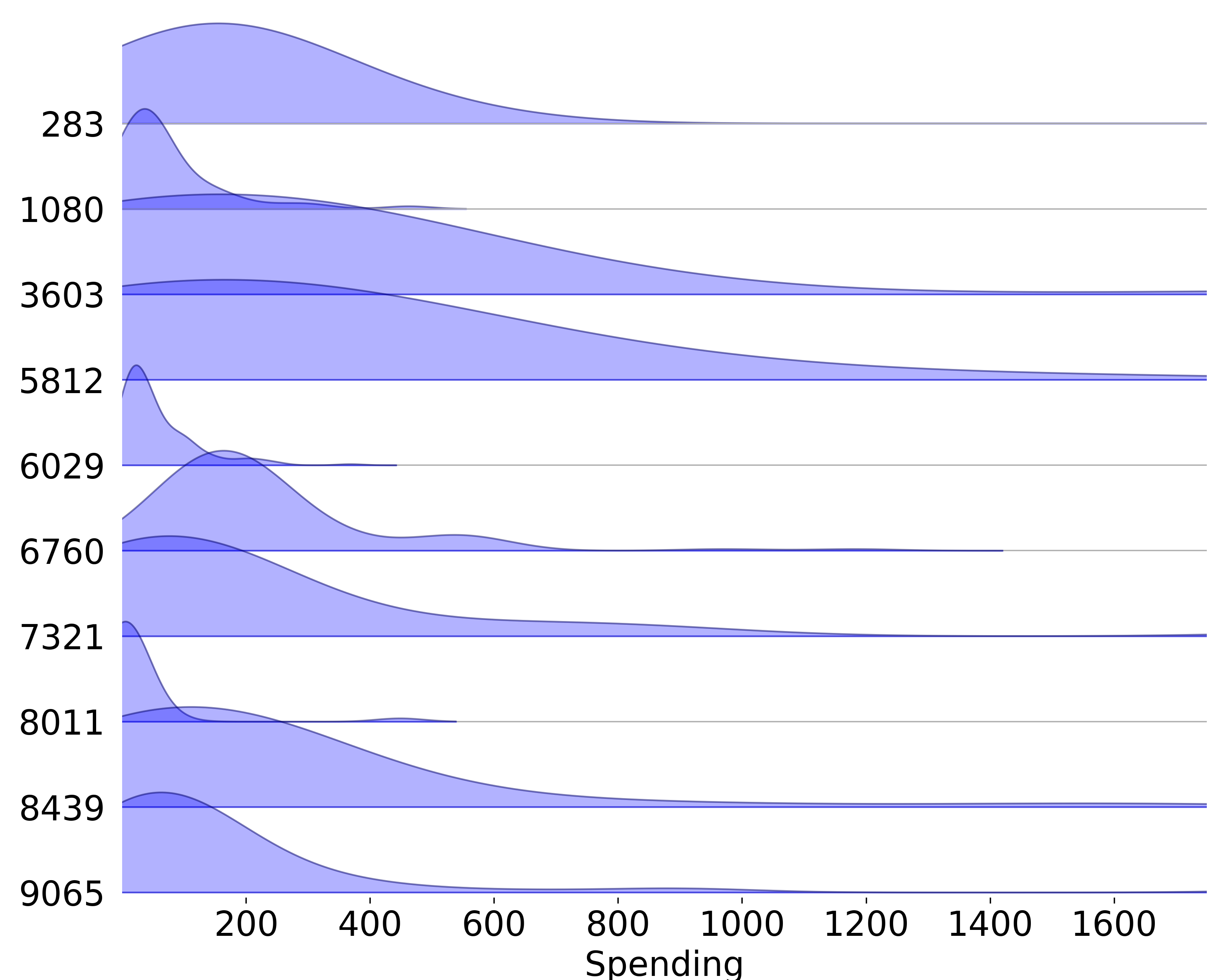}
        \caption{\captionsetup{size=small} The spending distributions of 10 consumers. \label{fig:instance_dist}}
    \end{subfigure}%
    ~ 
    \begin{subfigure}[t]{0.47\textwidth}
        \centering
        \includegraphics[scale=0.3]{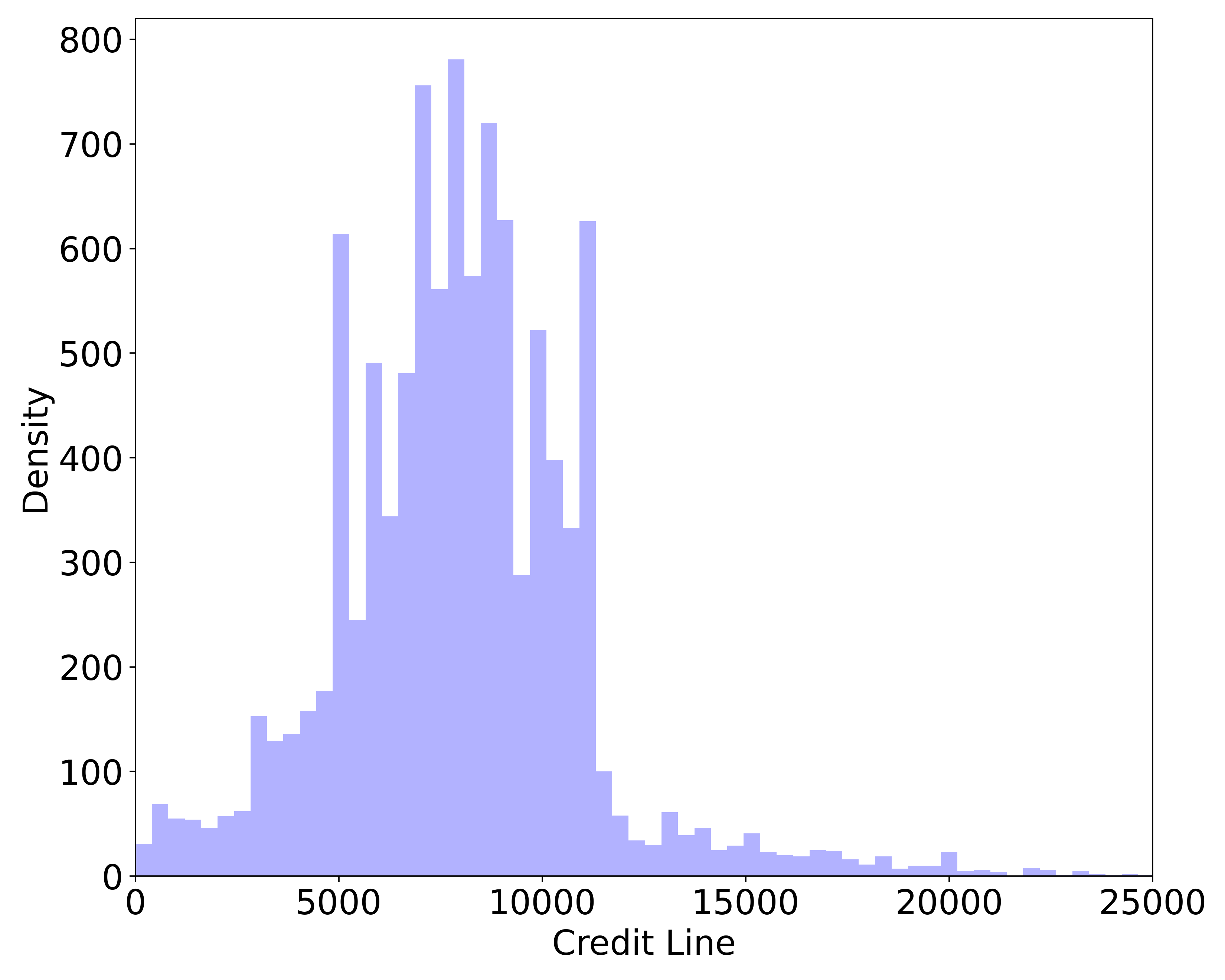}
        \caption{\captionsetup{size=small} The credit limit distribution of all consumers. \label{fig:credit_line}}
    \end{subfigure}
    \caption{The spending distribution and credit limit distribution.}
\end{figure*}

We randomly collect data from 10,220 consumers on the platform, spanning a 12-month period from January to December 2019, which encompasses a wide range of information, including demographic details such as gender, age, and geographic location, alongside comprehensive records of shopping and financial behavior. The shopping records include measures such as the total number of orders, pricing, discounts availed, payments for each order, etc. The financial records encompass the credit limits for each consumer, the credit status, and the borrowings, repayments, and refunds for each loan. Data from the first half of the year (January to June 2019) are used to construct the covariates. Subsequently, the impact of credit limits on spending distribution is analyzed using data from the latter half (July to December 2019). The spending distribution for each consumer is represented by all payments in individual orders during this period. Figure \eqref{fig:instance_dist} displays the spending distributions of ten randomly selected consumers, suggesting that the spending distribution varies greatly between consumers. Furthermore, Figure \eqref{fig:credit_line} presents the distribution of credit limits assigned to all users on the platform, revealing that the majority of users are assigned credit limits of around 8,000, while a small proportion of users are assigned higher credit limits, resulting in a skewed long-tailed distribution. For a detailed understanding of these variables and their distribution, Table \ref{tab:statistical descriptions} provides a statistical summary for some key variables.

\begin{table*}[!ht]
			\centering
			\caption{The statistical description for important variables.}
			\label{tab:statistical descriptions}
			\begin{tabular}{ccccccc}
				\toprule
				Category 	& Variables 					& mean 		& std 		& 25\% 		& 50\% 		& 75\%  \\ \midrule
				~			& age 							& 33.23 	& 6.86 		& 28 		& 32 		& 37  \\ 
				~ 			& gender 						& 0.64 		& 0.48 		& 0 		& 1 		& 1  \\ 
				~ 			& platform usage days  			& 2375.69 	& 460.11 	& 2094.75 	& 2291 		& 2499  \\ 
				~ 			& credit usage days 			& 1246.96 	& 325.53 	& 1003 		& 1147 		& 1466  \\ 
				~ 			& num of orders 				& 54.35 	& 28.96 	& 37 		& 46 		& 61  \\ 
				~ 			& num of products 				& 106.90 	& 66.26 	& 65 		& 89 		& 127  \\ 
				Covariates 	& averaged order price 			& 399.53 	& 336.27 	& 235.36 	& 338.50 	& 482.13  \\ 
				~ 			& averaged discount price 		& 111.73 	& 158.36 	& 62.43 	& 91.49 	& 134.03  \\ 
				~			& num of credit usage 			& 21.94 	& 26.59 	& 4 		& 14 		& 32  \\ 
				~ 			& amount of credit usage 		& 428.16 	& 785.81 	& 114.35 	& 211.00 	& 414.80  \\ 
				~ 			& num of credit repayment 		& 5.25 	    & 3.79   	& 3 		& 5 		& 6  \\ 
				~	 		& amount of credit repayment 	& 887.00 	& 1023.90 	& 243.80 	& 588.29 	& 1168.93  \\ \midrule
				Treatment 	& credit limit 					& 8042.93 	& 3184.64 	& 6161.63 	& 8000		& 9800  \\  \midrule
				~ 			& spending (Q=0.1)				& 29.84 	& 22.35 	& 12.98 	& 28.70 	& 41.79  \\ 
				~			& spending (Q=0.3)				& 62.15 	& 37.15 	& 35.9	 	& 59.9 		& 88.98  \\ 
				Outcome		& spending (Q=0.5)				& 104.1 	& 57.54 	& 68.99 	& 100.97 	& 130.00  \\ 
				~			& spending (Q=0.7)				& 180.94 	& 109.74 	& 109.90 	& 163.69	& 221.98  \\ 
				~			& spending (Q=0.9)				& 487.08 	& 401.06 	& 244	 	& 377.56	& 599.08  \\ \bottomrule
			\end{tabular}
\end{table*}
Specifically, the demographic profile of users on the E-commerce platform is relatively young, with an average age of 33.23 years, and is predominantly male, comprising 64\% of the consumer population. These individuals show strong loyalty and engagement, as evidenced by an average duration of platform use of nearly 2,375 days and an average involvement with credit services of approximately 1,246 days. Regarding purchasing behavior, consumers place an average of 54.35 orders comprising 106.9 products in the first half of the year. The average order value is 399.5, typically before applying an average discount of 111.73. From a financial perspective, the utilization of credit services is frequent among consumers, averaging 21.9 borrows with an average loan amount of 428.16. In addition, consumers often repay multiple loans concurrently, reflecting from the observations that the average number of credit repayments per consumer is 5.25, leading to an average total repayment amount of 887. The distribution of credit limits, with an average of 8,042 and a standard deviation of 3,184, reveals a considerable range in the credit capacity allocated to different consumers. The spending behavior of each consumer, as the outcome variable, is conceptualized as a distribution. We focus on the quantiles of these distributions, providing a detailed representation of spending patterns. Specifically, the average expenditures at the quantiles of 0.1, 0.3, 0.5, 0.7, and 0.9 are observed to be 29.84, 62.15, 104.1, 180.94, and 487.08, respectively.
\begin{figure}
\centering
\includegraphics[width=\columnwidth]{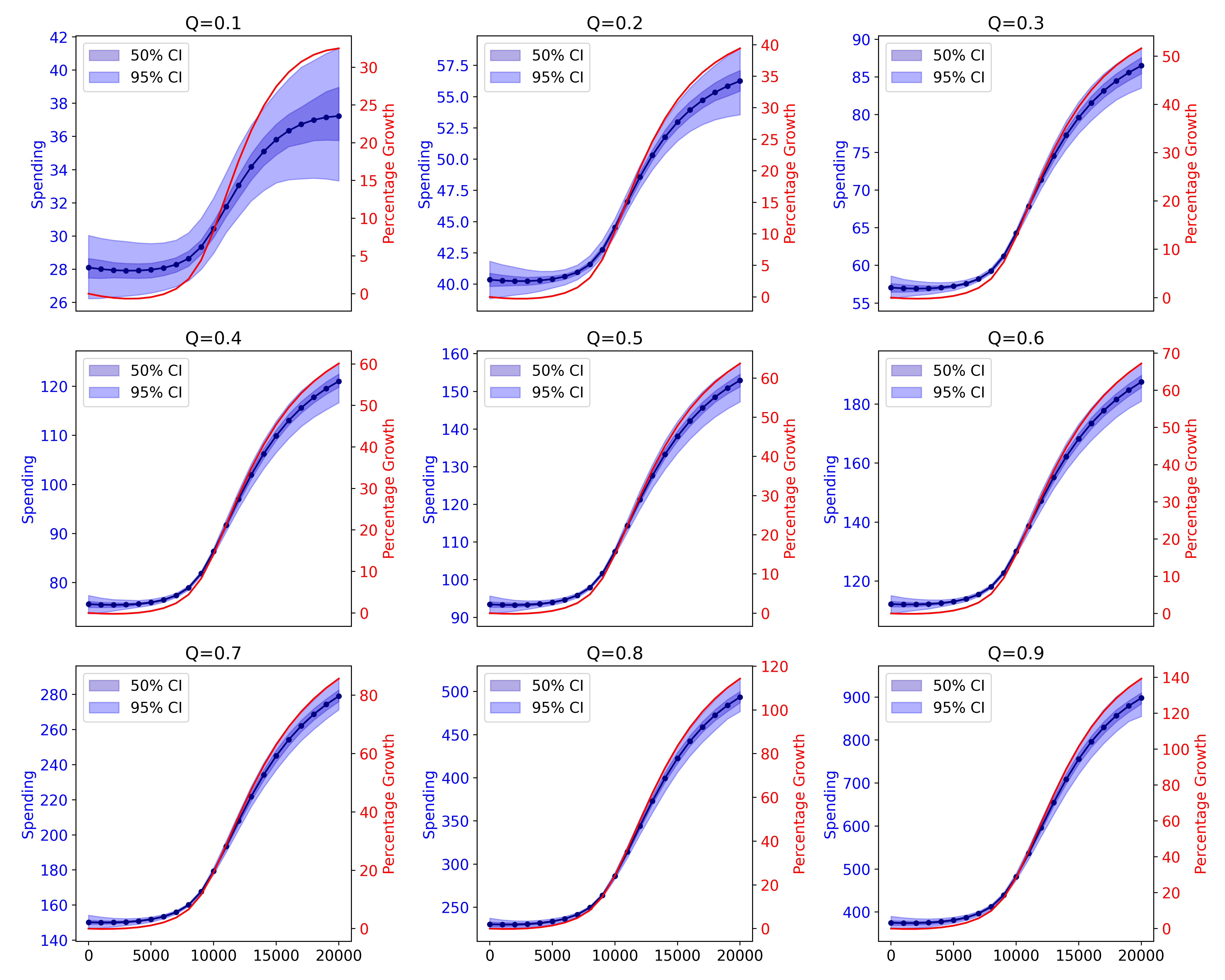}
\caption{The potential spending distribution outcome across credit limits from 0 to 20,000 \label{fig:JD res}}
\end{figure}
		
In line with our numerical experiment, we approximate the integral loss by discretizing it across 9 quantiles, ranging from 0.1 to 0.9. In this study, we explore potential shifts in spending distributions in response to a range of credit limits, extending from 0 to 20,000, incremented in steps of 1,000 (i.e., 0, 1,000, 2,000, $\cdots$, 20,000).  To ensure the reliability and robustness of our findings, we undertake multiple iterations of model training, repeating the process 100 times with both the NFR Net and the CNF Net. The results of our empirical experiments are visualized in Figure \ref{fig:JD res}. In these figures, each subfigure delineates the average potential spending at each quantile level (from $Q=0.1$ to $Q=0.9$) and its corresponding 95\% confidence interval across various credit limits. Generally, lower quantiles (e.g., $Q=0.1$) typically represent smaller amounts of expenditure, often associated with essential daily purchases like necessities. In contrast, the higher quantiles (e.g., $Q=0.9$) reflect larger spending amounts, usually indicative of discretionary purchases such as luxury items or services. 
		
In line with previous studies, our results demonstrate a positive correlation between credit limits and consumer spending, underscoring the role of credit as a catalyst for consumption \citep{aydin2022consumption}. Our analysis reveals a heterogeneous effect across different spending quantiles. In particular, as credit limits increase, we observe a substantial increase in spending at higher quantiles. For example, at the 0.9 quantile, spending increases from 375.1 to 897.9 with an increase in the credit line from 0 to 20,000, marking a growth of approximately 139\%. In contrast, spending at lower quantiles shows relatively modest growth. Specifically, at the 0.1 quantile, spending increases from 28.1 to 37.2 over the same range of increase in credit lines, reflecting growth of only 32\%. This trend suggests that consumers tend to disproportionately allocate additional credit toward the purchase of higher-priced items or services rather than distributing the credit uniformly across various spending categories. This discovery has practical implications for platforms considering increases in consumer credit limits. Specifically, by recommending higher-end products in conjunction with increased credit limits, platforms might tap into a market segment previously unexplored by consumers due to budget constraints.

\section{Conclusion} \label{sec:Conclusion}
In this paper, we tackle a significant challenge in the realm of causal inference: how to effectively estimate treatment-outcome relationships when the outcome of each individual is represented as distributions and the treatments are continuous. Our proposed causal inference framework utilizes the Wasserstein space that captures the underlying geometry of distributional outcomes, offering a more detailed understanding of complex behavioral patterns. 
		
We introduce two novel causal quantities, the Dist-APO and the Dist-ATE, designed specifically for the complexities inherent in distributional data. To accurately estimate these quantities, we have developed a machine learning-based robust estimator: the Dist-IPW estimator. Its statistical asymptotic properties have been rigorously established, laying a strong theoretical foundation for application.
		
To ensure a precise estimation of the necessary nuisance parameters in these estimators, we have developed a deep learning model comprising two main components: the NFR Net and the CNF Net. The NFR Net is highly effective in modeling complex, non-linear relationships, while the CNF Net excels in accurately estimating generalized propensity scores. Their combination provides a robust tool for handling high-dimensional data and intricate interactions among variables.
		
Through comprehensive numerical studies, we have demonstrated the effectiveness of our proposed Dist-DML estimator. In applying our approach to real-world data, we explored the causal effects of credit limit adjustments on consumer spending distributions. Our findings provide critical insights into consumer behavior, revealing a tendency to allocate increased credit towards purchasing more expensive items rather than uniformly increasing spending across all items. This behavior offers essential perspectives on consumer spending in response to changes in credit policy, contributing valuable knowledge to the financial and marketing sectors.


\bibliographystyle{apalike}

\bibliography{MNSC_Addition_sample-left} 

\appendix
\clearpage
\section{Causal assumptions} \label{appendix:causal assumptions}
\begin{description}
\item[\textbf{SUTVA}] Assumption \ref{ass:SUTVA}.\ref{SUTVA1} assures that the potential outcome of an individual is due to the level of treatment the individual receives, but not the assignment of treatment to other individuals. Assumption \ref{ass:SUTVA}.\ref{SUTVA2} ensures that each treatment level should be clearly characterized. Consider the case in which we are interested in the effects of taking Aspirin. If the treatment variable is binary (taking Aspirin or not), then every patient who takes Aspirin should take the same dose and the same type of Aspirin. 
\item[\textbf{Consistency}] It assures that the observed outcome is due to the assigned intervention that allows us to examine the target quantities from the observable data.
\item[\textbf{Ignorability/Unconfoundness}] It has two meanings. First, if two individuals have the same $\mathbf{X}$, then the joint distributions $\dutchcal{Y}(a)$ conditioning on the covariates $\mathbf{X}$ and the treatment assignment of the two individuals are the same. Second, if two individuals have the same $\mathbf{X}$, then the treatment assignment mechanism should be the same.
\item[\textbf{Overlap/Positivity}] It assures that every available combination of treatment and covariate levels has a positive density.
\end{description}

\section{Differences between the Wasserstein Mean and Euclidean Mean} \label{space comparison}
To further illustrate the superiority of the Wasserstein space in preserving the structural properties of distributions during operations, we present two examples in this section. Specifically, we examine the averaging of finite samples of probability distributions in the following contexts: (1) Gaussian samples; and (2) Exponential samples.

\textbf{Example 1: Gaussian samples}: Consider the case in which $\dutchcal{Y}$ is a random variable such that each realization is a normal distribution $\mathcal{N}(\lambda,1)$, where $\lambda$ follows a uniform distribution $\mathcal{U}(a,1+a)$. Denote $f_{\lambda}(u)$ as the density function of $\lambda$. 

The Euclidean mean is the point-wise average of all the distributions, which is equivalent to averaging all the probability density functions. We therefore have
\begin{equation*}
	\begin{aligned}
		&\frac{1}{1+a-a}\int_{a}^{1+a}\frac{1}{\sqrt{2\pi}}e^{-\frac{(x-u)^{2}}{2}}f_{\lambda}(u)du \\
		&=\int_{-\infty}^{1+a-x}\frac{1}{\sqrt{2\pi}}e^{-\frac{y^{2}}{2}}dy-\int_{-\infty}^{a-x}\frac{1}{\sqrt{2\pi}}e^{-\frac{y^{2}}{2}}du\\
		&=\Phi(1+a-x)-\Phi(a-x),
	\end{aligned}
\end{equation*}
where $\Phi(x)=\int_{-\infty}^{x}\frac{1}{\sqrt{2\pi}}e^{-\frac{y^{2}}{2}}dy$ and $\Phi(1+a-x)-\Phi(a-x)$ is a density function.

The Wasserstein mean is a distribution $\bar{\dutchcal{Y}}=\underset{v\in\mathcal{W}_{2}(\mathcal{I})}{\arg\min}\;\mathbb{E}_{\mathcal{P}}[\mathbb{D}_{2}(\dutchcal{Y}\|v)^{2}]$. In the given example, the distribution that makes $\mathbb{E}_{\mathcal{P}}[\mathbb{D}_{2}(\dutchcal{Y}\|v)^{2}]$ the smallest is the Gaussian distribution $\mathcal{N}\big(a+\frac{1}{2},1\big)$ which has the probability density function $\frac{1}{\sqrt{2\pi}}e^{-\frac{(x-a-\frac{1}{2})^{2}}{2}}$.

\textbf{Example 2: Exponential samples}: 
Consider the case in which $\dutchcal{Y}$ is a random variable such that each realization is an exponential distribution $\operatorname{Exp}(\lambda)$, where $\lambda$ is a random variable that follows a uniform distribution $\mathcal{U}(a,1+a)$ and $a>0$.

The point-wise average of all the distributions is a density function
\begin{equation*}
	\begin{aligned}
		&\frac{1}{1+a-a}\int_{a}^{1+a}ue^{-ux}f_{\lambda}(u)du\\
		&=-\frac{1}{x}\bigg[(1+a)e^{-(1+a)x}-ae^{-ax}\bigg]-\frac{1}{x^{2}}e^{-ux}\bigg|_{a}^{1+a}\\
		&=\frac{(1+ax)e^{-ax}-(1+(1+a)x)e^{-(1+a)x}}{x^{2}}. 
	\end{aligned}
\end{equation*}
The distribution that minimizes $\mathbb{E}_{\mathcal{P}}[\mathbb{D}_{2}(\dutchcal{Y}\|v)^{2}]$ is the exponential distribution with rate $\mu$, such that the mean of this distribution, $\frac{1}{\mu}$, should equal the mean of all exponential distributions $\operatorname{Exp}(\lambda)$ ($\lambda \sim \mathcal{U}(a,1+a)$) which equals $\int_{a}^{1+a}\frac{1}{\lambda}d\lambda=\ln\big(\frac{1+a}{a}\big)$. Thus, we have $\frac{1}{\mu}=\ln\left(\frac{1+a}{a}\right)$, implying that $\mu=\frac{1}{\ln\left(\frac{1+a}{a}\right)}$. Therefore, the probability density function of the Wasserstein mean is $\frac{1}{\ln\big(\frac{1+a}{a}\big)}\exp\bigg(-\frac{x}{\ln\big(\frac{1+a}{a}\big)}\bigg)$.

\section{Differences between Distributional/Scalar Outcome Framework}\label{comparison}
The distributional outcome causal framework represents a significant advancement in the field of causal inference, particularly by addressing scenarios where the outcome for each individual is a distribution, as opposed to a scalar value. This distinction is crucial because it allows for a more comprehensive analysis of causal effects in complex data.
			
\begin{figure*}
\centering
\includegraphics[scale=0.38]{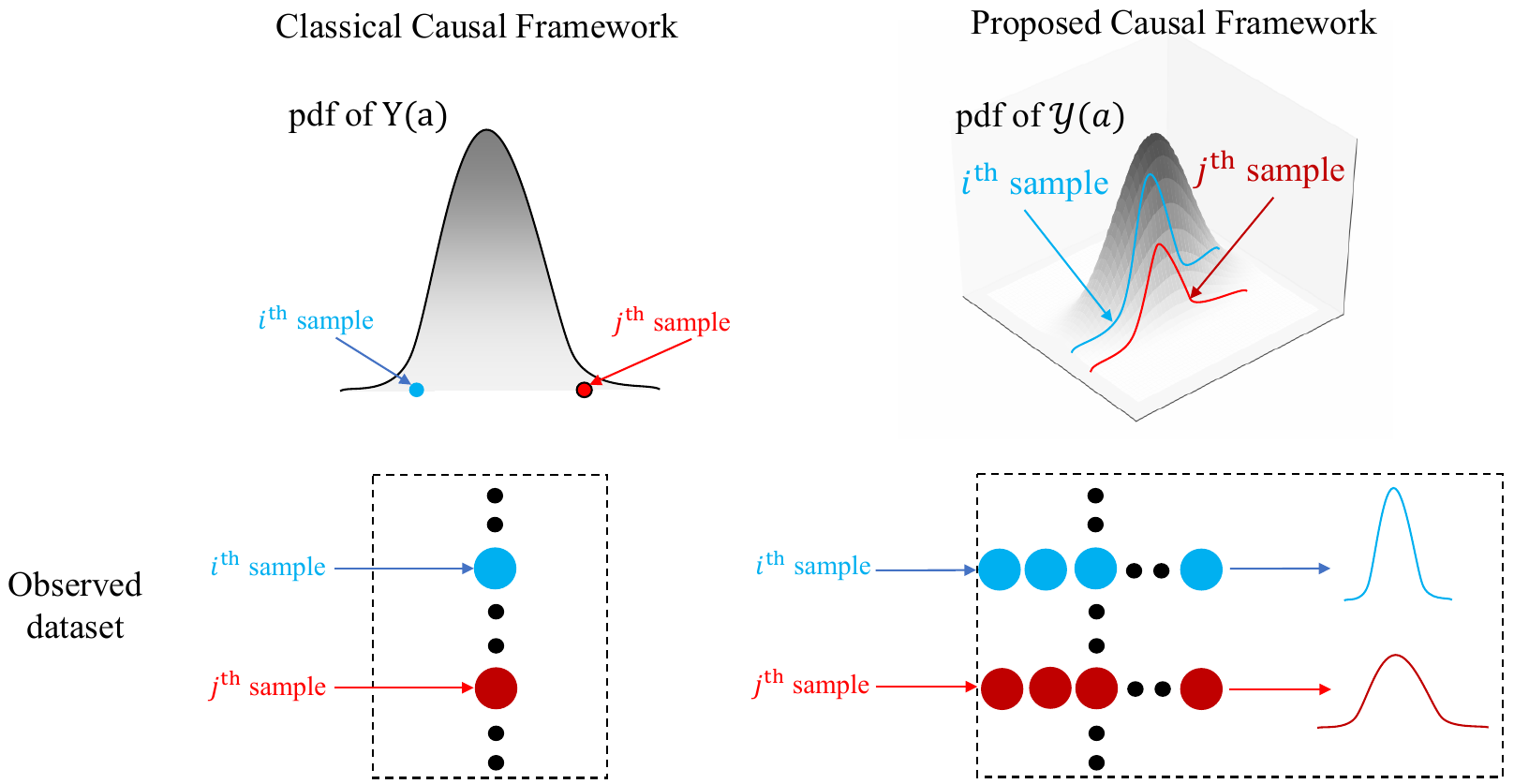}
\caption{Comparisons between the scalar outcome framework and distributional outcome framework. When the outcome is a scalar, the observed dataset contains a finite number of points. Each point represents a realization of one unit. When the outcome is a distribution, the observed dataset contains a finite number of collections. Each collection contains finitely many points, and each collection is a realization of one unit. \label{fig:comparisons_frameworks}}
\end{figure*}

We perform a comprehensive comparison between the scalar outcome framework and the distributional outcome framework in Table \ref{table:Comparisons_main} and Figure \ref{fig:comparisons_frameworks}. To distinguish the differences when the implementation of the outcome variable is a scalar, we use $\mathrm{Y}$, $\mathrm{Y}(a)$, $\mathbb{P}(\cdot)$ and $\mathbb{P}_{a}(\cdot)$ to represent the outcome, the outcome when treatment $A=a$, the probability measure of $\mathrm{Y}$, and the probability measure of $\mathrm{Y}(a)$, respectively. Specifically, the main differences can be summarized as three main points.
\begin{itemize}
\item  \textbf{Outcome/Potential outcome variable}. In scalar outcome frameworks, the potential outcome variable $\mathrm{Y}(a)$ for a given treatment $A=a$ is represented as a single scalar value. A scalar value (or a realization of $\mathrm{Y}(a)$) is drawn from a potential outcome distribution $\mathbb{P}_{a}(\cdot)$. For example, if we consider $\mathbb{P}_{a}(\cdot)$ to follow a normal distribution $\mathcal{N}(0, 1)$, then any realization of $\mathrm{Y}(a)$ would be a single point sampled from this normal distribution. On the other hand, the distributional outcome framework considers the potential outcome variable $\dutchcal{Y}(a)$ as a distribution in itself, rather than a single scalar value. This distribution is sampled from a high-dimensional potential outcome distribution $\mathcal{P}_a(\cdot)$. In this context, a realization of $\dutchcal{Y}(a)$ is an entire normal distribution, say $\mathcal{N}(\mu, \sigma^2)$. The parameters of this distribution, $(\mu,\log\sigma)$ in this case, might be drawn from a distribution, say $\mathcal{N}(0,1)$. Consequently, a single sample in our framework is not just a point but a collection of points. For example, an instance may obtain a collection of points drawn from $\mathcal{N}(0.5,0.25)$ where $(\mu,\log\sigma)$ is drawn from $\mathcal{N}(0,1)$ and equals $(0.5,\log 0.5)$, while another instance may obtain a collection of points drawn from $\mathcal{N}(0.3,0.16)$ where $(\mu,\log\sigma)$ is drawn from $\mathcal{N}(0,1)$ and equals $(0.3,\log 0.4)$. This conceptual shift is visually depicted in Figure \ref{fig:comparisons_frameworks}.
\item \textbf{Ambient space of outcome variable} $\mathbf{(\Omega)}$. In the scalar outcome framework, the outcomes are typically scalar values located within the ambient space of the Euclidean space, denoted as $\mathbb{R}$. However, our proposed framework considers the responses as distributions rather than scalar values. In this context, the ambient space for these outcomes is not the Euclidean space, but rather the Wasserstein space of distributions over a set $\mathcal{I}$, symbolized as $\mathcal{W}_{2}(\mathcal{I})$.
\item \textbf{Target quantity}. In the scalar outcome framework, the essential components are $\mathbb{E}_{\mathbb{P}_{a}}[\mathrm{Y}(a)]$ or $\mathbb{Q}_{\mathbb{P}_{a}}[\alpha, \mathrm{Y}(a)]$, representing the expected value or the $\alpha$-quantile value of the response when all individuals in a population receive a specific treatment $a$ respectively. The difference between $\mathbb{E}_{\mathbb{P}_{a}}[\mathrm{Y}(a)]$ and $\mathbb{E}_{\mathbb{P}_{a^{\prime}}}[\mathrm{Y}(a^{\prime})]$ (or $\mathbb{Q}_{\mathbb{P}_{a}}[\alpha, \mathrm{Y}(a)]$ and $\mathbb{Q}_{\mathbb{P}_{a^{\prime}}}[\alpha, \mathrm{Y}(a^{\prime})]$), where $a^\prime$ represents an alternative treatment, is often used to quantify ATE and QTE in the population. In contrast, in the distributional outcome, the outcome of each individual is characterized as a distribution. The key component in this framework is $\Theta(a)$, which represents the Dist-APO (i.e., quantile function of the barycenter) in the Wasserstein space $\mathcal{W}_{2}(\mathcal{I})$ when assuming that all individuals receive treatment $A=a$. Furthermore, the difference between $\Theta(a)$ and $\Theta(a^{\prime})$, denoted as $\Theta(aa^{\prime})$, is referred to as the quantile differences of Dist-ATE. This measure captures the variation in treatment effects in different quantiles of the outcome distribution, providing a detailed understanding of how treatment effects vary across the spectrum of potential outcome distributions. 
\end{itemize}

\section{Dist-DR and Dist-IPW estimator}
\label{appendix:Dist-DR and Dist-IPW}
	\subsection{Dist-DR} 
    
The core concept of the Dist-DR form involves treating the distributional outcome variable as a functional response and modeling a functional relationship among the outcome, the treatment, and the covariates. Based on Assumptions \ref{ass:Consistency} and \ref{ass:Ignorability}, the Dist-DR form can be derived as follows
{\small
\begin{align}
&\Theta(a)=\mathbb{E}_{\mathcal{P}_{a}}[\dutchcal{Y}(a)^{-1}]=\mathbb{E}_{\mathbb{P}(\mathbf{X})}[\mathbb{E}_{\mathcal{P}_{a}|\mathbb{P}(\mathbf{X})}[\dutchcal{Y}(a)^{-1}|\mathbf{X}]]\nonumber\\ 
&\overset{\ast}{=}\mathbb{E}_{\mathbb{P}(\mathbf{X})}[\mathbb{E}_{\mathcal{P}_{a}|\mathbb{P}(\mathbf{X})}[\dutchcal{Y}(a)^{-1}|A=a,\mathbf{X}]]\nonumber\\
&\overset{\star}{=}\mathbb{E}_{\mathbb{P}(\mathbf{X})}[\mathbb{E}_{\mathcal{P}|\mathbb{P}(\mathbf{X})}[\dutchcal{Y}^{-1}|A=a,\mathbf{X}]]\coloneqq \mathbb{E}_{\mathbb{P}(\mathbf{X})}[m(a;\mathbf{X})].\label{eqt:DR expectation form}
\end{align}
}\noindent
Here, $\star$ follows from Assumption \ref{ass:Consistency}, while $\ast$ follows from Assumption \ref{ass:Ignorability}. This form only requires estimating $m(a;\mathbf{X})=\mathbb{E}_{\mathcal{P}|\mathbb{P}(\mathbf{X})}[\dutchcal{Y}^{-1}|A=a,\mathbf{X}]$ from the observed data using an appropriate regression model. We can then construct the Dist-DR estimator, termed $\hat{\Theta}^{DR}(a)$, according to the Dist-DR form given in Eqn. \eqref{eqt:DR expectation form}. This estimator is derived by averaging all $N$ individuals, depending on the regression of the distributional outcome $\dutchcal{Y}^{-1}$ on the treatment and covariate variables $(A,\mathbf{X})$. The explicit formulation of the Dist-DR estimator is encapsulated in the following equation:
{\small
\begin{align}\label{eqt:DR estimator}
\hat{\Theta}^{DR}(a)=\frac{1}{N}\underset{i=1}{\overset{N}{\sum}}m(a;\mathbf{X}_{i}).
\end{align}
}\noindent
However, a potential limitation of the Dist-DR form is that it overlooks the potential influence of the covariates $\mathbf{X}$ on the treatment variable $A$. Thus, the corresponding estimator is highly dependent on the accurate estimation of the regression function. The results could be biased if the functional relationship between variables is misspecified. As such, we consider to express $\mathbb{E}_{\mathcal{P}_{a}}[\dutchcal{Y}^{-1}(a)]$ in other forms.

\subsection{Dist-IPW} 
    The Dist-IPW form is an alternative approach to estimate the Dist-APO $\Theta(a)$ according to the Horvitz–Thompson Theorem \citep{horvitz1952generalization, overton1995horvitz}. The essence of the Dist-IPW form lies in the creation of a pseudo-population from the observed dataset by assigning specific weights to each unit. These weights are strategically designed to balance the representation of various groups within the dataset, mirroring the conditions of an RCT. In this pseudo-population, groups with a smaller portion in the observed dataset are assigned larger weights, while groups with a larger portion receive smaller weights. The calculation of these weights involves the use of (generalized) propensity scores that quantify the likelihood that an individual will receive a particular treatment based on its covariates. The formulation of the Dist-IPW form is presented in Proposition \ref{prop:IPW expectation form}.
	
	\begin{proposition}\label{prop:IPW expectation form}
		Given that Assumptions \ref{ass:SUTVA} - \ref{ass:Positivity} hold, 
		{\small
			\begin{align}
				\Theta(a)=\mathbb{E}_{(\mathbb{P}(A),\mathbb{P}(\mathbf{X}),\mathcal{P})}\bigg[\frac{\delta(A-a)}{p(a|\mathbf{X})}\dutchcal{Y}^{-1}\bigg].\label{eqt:IPW expectation form}
			\end{align}
		}
		
	\end{proposition}
	
The detailed proof of the Dist-IPW form is provided in Appendix \ref{appendix:proofs of IPW expectation form}. A primary advantage of Dist-IPW is that it does not require modeling the distributional outcome as a function of treatment and covariates. Instead, it focuses on modeling the process of treatment assignment, which can offer greater robustness against model misspecification compared to the Dist-DR form. However, the Dist-IPW form can be susceptible to issues of high variance. This situation typically arises in instances where certain subjects in the study have exceptionally low or high propensity scores. Such extremities in propensity scores result in the assignment of extreme weights to these subjects in the pseudo-population. The consequence of these extreme weights is an increased variance in the estimates derived from the Dist-IPW form. 

The construction of estimators based on the Dist-IPW form (i.e., Eqn. \eqref{eqt:IPW expectation form}), denoted as $\hat{\Theta}^{IPW}(a)$, is thus formulated by sample averaging:
{\small
\begin{equation}\label{eqt:continuoous IPW estimator}
\begin{aligned}
\hat{\Theta}^{IPW}(a)=\frac{1}{N}\underset{i=1}{\overset{N}{\sum}}\frac{K_{h}(A_{i}-a)}{p(a|\mathbf{X}_{i})}\dutchcal{Y}_{i}^{-1}.
\end{aligned}
\end{equation}
}\noindent

\section{Kernel Functions}\label{appendix:tables and figures}
Table \ref{tab:kernel functions} summarizes the common kernel functions of order 2 that exist in the literature.
\begin{table}[h]
\centering
\caption{Some common kernel functions of order $2$ that exist in the literature \label{tab:kernel functions}}
\begin{tabular}{ccccccccccccccccc}
& \multicolumn{3}{c}{Kernel Function $K(u)$} & \multicolumn{3}{c}{Support} \\
\midrule
\multicolumn{1}{l|}{Uniform} & $K(u)=\frac{1}{2}$ & $|u|\leq 1$\\
\midrule
\multicolumn{1}{l|}{Triangular} & $K(u)=(1-|u|)$ & $|u|\leq 1$\\
\midrule
\multicolumn{1}{l|}{Epanechnikov} & $K(u)=\frac{3}{4}(1-u^{2})$ & $|u|\leq 1$\\
\midrule
\multicolumn{1}{l|}{Quartic} & $K(u)=\frac{15}{16}(1-u^{2})^{2}$ & $|u|\leq 1$\\
\midrule
\multicolumn{1}{l|}{Triweight} & $K(u)=\frac{35}{32}(1-u^{2})^{3}$ & $|u|\leq 1$\\
\midrule
\multicolumn{1}{l|}{Tricube} & $K(u)=\frac{70}{81}(1-|u|^{3})^{3}$ & $|u|\leq 1$\\
\midrule
\multicolumn{1}{l|}{Gaussian} & $K(u)=\frac{1}{\sqrt{2\pi}}e^{-\frac{u^{2}}{2}}$ & $u\in\mathbb{R}$\\
\midrule
\multicolumn{1}{l|}{Cosine} & $K(u)=\frac{\pi}{4}\cos\big(\frac{\pi}{2}u\big)$ & $|u|\leq 1$\\
\midrule
\multicolumn{1}{l|}{Logistic} & $K(u)=\frac{1}{e^{u}+2+e^{-u}}$ & $u\in\mathbb{R}$\\
\midrule
\multicolumn{1}{l|}{Sigmoid} & $K(u)=\frac{2}{\pi}\frac{1}{e^{u}+e^{-u}}$ & $u\in\mathbb{R}$\\
\bottomrule
\end{tabular}
\end{table}%
			
\section{Proofs of Theorems, Propositions, and Corollaries}\label{appendix:proofs of theorems and propositions}
\subsection{Proofs of Proposition \ref{prop:no optimization}}\label{appendix:proofs of no optimization}
The proof requires Theorem 2.18 of \cite{villani2021topics}. We state the theorem here:
\begin{theorem}
Let $\lambda_{1}(\cdot)$ and $\lambda_{2}(\cdot)$ be two cumulative distribution functions defined on $\mathcal{I}\subseteq\mathbb{R}$ of variables $V_{1}$ and $V_{2}$ respectively. Let $\bar{\lambda}$ be the joint cumulative distribution function such that
\begin{equation*}
\begin{aligned}
\bar{\lambda}(s,t)=\min\{\lambda_{1}(s),\lambda_{2}(t)\},\quad (s,t)\in\mathcal{I}\times\mathcal{I}.
\end{aligned}
\end{equation*}
Then $\bar{\lambda}\in\Lambda(\lambda_{1},\lambda_{2})$ ($\Lambda(\lambda_{1},\lambda_{2})$ is the set containing all the joint distributions which have $\lambda_{1}$ and $\lambda_{2}$ as the marginal distributions) and
\begin{equation*}
\begin{aligned}
\underset{\tilde{\lambda}\in\Lambda}{\inf}\int_{\mathcal{I}\times\mathcal{I}}|s-t|^{2}d\tilde{\lambda}(s,t)=\int_{\mathcal{I}\times\mathcal{I}}|s-t|^{2}d\bar{\lambda}(s,t).
\end{aligned}
\end{equation*}
Furthermore, we have
\begin{equation*}
\begin{aligned}
\int_{\mathcal{I}\times\mathcal{I}}|s-t|^{2}d\bar{\lambda}(s,t)=\int_{0}^{1}|\lambda_{1}^{-1}(t)-\lambda_{2}^{-1}(t)|^{2}dt
\end{aligned}
\end{equation*}
\end{theorem}
For the detailed proof, please refer to \cite{villani2021topics}.
\noindent

\begin{proof}\
\noindent Proof of Assertion 1: Our goal is proving $\mathbb{E}_{\mathcal{P}_{a}}\big[\dutchcal{Y}(a)^{-1}\big]=\bar{\dutchcal{Y}}^{-1}(a)$. Let $\mathcal{Q}$ be the set containing all the left-continuous non-decreasing functions on $(0,1)$. If we view $\mathcal{Q}$ as a subspace of $\mathcal{L}^{2}([0,1];\lambda)$ where $\lambda$ represents the Lebesgue measure, then it is isometric to $\mathcal{W}_{2}(\mathcal{I})$ (e.g., see \cite{panaretos2020invitation}). Indeed, $\mu_{a}=\underset{\nu\in\mathcal{W}_{2}(\mathcal{I})}{\arg\min}\;\mathbb{E}_{\mathcal{P}_{a}}\big[\mathbb{D}_{2}(\dutchcal{Y}(a),\nu)^{2}\big]\overset{\bullet}{=}\underset{\nu\in\mathcal{Q}}{\arg\min}\;\mathbb{E}_{\mathcal{P}_{a}}\big[\int_{0}^{1}|\dutchcal{Y}(a)^{-1}(t)-\nu^{-1}(t)|^{2}dt\big]$. Here, $\overset{\bullet}{=}$ follows from Theorem 2.18 of \cite{villani2021topics}. Since we can interchange the integral sign $\int$ and $\mathbb{E}_{\mathcal{P}_{a}}$, we notice that 
{\small
\begin{equation*}
\begin{aligned}
&\mathbb{E}_{\mathcal{P}_{a}}\big[\int_{0}^{1}|\dutchcal{Y}(a)^{-1}(t)-\nu^{-1}(t)|^{2}dt\big]\\
=&\int_{0}^{1}\mathbb{E}_{\mathcal{P}_{a}}\big[|\dutchcal{Y}(a)^{-1}(t)-\nu^{-1}(t)|^{2}\big]dt\\
=&\int_{0}^{1}\mathbb{E}_{\mathcal{P}_{a}}\big[|\dutchcal{Y}(a)^{-1}(t)-\mathbb{E}_{\mathcal{P}_{a}}\big[\dutchcal{Y}(a)^{-1}(t)\big]+\mathbb{E}_{\mathcal{P}_{a}}\big[\dutchcal{Y}(a)^{-1}(t)\big]-\nu^{-1}(t)|^{2}\big]dt\\
=&\int_{0}^{1}(\mathbb{E}_{\mathcal{P}_{a}}\big[\dutchcal{Y}(a)^{-1}(t)\big]-\nu^{-1}(t))^{2}dt\\
&+2\int_{0}^{1}(\mathbb{E}_{\mathcal{P}_{a}}[(\mathbb{E}_{\mathcal{P}_{a}}\big[\dutchcal{Y}(a)^{-1}(t)\big]-\dutchcal{Y}(a)^{-1}(t))])\times(\mathbb{E}_{\mathcal{P}_{a}}\big[\dutchcal{Y}(a)^{-1}(t)\big]-\nu^{-1}(t))dt\\
&+\int_{0}^{1}\mathbb{E}_{\mathcal{P}_{a}}[(\mathbb{E}_{\mathcal{P}_{a}}\big[\dutchcal{Y}(a)^{-1}(t)\big]-\dutchcal{Y}(a)^{-1}(t))^{2}]dt\\
\overset{\ddagger}{=}&\int_{0}^{1}(\mathbb{E}_{\mathcal{P}_{a}}\big[\dutchcal{Y}(a)^{-1}(t)\big]-\nu^{-1}(t))^{2}dt+\int_{0}^{1}\mathbb{E}_{\mathcal{P}_{a}}[(\mathbb{E}_{\mathcal{P}_{a}}\big[\dutchcal{Y}(a)^{-1}(t)\big]-\dutchcal{Y}(a)^{-1}(t))^{2}]dt.
\end{aligned}
\end{equation*}
}\noindent
$\ddagger$ follows since 
\begin{equation*}
\begin{aligned}
&\mathbb{E}_{\mathcal{P}_{a}}[(\mathbb{E}_{\mathcal{P}_{a}}\big[\dutchcal{Y}(a)^{-1}(t)\big]-\dutchcal{Y}(a)^{-1}(t))]=\mathbb{E}_{\mathcal{P}_{a}}\big[\dutchcal{Y}(a)^{-1}(t)\big]-\mathbb{E}_{\mathcal{P}_{a}}[\dutchcal{Y}(a)^{-1}(t)]=0.
\end{aligned}
\end{equation*}
Thus, $\mathbb{E}_{\mathcal{P}_{a}}\big[\int_{0}^{1}|\dutchcal{Y}(a)^{-1}(t)-\nu^{-1}(t)|^{2}dt\big]$ attains its minimum when $\int_{0}^{1}(\mathbb{E}_{\mathcal{P}_{a}}\big[\dutchcal{Y}(a)^{-1}(t)\big]-\nu^{-1}(t))^{2}dt$ attains its minimum. Equivalently, we must have $\int_{0}^{1}(\mathbb{E}_{\mathcal{P}_{a}}\big[\dutchcal{Y}(a)^{-1}(t)\big]-\nu^{-1}(t))^{2}dt=0$, implying that $\nu^{-1}(t)=\mathbb{E}_{\mathcal{P}_{a}}\big[\dutchcal{Y}(a)^{-1}(t)\big]$. We can conclude that $\bar{\dutchcal{Y}}(a)=\big(\mathbb{E}_{\mathcal{P}_{a}}\big[\dutchcal{Y}(a)^{-1}\big]\big)^{-1}\Rightarrow \Theta(a)=\bar{\dutchcal{Y}}(a)^{-1}=\mathbb{E}_{\mathcal{P}_{a}}\big[\dutchcal{Y}(a)^{-1}\big]$.
\end{proof}
			
\subsection{Proofs of Proposition \ref{prop:IPW expectation form}}\label{appendix:proofs of IPW expectation form}
\begin{proof}\
\noindent The derivations are as follows:
{\small
\begin{align*}
&\mathbb{E}_{(\mathbb{P}(A),\mathbb{P}(\mathbf{X}),\mathcal{P})}\bigg[\frac{\delta(A-a)}{p(a|\mathbf{X})}\dutchcal{Y}^{-1}\bigg]=\mathbb{E}_{\mathbb{P}(\mathbf{X})}\bigg[\frac{1}{p(a|\mathbf{X})}\mathbb{E}_{(\mathbb{P}(A),\mathcal{P})|\mathbb{P}(\mathbf{X})}[\delta(A-a)\dutchcal{Y}^{-1}|\mathbf{X}]\bigg]\\
=&\mathbb{E}_{\mathbb{P}(\mathbf{X})}\bigg[\frac{1}{p(a|\mathbf{X})}\mathbb{E}_{\mathcal{P}_{a}|\mathbb{P}(\mathbf{X})}[\dutchcal{Y}^{-1}|A=a,\mathbf{X}]p(a|\mathbf{X})\bigg]=\mathbb{E}_{\mathbb{P}(\mathbf{X})}[\mathbb{E}_{\mathcal{P}_{a}|\mathbb{P}(\mathbf{X})}[\dutchcal{Y}^{-1}|A=a,\mathbf{X}]]\\
\overset{\star}{=}&\mathbb{E}_{\mathbb{P}(\mathbf{X})}[\mathbb{E}_{\mathcal{P}_{a}|\mathbb{P}(\mathbf{X})}[\dutchcal{Y}(a)^{-1}|A=a,\mathbf{X}]]\overset{\ast}{=}\mathbb{E}_{\mathbb{P}(\mathbf{X})}[\mathbb{E}_{\mathcal{P}_{a}|\mathbb{P}(\mathbf{X})}[\dutchcal{Y}(a)^{-1}|\mathbf{X}]]=\mathbb{E}_{\mathcal{P}_{a}}[\dutchcal{Y}(a)^{-1}]=\Theta(a).
\end{align*}
}\noindent

Again, $\star$ is due to Assumption \ref{ass:Consistency} and $\ast$ is due to Assumption \ref{ass:Ignorability}. 
\end{proof}
			
\subsection{Proofs of Proposition \ref{prop:DML expectation form}}\label{appendix:proofs of DML expectation form}
\begin{proof}\
We have proven that $\mathbb{E}_{\mathbb{P}(\mathbf{X})}[m(a;\mathbf{X})]=\Theta(a)$ in Eqn. \eqref{eqt:DR expectation form} under given Assumptions. Additionally, we have proven that $\mathbb{E}_{(\mathbb{P}(A),\mathbb{P}(\mathbf{X}),\mathcal{P})}\bigg[\frac{\delta(A-a)}{p(a|\mathbf{X})}\dutchcal{Y}^{-1}\bigg]=\Theta(a)$ in Proposition \ref{prop:IPW expectation form}. It suffices to prove that $\mathbb{E}_{(\mathbb{P}(A),\mathbb{P}(\mathbf{X}),\mathcal{P})}\bigg[\frac{\delta(A-a)}{p(a|\mathbf{X})}m(a;\mathbf{X})\bigg]=\Theta(a)$. Indeed, we have
{\small
\begin{align*}
&\mathbb{E}_{(\mathbb{P}(A),\mathbb{P}(\mathbf{X}))}\bigg[\frac{\delta(A-a)}{p(a|\mathbf{X})}m(a;\mathbf{X})\bigg]=\mathbb{E}_{\mathbb{P}(\mathbf{X})}\bigg[\frac{m(a;\mathbf{X})}{p(a|\mathbf{X})}\mathbb{E}_{\mathbb{P}(A)|\mathbb{P}(\mathbf{X})}[\delta(A-a)|\mathbf{X}]\bigg]\\
=&\mathbb{E}_{\mathbb{P}(\mathbf{X})}\bigg[\frac{m(a;\mathbf{X})}{p(a|\mathbf{X})}\int_{\bar{a}\in\mathcal{A}}\delta_{a}(\bar{a})p(\bar{a}|\mathbf{X})d\bar{a}\bigg]=\mathbb{E}_{\mathbb{P}(\mathbf{X})}\bigg[\frac{m(a;\mathbf{X})}{p(a|\mathbf{X})}p(a|\mathbf{X})\bigg]=\mathbb{E}_{\mathbb{P}(\mathbf{X})}[m(a;\mathbf{X})]=\Theta(a).
\end{align*}
}\noindent
\end{proof}
			
\subsection{Proof of Theorem \ref{thm:asymptotic property}} \label{Proof of Theorem 1}
Before presenting the proofs of Theorem \ref{thm:asymptotic property}, we present two lemmas that are useful in proofing Theorem \ref{thm:asymptotic property}.
\begin{lemma}\label{lemma:1Lemma}
For $G_{1},\;G_{2}\in\mathcal{W}_{2}(\mathcal{I})$, $G_{1}^{-1},\;G_{2}^{-1}$ can be treated as elements in $\mathcal{L}^{2}([0,1];\lambda)$ where $\lambda$ here represents the Lebesgue measure. Hence, we can calculate $\mathbb{D}_{2}(G_{1},G_{2})$ in $\mathcal{W}_{2}(\mathcal{I})$ and $\|G_{1}^{-1}-G_{2}^{-1}\|$ in $\mathcal{L}^{2}([0,1];\lambda)$, and conclude that $\mathbb{D}_{2}(G_{1},G_{2})=\mathcal{L}^{2}([0,1];\lambda)$.
\end{lemma}
\begin{lemma}\label{lemma:2Lemma}
Given that $(\hat{\dutchcal{Y}}_{i})_{i=1}^{N}$ are estimates of $(\dutchcal{Y}_{i})_{i=1}^{N}$. Suppose that $(\hat{\dutchcal{Y}}_{i})_{i=1}^{N}$ are independent of each other and the Convergence Assumption \ref{ass:assumption1} holds. Then $(\hat{\dutchcal{Y}}_{i})_{i=1}^{N}$ and $(\dutchcal{Y}_{i})_{i=1}^{N}$ are in $\mathcal{L}^{2}([0,1];\lambda)$ and we have $\frac{1}{N}\underset{i=1}{\overset{N}{\sum}}\|\hat{\dutchcal{Y}}_{i}^{-1}-\dutchcal{Y}_{i}^{-1}\|^{2}=O_{P}(\alpha_{N}^{2}+\nu_{N}^{2})$.
\end{lemma}

Before presenting the proofs of Theorem \ref{thm:asymptotic property}. We restate it here. In addition, the results given in Theorem \ref{thm:asymptotic property} focus on \(\hat{\Theta}^{DML}(a)\), the restated version also incorporates the results related to \(\hat{\Theta}^{IPW}(a)\).
\begin{theorem}\label{thm:asymptotic property-full}
Let $h\rightarrow 0$, $Nh \rightarrow \infty$, and $Nh^{5} \rightarrow C\in[0,\infty)$. Suppose that $p(a|\mathbf{x})\in\mathcal{C}^{3}$ on $\mathcal{A}$ such that the derivatives (including the derivative of $0$ order) are uniformly bounded in the sample space for any $\mathbf{x}$. Furthermore, we assume that $\mathbb{E}_{\mathcal{P}|\mathbb{P}(\mathbf{X})}\big[\dutchcal{Y}^{-1}|A=a,\mathbf{X}\big]\in\mathcal{C}^{3}$ in $[0,1]\times\mathcal{A}$ and $\mathbb{E}_{\mathcal{P}|\mathbb{P}(\mathbf{X})}\big[\|\dutchcal{Y}^{-1}\||A=a,\mathbf{X}\big]\in\mathcal{C}^{3}$ in $\mathcal{A}$ are uniformly bounded in the sample spaces. For any $w\in\{IPW,DML\}$, under the convergence assumptions, we have
{\small
\begin{equation}
\begin{aligned}\label{eqt:asymptotic result continuous-full}
&\sqrt{Nh}\big(\hat{\Theta}^{w}(a)-\Theta(a)\big)=\sqrt{Nh}\bigg[\mathbb{P}_{N}\{\varphi(A,\mathbf{X},\dutchcal{Y})\}-\Theta(a)\bigg]+o_{P}(1),
\end{aligned}
\end{equation}
}\noindent
\begin{enumerate}
\item where $\varphi(A,\mathbf{X},\dutchcal{Y}):=\varphi(A,\mathbf{X},\dutchcal{Y})(t)=\frac{K_{h}(A=a)\dutchcal{Y}^{-1}(t)}{p(a|\mathbf{X})}$ if $w=IPW$ and $\rho_{p}=o(N^{-\frac{1}{2}})$;\label{thm:root N consistent IPW_continuous-full}
\item where $\varphi(A,\mathbf{X},\dutchcal{Y}):=\varphi(A,\mathbf{X},\dutchcal{Y})(t)=\frac{K_{h}(A-a)\{\dutchcal{Y}^{-1}(t)-m(a;\mathbf{X})(t)\}}{p(a|\mathbf{X})}+m(a;\mathbf{X})(t)$ if $w=DML$ and $\rho_{m}\rho_{p}=o(N^{-\frac{1}{2}})$, $\rho_{m}=o(1)$, $\rho_{p}=o(1)$.\label{thm:root N consistent DML_continuous-full}
\end{enumerate}
Additionally, we have that
\begin{subequations}
\begin{align}\label{eqt:Gaussian IPW_continuous-full}
\sqrt{Nh}\{{\hat{\Theta}^{w}(a)}-{\Theta(a)}-h^{2}B_{a}\}
\end{align}
converges weakly to a centred Gaussian process in $\mathcal{L}^{2}([0,1];\lambda)$ such that when $w=IPW$, we have
{\small 
\begin{equation*}
\begin{aligned}
B_{a}=&\frac{\int u^{2}K(u)du}{2}\times\bigg(\mathbb{E}_{\mathbb{P}(\mathbf{X})}\bigg[\partial_{aa}^{2}{m(a;\mathbf{X})}+\frac{{m(a;\mathbf{X})}\partial_{aa}^{2}p(a|\mathbf{X})}{p(a|\mathbf{X})}+\frac{2\partial_{a}{m(a;\mathbf{X})}\partial_{a}p(a|\mathbf{X})}{p(a|\mathbf{X})}\bigg]\bigg).
\end{aligned}
\end{equation*}
}\noindent
On the other hand, when $w=DML$, we have
{\small 
\begin{equation*}
\begin{aligned}
B_{a}=&\frac{\int u^{2}K(u)du}{2}\times\bigg(\mathbb{E}_{\mathbb{P}(\mathbf{X})}\bigg[\partial_{aa}^{2}{m(a;\mathbf{X})}+\frac{2\partial_{a}{m(a;\mathbf{X})}\partial_{a}p(a|\mathbf{X})}{p(a|\mathbf{X})}\bigg]\bigg).
\end{aligned}
\end{equation*}
}\noindent
\end{subequations}
\end{theorem}
\noindent
\begin{proof}\ 
\noindent[Proof of Theorem \ref{thm:asymptotic property-full}] We are going to prove the case when the estimators are chosen as $\hat{\Theta}^{w}(a)$, where $w\in\{IPW,DML\}$. We present the proofs for the estimator $\hat{\Theta}^{DML}(a)$. 
                
We consider the case when $\mathcal{K}=2$ for simplicity; the general case can be proven in a similar fashion. In the sequel, given that $W$ is a random function of $(A,\mathbf{X},\dutchcal{Y})$, we denote $\mathbb{P}_{N}W=\frac{1}{N}\underset{i=1}{\overset{N}{\sum}}W_{i}$ and $\mathbb{E}_{N}W=\frac{1}{N}\underset{i=1}{\overset{N}{\sum}}\mathbb{E}_{(\mathbb{P}(A),\mathbb{P}(\mathbf{X}),\mathcal{P})}[W_{i}]$. Let $Z=\dutchbcal{L}\dutchcal{Y}$ and $\hat{Z}=\dutchbcal{L}\hat{\dutchcal{Y}}$, where $\dutchbcal{L}\dutchcal{Y}=\dutchcal{Y}^{-1}$. Write $R_{i}=\hat{Z}_{i}-Z_{i}$ and $D_{a}^{k}(\mathbf{x})=\hat{m}^{k}(a;\mathbf{x})-\tilde{m}^{k}(a;\mathbf{x})$. Define
{\small
\begin{align}
&\psi_{a}=\mathbb{E}_{(\mathbb{P}(A),\mathbb{P}(\mathbf{X}),\mathcal{P})}\bigg[\frac{K_{h}(A-a)Z}{p(a|\mathbf{X})}\bigg]-\mathbb{E}_{(\mathbb{P}(A),\mathbb{P}(\mathbf{X}),\mathcal{P})}\bigg[\big\{\frac{K_{h}(A-a)}{p(a|\mathbf{X})}-1\big\}m(a;\mathbf{X})\bigg],\label{eqt:true kernel}\\
&\hat{\psi}_{a,k}=\mathbb{P}_{N_{k}}\bigg[\frac{K_{h}(A-a)\hat{Z}}{\hat{p}^{k}(a|\mathbf{X})}\bigg]-\mathbb{P}_{N_{k}}\bigg[\big\{\frac{K_{h}(A-a)}{\hat{p}^{k}(a|\mathbf{X})}-1\big\}\hat{m}^{k}(a;\mathbf{X})\bigg].\label{eqt:estimated kernel}
\end{align}
}\noindent
Hence, we have 
{\small
\begin{align*}
\hat{\Theta}^{DML}(a)&=\frac{1}{N}(N_{1}\hat{\psi}_{a,1}+N_{2}\hat{\psi}_{a,2}).
\end{align*}
}\noindent			
Moreover, since $h\rightarrow 0$, W.L.O.G., we assume that $h<1$. Hence, we have $0<\sqrt{h}<1$ and $0<\frac{\sqrt{Nh}}{N}<\frac{\sqrt{N}}{N}=\frac{1}{\sqrt{N}}$. Note that from Eqn. \eqref{eqt:true kernel}, we have
{\small
\begin{align*}
\psi_{a}&=\mathbb{E}_{(\mathbb{P}(A),\mathbb{P}(\mathbf{X}),\mathcal{P})}\bigg[\frac{K_{h}(A-a)(Z-m(a;\mathbf{X}))}{p(a|\mathbf{X})}\bigg]+\Theta(a).
\end{align*}
}\noindent
As a result, we have
\begin{align*}
&\sqrt{Nh}\big(\hat{\Theta}^{DML}(a)-\Theta(a)\big)\\
=&\sqrt{Nh}\big(\frac{1}{N}(N_{1}\hat{\psi}_{a,1}+N_{2}\hat{\psi}_{a,2})-\psi_{a}\big)+\sqrt{Nh}\mathbb{E}_{(\mathbb{P}(A),\mathbb{P}(\mathbf{X}),\mathcal{P})}\bigg[\frac{K_{h}(A-a)(Z-m(a;\mathbf{X}))}{p(a|\mathbf{X})}\bigg].
\end{align*}
We can then decompose $\sqrt{Nh}\big(\frac{1}{N}(N_{1}\hat{\psi}_{a,1}+N_{2}\hat{\psi}_{a,2})-\psi_{a}\big)$ into the sum of five terms as follows:
{\small
\begin{align*}
&\sqrt{Nh}\big(\frac{1}{N}(N_{1}\hat{\psi}_{a,1}+N_{2}\hat{\psi}_{a,2})-\psi_{a}\big)\\
=&\sqrt{N}\underset{k=1,2}{\sum}\frac{N_{k}}{N}\text{I}+\sqrt{N}\underset{k=1,2}{\sum}\frac{N_{k}}{N}\text{II}+\sqrt{N}\underset{k=1,2}{\sum}\frac{N_{k}}{N}\text{III}+\sqrt{N}\underset{k=1,2}{\sum}\frac{N_{k}}{N}\text{IV}+\sqrt{N}\underset{k=1,2}{\sum}\frac{N_{k}}{N}\text{V},
\end{align*}
}\noindent
where
{\small
\begin{align*}
\text{I}&=\sqrt{h}(\mathbb{P}_{N_{k}}-\mathbb{E}_{N_{k}})\bigg[\frac{K_{h}(A-a)(Z-\tilde{m}^{k}(a;\mathbf{X}))}{\hat{p}^{k}(a|\mathbf{X})}-\frac{K_{h}(A-a)(Z-m(a;\mathbf{X}))}{p(a|\mathbf{X})}+\tilde{m}^{k}(a;\mathbf{X})-m(a;\mathbf{X})\bigg],\\
\text{II}&=\sqrt{h}(\mathbb{P}_{N_{k}}-\mathbb{E}_{N_{k}})\bigg[m(a;\mathbf{X})+\frac{K_{h}(A-a)(Z-m(a;\mathbf{X}))}{p(a|\mathbf{X})}\bigg]\\
&=\sqrt{h}(\mathbb{P}_{N_{k}}-\mathbb{E}_{N_{k}})\varphi(A,\mathbf{X},\dutchcal{Y}),\\
\text{III}&=\sqrt{h}\mathbb{E}_{N_{k}}\bigg[K_{h}(A-a)(Z-m(a;\mathbf{X}))\times\frac{(p(a|\mathbf{X})-\hat{p}^{k}(a|\mathbf{X}))}{\hat{p}^{k}(a|\mathbf{X})p(a|\mathbf{X})}\bigg]\\
&\quad+\sqrt{h}\mathbb{E}_{N_{k}}\bigg[\{\tilde{m}^{k}(a;\mathbf{X})-m(a;\mathbf{X})\}\times\frac{\{\hat{p}^{k}(a|\mathbf{X})-K_{h}(A-a)\}}{\hat{p}^{k}(a|\mathbf{X})}\bigg],\\
\text{IV}&=\sqrt{h}\mathbb{P}_{N_{k}}\bigg[\big\{1-\frac{K_{h}(A-a)}{\hat{p}^{k}(a|\mathbf{X})}\big\}\{D_{a}^{k}(\mathbf{X})\}\bigg],\\
\text{V}&=\sqrt{h}\mathbb{P}_{N_{k}}\bigg[\frac{K_{h}(A-a)R}{\hat{p}^{k}(a|\mathbf{X})}\bigg].
\end{align*}
}\noindent
Define three quantities $H_{1}(A,\mathbf{X},Z)$, $H_{2}(A,\mathbf{X},Z)$, and $H_{3}(A,\mathbf{X},Z)$ such that we have {\small $H_{1}(A,\mathbf{X},Z)=\frac{K_{h}(A-a)Z\{p(a|\mathbf{X})-\hat{p}^{k}(a|\mathbf{X})\}}{\hat{p}^{k}(a|\mathbf{X})p(a|\mathbf{X})}$}, {\small $H_{2}(A,\mathbf{X},Z)=K_{h}(A-a)\frac{\{\hat{p}^{k}(a|\mathbf{X})m(a;\mathbf{X})-p(a|\mathbf{X})\tilde{m}^{k}(a;\mathbf{X})\}}{\hat{p}^{k}(a|\mathbf{X})p(a|\mathbf{X})}$}, and {\small $H_{3}(A,\mathbf{X},Z)=\tilde{m}_{k}(a;\mathbf{X})-m(a;\mathbf{X})$}. Also, we write $H(A,\mathbf{X},Z)=H_{1}(A,\mathbf{X},Z)+H_{2}(A,\mathbf{X},Z)+H_{3}(A,\mathbf{X},Z)$.
It suffices to show that $\text{I}$, $\text{III}$, $\text{IV}$, and $\text{V}$ are $o_{P}(1)$.
				
\noindent Consider term \text{I}. Note that
\begin{align*}
&\frac{K_{h}(A-a)(Z-\tilde{m}^{k}(a;\mathbf{X}))}{\hat{p}^{k}(a|\mathbf{X})}+\tilde{m}^{k}(a;\mathbf{X})-\frac{K_{h}(A-a)(Z-m(a;\mathbf{X}))}{p(a|\mathbf{X})}-m(a;\mathbf{X})\\
=&H_{1}(A,\mathbf{X},Z)+H_{2}(A,\mathbf{X},Z)+H_{3}(A,\mathbf{X},Z)=H(A,\mathbf{X},Z).
\end{align*}
We compute $\mathbb{E}_{(\mathbb{P}(A),\mathbb{P}(\mathbf{X}),\mathcal{P})}[\|\text{I}\|^{2}]$. Indeed, it is equal to 
$\mathbb{E}_{(\mathbb{P}(A),\mathbb{P}(\mathbf{X}),\mathcal{P})}[\|\sqrt{h}(\mathbb{P}_{N_{k}}-\mathbb{E}_{(\mathbb{P}(A),\mathbb{P}(\mathbf{X}),\mathcal{P})})H\|^{2}]$. Now, we can decompose it into the sum of $\text{I}_{1}$ and $\text{I}_{2}$
where 
\begin{equation*}
\begin{aligned}
\text{I}_{1}&=\frac{h}{N_{k}^{2}}\underset{i\in\mathcal{D}_{k}}{\overset{}{\sum}}\mathbb{E}_{{(\mathbb{P}(A),\mathbb{P}(\mathbf{X}),\mathcal{P})}}\big[\Vert H(A_{i},\mathbf{X}_{i},Z_{i})-\mathbb{E}_{(\mathbb{P}(A),\mathbb{P}(\mathbf{X}),\mathcal{P})}\big[H(A_{i},\mathbf{X}_{i},Z_{i})\big]\Vert^{2}\big]
\end{aligned}
\end{equation*}
and
\begin{equation*}
\begin{aligned}
\text{I}_{2}&=\frac{h}{N_{k}^{2}}\underset{\substack{i,j\in\mathcal{D}_{k}\\i\neq j}}{\overset{}{\sum}}\mathbb{E}_{{(\mathbb{P}(A),\mathbb{P}(\mathbf{X}),\mathcal{P})}}\big[\langle H(A_{i},\mathbf{X}_{i},Z_{i})-\mathbb{E}_{{(\mathbb{P}(A),\mathbb{P}(\mathbf{X}),\mathcal{P})}}\big[H(A_{i},\mathbf{X}_{i},Z_{i})\big],\\
&\quad\quad\quad\quad\quad\quad\quad\quad H(A_{j},\mathbf{X}_{j},Z_{j})-\mathbb{E}_{{(\mathbb{P}(A),\mathbb{P}(\mathbf{X}),\mathcal{P})}}\big[H(A_{j},\mathbf{X}_{j},Z_{j})\big]\rangle\big].
\end{aligned}
\end{equation*}
We first bound $\text{I}_{1}$. Note that, since $H-\mathbb{E}_{(\mathbb{P}(A),\mathbb{P}(\mathbf{X}),\mathcal{P})}[H]=H_{1}-\mathbb{E}_{(\mathbb{P}(A),\mathbb{P}(\mathbf{X}),\mathcal{P})}[H_{1}]+H_{2}-\mathbb{E}_{(\mathbb{P}(A),\mathbb{P}(\mathbf{X}),\mathcal{P})}[H_{2}]+H_{3}-\mathbb{E}_{(\mathbb{P}(A),\mathbb{P}(\mathbf{X}),\mathcal{P})}[H_{3}]$, we have
\begin{align*}
\text{I}_{1}\lesssim&\frac{h}{N_{k}^{2}}\underset{p=1}{\overset{3}{\sum}}\;\underset{i\in\mathcal{D}_{k}}{\overset{}{\sum}}\mathbb{E}_{(\mathbb{P}(A),\mathbb{P}(\mathbf{X}),\mathcal{P})}\big[\|H_{p}(A_{i},\mathbf{X}_{i},Z_{i})-\mathbb{E}_{(\mathbb{P}(A),\mathbb{P}(\mathbf{X}),\mathcal{P})}\big[H_{p}(A_{i},\mathbf{X}_{i},Z_{i})\big]\|^{2}\big]\\
\lesssim&\text{I}_{1-1}+\text{I}_{1-2}+\text{I}_{1-3}.
\end{align*}
Here, $\text{I}_{1-j}=\frac{h}{N_{k}^{2}}\underset{i\in\mathcal{D}_{k}}{\overset{}{\sum}}\mathbb{E}_{(\mathbb{P}(A),\mathbb{P}(\mathbf{X}),\mathcal{P})}\big[\|H_{j}(A_{i},\mathbf{X}_{i},Z_{i})\|^{2}\big]$ such that $j=1,\;2,\;3$. 

\noindent We first consider $h\mathbb{E}_{(\mathbb{P}(A),\mathbb{P}(\mathbf{X}),\mathcal{P})}[\left\|H_{1}(A,\mathbf{X},Z)\right\|^{2}]$. Note that it is equal to $h\mathbb{E}_{(\mathbb{P}(A),\mathbb{P}(\mathbf{X}),\mathcal{P})}\bigg[K_{h}(A-a)^{2}\left\|\frac{Z\{p(a|\mathbf{X})-\hat{p}^{k}(a|\mathbf{X})\}}{\hat{p}^{k}(a|\mathbf{X})p(a|\mathbf{X})}\right\|^{2}\bigg]$. Our next objective is showing that the quantity is bounded above by $ch\mathbb{E}_{\mathbb{P}(\mathbf{X})}[\left|p(a|\mathbf{x})-\hat{p}^{k}(a|\mathbf{X})\right|^{2}\mathbb{E}_{\mathbb{P}(A),\mathcal{P}|\mathbb{P}(\mathbf{X}))}[K_{h}(A-a)^{2}\left\|Z\right\|^{2}|\mathbf{X}]]$ for some constant $c$.
Although $Z=\dutchcal{Y}^{-1}$ is a function, $\|Z\|$ is a scalar. Hence, $\mathbb{E}_{\mathcal{P}|\mathbb{P}(\mathbf{X})}[\|Z\|^{2}|A=a,\mathbf{X}]$ can be treated as a function of $a$. Hence, we may express
{\small
\begin{align*}
&\mathbb{E}_{\mathcal{P}|\mathbb{P}(\mathbf{X})}\big[\left\|Z\right\|^{2}\mid A=a+uh,\mathbf{X}\big]\\
=&\mathbb{E}_{\mathcal{P}|\mathbb{P}(\mathbf{X})}\big[\left\|Z\right\|^{2}\mid A=a,\mathbf{X}\big]+\partial_{a}\mathbb{E}_{\mathcal{P}|\mathbb{P}(\mathbf{X})}\big[\left\|Z\right\|^{2}\mid A=a,\mathbf{X}\big]uh+\frac{\partial_{aa}^{2}\mathbb{E}_{\mathcal{P}|\mathbb{P}(\mathbf{X})}\big[\left\|Z\right\|^{2}\mid A=a,\mathbf{X}\big]u^{2}h^{2}}{2}+O_{P}(h^{3}).
\end{align*}
}\noindent
Further, since 
{\small
\begin{equation*}
\begin{aligned}
p(a+uh|\mathbf{X})=p(a|\mathbf{X})+\partial_{a}p(a|\mathbf{X})uh+\frac{\partial_{aa}^{2}p(a|\mathbf{X})u^{2}h^{2}}{2}+O_{P}(h^{3}),
\end{aligned}
\end{equation*}
}\noindent
we have
{\small
\begin{align*}
&\mathbb{E}_{\mathbb{P}(A),\mathcal{P}|\mathbb{P}(\mathbf{X})}[K_{h}(A-a)^{2}\left\|Z\right\|^{2}|\mathbf{X}]\\
&=\int \mathbb{E}_{\mathcal{P}|\mathbb{P}(\mathbf{X})}[K_{h}(A-a)^{2}\left\|Z\right\|^{2}|A=s,\mathbf{X}]p(s|\mathbf{X})ds\\
&=\frac{1}{h}\bigg(\int K(u)^{2}du\bigg)\mathbb{E}_{\mathcal{P}|\mathbb{P}(\mathbf{X})}\big[\left\|Z\right\|^{2}| A=a,\mathbf{X}\big]p(a|\mathbf{X})\\
&\quad+\frac{h^{2}}{h}\bigg(\int K(u)^{2}u^{2}\;du\bigg)\times\\
&\quad\biggl\{\mathbb{E}_{\mathcal{P}|\mathbb{P}(\mathbf{X})}\big[\left\|Z\right\|^{2}| A=a,\mathbf{X}\big]\frac{\partial_{aa}^{2}p(a|\mathbf{X})}{2}+\partial_{a}\mathbb{E}_{\mathcal{P}|\mathbb{P}(\mathbf{X})}\big[\left\|Z\right\|^{2}| A=a,\mathbf{X}\big]\partial_{a}p(a|\mathbf{X})\\
&\quad\quad+\frac{\partial_{aa}^{2}\mathbb{E}_{\mathcal{P}|\mathbb{P}(\mathbf{X})}\big[\left\|Z\right\|^{2}| A=a,\mathbf{X}\big]}{2}p(a|\mathbf{X})\biggl\}+O_{P}(h^{2}).
\end{align*}
}\noindent
Hence, we have
{\small
\begin{align*}
&h\mathbb{E}_{(\mathbb{P}(A),\mathbb{P}(\mathbf{X}),\mathcal{P})}\big[\left\|H_{1}(A,\mathbf{X},Z)\right\|^{2}\big]\\
\lesssim& h\mathbb{E}_{\mathbb{P}(\mathbf{X})}\bigg[\left|p(a|\mathbf{X})-\hat{p}^{k}(a|\mathbf{X})\right|^{2}\times\mathbb{E}_{\mathbb{P}(A),\mathcal{P}|\mathbb{P}(\mathbf{X})}\big[K_{h}(A-a)^{2}\left\|Z\right\|^{2}|\mathbf{X}\big]\bigg]\\
=&\int K(u)^{2}du\times\text{I}_{1-1a}+h^{2} (\int K(u)^{2}u^{2}\;du)({\text{I}_{1-1b}+\text{I}_{1-1c}+\text{I}_{1-1d}})+O(h^{3}).
\end{align*}
}\noindent
where
{\small
\begin{align*}
\text{I}_{1-1a}=&\mathbb{E}_{\mathbb{P}(\mathbf{X})}\bigg[\left|p(a|\mathbf{X})-\hat{p}^{k}(a|\mathbf{X})\right|^{2}\times\mathbb{E}_{\mathcal{P}|\mathbb{P}(\mathbf{X})}\big[\left\|Z\right\|^{2}| A=a,\mathbf{X}\big]p(a|\mathbf{X})\bigg],\\
\text{I}_{1-1b}=&\mathbb{E}_{\mathbb{P}(\mathbf{X})}\bigg[\left|p(a|\mathbf{x})-\hat{p}^{k}(a|\mathbf{X})\right|^{2}\times\mathbb{E}_{\mathcal{P}|\mathbb{P}(\mathbf{X})}\big[\left\|Z\right\|^{2}| A=a,\mathbf{X}\big]\frac{\partial_{aa}^{2}p(a|\mathbf{X})}{2}\bigg],\\
\text{I}_{1-1c}=&\mathbb{E}_{\mathbb{P}(\mathbf{X})}\bigg[\left|p(a|\mathbf{x})-\hat{p}^{k}(a|\mathbf{X})\right|^{2}\times\partial_{a}\mathbb{E}_{\mathcal{P}|\mathbb{P}(\mathbf{X})}\big[\left\|Z\right\|^{2}| A=a,\mathbf{X}\big]\partial_{a}p(a|\mathbf{X})\bigg],\\
\text{I}_{1-1d}=&\mathbb{E}_{\mathbb{P}(\mathbf{X})}\bigg[\left|p(a|\mathbf{x})-\hat{p}^{k}(a|\mathbf{X})\right|^{2}\times\frac{\partial_{aa}^{2}\mathbb{E}_{\mathcal{P}|\mathbb{P}(\mathbf{X})}\big[\left\|Z\right\|^{2}| A=a,\mathbf{X}\big]}{2}p(a|\mathbf{X})\bigg].
\end{align*}
}\noindent

We find the bounds of $\text{I}_{1-1a}$, $\text{I}_{1-1b}$, $\text{I}_{1-1c}$, and $\text{I}_{1-1d}$. Note that, according to the given conditions, we have
\begin{align*}
&\text{I}_{1-1a},\;\text{I}_{1-1b},\;\text{I}_{1-1c},\;\text{I}_{1-1d}\\
&\lesssim\mathbb{E}_{\mathbb{P}(\mathbf{X})}[|p(a|\mathbf{X})-\hat{p}^{k}(a|\mathbf{X})|^{2}]\leq(\mathbb{E}_{\mathbb{P}(\mathbf{X})}[|p(a|\mathbf{X})-\hat{p}^{k}(a|\mathbf{X})|^{4}])^{\frac{1}{2}}\leq \rho_{p}^{2}.
\end{align*}
As a result, we conclude that
{\small
\begin{equation*}
\begin{aligned}
\text{I}_{1-1}&\lesssim\mathbb{E}_{\mathbb{P}(\mathbf{X})}[|p(a|\mathbf{X})-\hat{p}^{k}(a|\mathbf{X})|^{2}]+O(h^{3})\leq(1+h^{2})\rho_{p}^{2}+O(h^{3})).
\end{aligned}
\end{equation*}
}\noindent
We therefore have
{\small
\begin{align*}
\text{I}_{1-1}=O\bigg(\frac{1}{N_{k}}\rho_{p}^{2}+\frac{h^{2}}{N_{k}}\rho_{p}^{2}+h^{3}\bigg).
\end{align*}
}\noindent 
We bound $\text{I}_{1-2}$. To start with, we consider $h\mathbb{E}_{(\mathbb{P}(A),\mathbb{P}(\mathbf{X}),\mathcal{P})}\big[\left\|H_{2}(A,\mathbf{X},Z)\right\|^{2}\big]$, and we have
{\small
\begin{align*}
&h\mathbb{E}_{(\mathbb{P}(A),\mathbb{P}(\mathbf{X}),\mathcal{P})}\big[\left\|H_{2}(A,\mathbf{X},Z)\right\|^{2}\big]\\
=&h\mathbb{E}_{(\mathbb{P}(A),\mathbb{P}(\mathbf{X}))}\bigg[K_{h}(A-a)^{2}\times\left\|\frac{\hat{p}^{k}(a|\mathbf{X})m(a;\mathbf{X})-p(a|\mathbf{X})\tilde{m}^{k}(a;\mathbf{X})}{\hat{p}^{k}(a|\mathbf{X})p(a|\mathbf{X})}\right\|^{2}\bigg]\\
\leq& ch\mathbb{E}_{\mathbb{P}(\mathbf{X})}\big[\left\|\hat{p}^{k}(a|\mathbf{X})m(a;\mathbf{X})-p(a|\mathbf{X})m(a;\mathbf{X})\right\|^{2}\times\mathbb{E}_{\mathbb{P}(A)|\mathbb{P}(\mathbf{X})}[K_{h}(A-a)^{2}|\mathbf{X}]\big]\\
&+ch\mathbb{E}_{\mathbb{P}(\mathbf{X})}\big[\left\|p(a|\mathbf{X})m(a;\mathbf{X})-p(a|\mathbf{X})\tilde{m}^{k}(a;\mathbf{X})\right\|^{2}\times\mathbb{E}_{\mathbb{P}(A)|\mathbb{P}(\mathbf{X})}[K_{h}(A-a)^{2}|\mathbf{X}]\big].
\end{align*}
}\noindent
Standard algebraic derivations also give that $\mathbb{E}_{\mathbb{P}(A)|\mathbb{P}(\mathbf{X})}[K_{h}(A-a)^{2}|\mathbf{X}]=\frac{\big(\int K(u)^{2}du\big)p(a|\mathbf{X})}{h}+\frac{\big(\int u^{2}K(u)^{2}du\big)\partial_{aa}^{2}p(a|\mathbf{X})h}{2}+O_{P}(h^{2})$. Thus, we have
{\small
\begin{align*}
&h\mathbb{E}_{(\mathbb{P}(A),\mathbb{P}(\mathbf{X}),\mathcal{P})}\big[\left\|H_{2}(A,\mathbf{X},Z)\right\|^{2}\big]\\
&\leq c\mathbb{E}_{\mathbb{P}(\mathbf{X})}\bigg[\left\|\hat{p}^{k}(a|\mathbf{X})m(a;\mathbf{X})-p(a|\mathbf{X})m(a;\mathbf{X})\right\|^{2}\times\bigg(\int K(u)^{2}du\bigg)p(a|\mathbf{X})\bigg]\\
&\;+ch^{2}\mathbb{E}_{\mathbb{P}(\mathbf{X})}\bigg[\|\hat{p}^{k}(a|\mathbf{X})m(a;\mathbf{X})-p(a|\mathbf{X})m(a;\mathbf{X})\|^{2}\times\frac{\bigg(\int u^{2}K(u)^{2}du\bigg)\partial_{aa}^{2}p(a|\mathbf{X})}{2}\bigg]\\
&\;+c\mathbb{E}_{\mathbb{P}(\mathbf{X})}\bigg[\|p(a|\mathbf{X})m(a;\mathbf{X})-p(a|\mathbf{X})\tilde{m}^{k}(a;\mathbf{X})\|^{2}\times\bigg(\int K(u)^{2}du\bigg)p(a|\mathbf{X})\bigg]\\
&\;+ch^{2}\mathbb{E}_{\mathbb{P}(\mathbf{X})}\bigg[\left\|p(a|\mathbf{X})m(a;\mathbf{X})-p(a|\mathbf{X})\tilde{m}^{k}(a;\mathbf{X})\right\|^{2}\times\frac{\bigg(\int u^{2}K(u)^{2}du\bigg)\partial_{aa}^{2}p(a|\mathbf{X})}{2}\bigg]+O(h^{3}).
\end{align*}
}\noindent
Therefore, we have
{\small
\begin{align*}
\text{I}_{1-2}&\lesssim \frac{1+h^{2}}{N_{k}}\mathbb{E}_{\mathbb{P}(\mathbf{X})}\big[|\hat{p}^{k}(a|\mathbf{X})-p(a|\mathbf{X})|^{2}\big]+\frac{1+h^{2}}{N_{k}}\mathbb{E}_{\mathbb{P}(\mathbf{X})}[\|m(a;\mathbf{X})-\tilde{m}^{k}(a;\mathbf{X})\|^{2}]+O(h^{3})\\
&\leq \frac{1+h^{2}}{N_{k}}(\mathbb{E}_{\mathbb{P}(\mathbf{X})}\big[|\hat{p}^{k}(a|\mathbf{X})-p(a|\mathbf{X})|^{4}\big])^{\frac{1}{2}}+\frac{1+h^{2}}{N_{k}}(\mathbb{E}_{\mathbb{P}(\mathbf{X})}[\|m(a;\mathbf{X})-\tilde{m}^{k}(a;\mathbf{X})\|^{4}])^{\frac{1}{2}}+O(h^{3}).
\end{align*}
}\noindent
Thus, we have
\begin{align*}
\text{I}_{1-2}=O\bigg(\frac{1+h^{2}}{N_{k}}\rho_{p}^{2}+\frac{1+h^{2}}{N_{k}}\rho_{m}^{2}+h^{3}\bigg).
\end{align*}
We now bound $\text{I}_{1-3}$. Note that 
{\small
\begin{equation*}
\begin{aligned}
h\mathbb{E}_{(\mathbb{P}(A),\mathbb{P}(\mathbf{X}),\mathcal{P})}\big[\|H_{3}(A,\mathbf{X},Z)\|^{2}\big]\lesssim h\mathbb{E}\big[\|\tilde{m}^{k}(a;\mathbf{X})-m(a;\mathbf{X})\|^{2}\big],
\end{aligned}
\end{equation*}
}\noindent
we therefore have
{\small
\begin{equation*}
\begin{aligned}
\text{I}_{1-3}\lesssim \frac{h}{N_{k}}\mathbb{E}_{\mathbb{P}(\mathbf{X})}\big[\left\|\tilde{m}^{k}(a;\mathbf{X})-m(a;\mathbf{X})\right\|^{2}\big]\leq\frac{h}{N_{k}}\rho_{m}^{2}.
\end{aligned}
\end{equation*}
}\noindent
Thus, we have $\text{I}_{1-3}=O\bigg(\frac{h}{N_{k}}\rho_{m}^{2}\bigg)$.

\noindent Next, we bound $\text{I}_{2}$. Define 				
\begin{align*}
G(A,\mathbf{X},Z)&:=\frac{K_{h}(A-a)\{Z-\tilde{m}^{k}(a;\mathbf{X})\}}{\hat{p}^{k}(a|\mathbf{X})}+\tilde{m}^{k}(a;\mathbf{X})-m(a;\mathbf{X})\\
F(A,\mathbf{X},Z)&:=-\frac{K_{h}(A-a)\{Z-m(a;\mathbf{X})\}}{p(a|\mathbf{X})}.
\end{align*}
\noindent We notice that $H(A,\mathbf{X},Z)=G(A,\mathbf{X},Z)+F(A,\mathbf{X},Z)$. In addition, we denote
\begin{equation*}
\begin{aligned}
\gamma^{w}&=\mathbb{E}_{(\mathbb{P}(A),\mathbb{P}(\mathbf{X}),\mathcal{P})}[w(A,\mathbf{X},Z)],\\
\gamma_{k}^{w}&=\mathbb{E}_{(\mathbb{P}(A),\mathbb{P}(\mathbf{X}),\mathcal{P})}[w(A_{k},\mathbf{X}_{k},Z_{k})],\\
w_{k}&=w(A_{k},\mathbf{X}_{k},Z_{k})
\end{aligned}
\end{equation*}
for $w\in\{G,F,H\}$. As a result, we have
{\small
\begin{align*}
&\left|\mathbb{E}_{(\mathbb{P}(A),\mathbb{P}(\mathbf{X}),\mathcal{P})}\langle H_{i}-\gamma_{i}^{H},\right.\left.H_{j}-\gamma_{j}^{H}\rangle\right|\\
\leq&\left|\mathbb{E}_{(\mathbb{P}(A),\mathbb{P}(\mathbf{X}),\mathcal{P})}\langle G_{i},G_{j}\rangle-\langle\gamma_{i}^{G},\gamma_{j}^{G}\rangle\right|+\left|\mathbb{E}_{(\mathbb{P}(A),\mathbb{P}(\mathbf{X}),\mathcal{P})}\langle G_{i},F_{j}\rangle\right.\left.-\langle\gamma_{i}^{G},\gamma_{j}^{F}\rangle\right|\\
&+\left|\mathbb{E}_{(\mathbb{P}(A),\mathbb{P}(\mathbf{X}),\mathcal{P})}\langle G_{j},F_{i}\rangle\right.\left.-\langle\gamma_{j}^{G},\gamma_{i}^{F}\rangle\right|+\left|\mathbb{E}_{(\mathbb{P}(A),\mathbb{P}(\mathbf{X}),\mathcal{P})}\langle F_{i},F_{j}\rangle\right.\left.-\langle\gamma_{i}^{F},\gamma_{j}^{F}\rangle\right|.
\end{align*}
}\noindent
Consider $|\mathbb{E}_{(\mathbb{P}(A),\mathbb{P}(\mathbf{X}),\mathcal{P})}\langle G_{i},G_{j}\rangle-\langle\gamma_{i}^{G},\gamma_{j}^{G}\rangle|$.
We have
{\small
\begin{align*}
&\left|\mathbb{E}_{(\mathbb{P}(A),\mathbb{P}(\mathbf{X}),\mathcal{P})}\langle G_{i},G_{j}\rangle\right.\left.-\langle\gamma_{i}^{G},\gamma_{j}^{G}\rangle\right|\\
\leq&|\mathbb{E}_{(\mathbb{P}(A),\mathbb{P}(\mathbf{X}),\mathcal{P})}\langle G_{i},G_{j}\rangle|+|\langle\gamma_{i}^{G},\gamma_{j}^{G}\rangle|\overset{\diamond}{\leq}\|\gamma_{i}^{G}\|\|\gamma_{j}^{G}\|+\|\gamma_{i}^{G}\|\|\gamma_{j}^{G}\|=2\|\gamma^{G}\|^{2}.
\end{align*}
}\noindent
$\overset{\diamond}{=}$ holds by using the Cauchy Schwartz inequality and the fact that $(A_{i},\mathbf{X}_{i},Z_{i})$ and $(A_{j},\mathbf{X}_{j},Z_{j})$ are independent of each other. Similarly, we have
{\small
\begin{align*}
&\left|\mathbb{E}_{(\mathbb{P}(A),\mathbb{P}(\mathbf{X}),\mathcal{P})}\langle G_{i},F_{j}\rangle\right.\left.-\langle\gamma_{i}^{G},\gamma_{j}^{F}\rangle\right|\leq 2\|\gamma^{G}\|\|\gamma^{F}\|,
\end{align*}
}\noindent
{\small
\begin{align*}
&\left|\mathbb{E}_{(\mathbb{P}(A),\mathbb{P}(\mathbf{X}),\mathcal{P})}\langle F_{i},G_{j}\rangle\right.\left.-\langle\gamma_{i}^{F},\gamma_{j}^{G}\rangle\right|\leq 2\|\gamma^{F}\|\|\gamma^{G}\|,
\end{align*}
}\noindent
and
{\small
\begin{align*}
&\left|\mathbb{E}_{(\mathbb{P}(A),\mathbb{P}(\mathbf{X}),\mathcal{P})}\langle F_{i},F_{j}\rangle\right.\left.-\langle\gamma_{i}^{F},\gamma_{j}^{F}\rangle\right|\leq 2\|\gamma^{F}\|^{2}.
\end{align*}
}\noindent
Thus, we have
{\small
\begin{align*}
&\left|\mathbb{E}_{(\mathbb{P}(A),\mathbb{P}(\mathbf{X}),\mathcal{P})}\langle H_{i}-\gamma_{i}^{H},\right.\left.H_{j}-\gamma_{j}^{H}\rangle\right|\\
&\leq 2\|\gamma^{G}\|^{2}+2\|\gamma^{F}\|^{2}+4\|\gamma^{F}\|\|\gamma^{G}\|= 2\big(\|\gamma^{G}\|+\|\gamma^{F}\|\big)^{2}\lesssim \|\gamma^{G}\|^{2}+\|\gamma^{F}\|^{2}.
\end{align*}
}\noindent
Note that
\begin{align*}
&\|\mathbb{E}_{(\mathbb{P}(A),\mathbb{P}(\mathbf{X}),\mathcal{P})}[G(A,\mathbf{X},Z)]\|\\
&=\|\mathbb{E}_{\mathbb{P}(\mathbf{X})}[\mathbb{E}_{\mathbb{P}(A),\mathcal{P}|\mathbb{P}(\mathbf{X})}[G(A,\mathbf{X},Z)|\mathbf{X}]]\| \leq\mathbb{E}_{\mathbb{P}(\mathbf{X})}[\|\mathbb{E}_{\mathbb{P}(A),\mathcal{P}|\mathbb{P}(\mathbf{X})}[G(A,\mathbf{X},Z)|\mathbf{X}]\|],
\end{align*}
we have
{\small
\begin{align*}
&\|\mathbb{E}_{(\mathbb{P}(A),\mathbb{P}(\mathbf{X}),\mathcal{P})}[G(A,\mathbf{X},Z)]\|^{2}\\
&\leq (\mathbb{E}_{\mathbb{P}(\mathbf{X})}[\|\mathbb{E}_{\mathbb{P}(A),\mathcal{P}|\mathbb{P}(\mathbf{X})}[G(A,\mathbf{X},Z)|\mathbf{X}]\|])^{2}\leq \mathbb{E}_{\mathbb{P}(\mathbf{X})}[\|\mathbb{E}_{\mathbb{P}(A),\mathcal{P}|\mathbb{P}(\mathbf{X})}[G(A,\mathbf{X},Z)|\mathbf{X}]\|^{2}].
\end{align*}
}\noindent
It remains to consider {\small $\|\mathbb{E}_{\mathbb{P}(A),\mathcal{P}|\mathbb{P}(\mathbf{X})}\big[G(A,\mathbf{X},Z)|\mathbf{X}\big]\|$} and {\small $\|\mathbb{E}_{\mathbb{P}(A),\mathcal{P}|\mathbb{P}(\mathbf{X})}\big[F(A,\mathbf{X},Z)|\mathbf{X}\big]\|$}. Now, from the definition of $G(A,\mathbf{X},Z)$, we have
{\small
\begin{align*}
&\mathbb{E}_{\mathbb{P}(A),\mathcal{P}|\mathbb{P}(\mathbf{X})}\big[G(A,\mathbf{X},Z)|\mathbf{X}\big]\\
&=\frac{(m(a;\mathbf{X})-\tilde{m}^{k}(a;\mathbf{X}))(p(a|\mathbf{X})-\hat{p}^{k}(a|\mathbf{X}))}{\hat{p}^{k}(a|\mathbf{X})}\\
&\quad+(\int u^{2}K(u)du)\times\\
&\quad\quad\biggl\{\frac{(m(a;\mathbf{X})-\tilde{m}^{k}(a;\mathbf{X}))\partial_{aa}^{2}p(a|\mathbf{X})h^{2}}{2\hat{p}^{k}(a|\mathbf{X})}+\frac{\partial_{a}\mathbb{E}_{\mathcal{P}|\mathbb{P}(\mathbf{X})}[Z|A=a,\mathbf{X}]\partial_{a}p(a|\mathbf{X})h^{2}}{\hat{p}^{k}(a|\mathbf{X})}\\
&\quad\quad+\frac{p(a|\mathbf{x})\partial_{aa}^{2}\mathbb{E}_{\mathcal{P}|\mathbb{P}(\mathbf{X})}[Z|A=a,\mathbf{X}]h^{2}}{2\hat{p}^{k}(a|\mathbf{X})}\biggl\}+O_{P}(h^{3}).
\end{align*}
}\noindent
Thus, we have
{\small
\begin{equation*}
\begin{aligned}
&\|\mathbb{E}_{\mathbb{P}(A),\mathcal{P}|\mathbb{P}(\mathbf{X})}\big[G(A,\mathbf{X},Z)|\mathbf{X}\big]\|\\
\lesssim &\|(m(a;\mathbf{X})-\tilde{m}^{k}(a;\mathbf{X}))\||(p(a|\mathbf{X})-\hat{p}^{k}(a|\mathbf{X}))|+\|m(a;\mathbf{X})-\tilde{m}^{k}(a;\mathbf{X})\||\partial_{aa}^{2}p(a|\mathbf{X})|h^{2}\\
&+\|\partial_{a}\mathbb{E}_{\mathcal{P}|\mathbb{P}(\mathbf{X})}[Z|A=a,\mathbf{X}]\||\partial_{a}p(a|\mathbf{X})|h^{2}+|p(a|\mathbf{X})|\|\partial_{aa}^{2}\mathbb{E}_{\mathcal{P}|\mathbb{P}(\mathbf{X})}[Z|A=a,\mathbf{X}]\|h^{2}+O_{P}(h^{3}).
\end{aligned}
\end{equation*}
}\noindent
Similarly, we have
{\small
\begin{align*}
&\mathbb{E}_{\mathbb{P}(A),\mathcal{P}|\mathbb{P}(\mathbf{X})}\big[F(A,\mathbf{X},Z)|\mathbf{X}\big]\\
&=(\int u^{2}K(u)du)\times\\
&\quad\bigg\{-\frac{\partial_{a}\mathbb{E}_{\mathcal{P}|\mathbb{P}(\mathbf{X})}[Z|A=a,\mathbf{X}]\partial_{a}p(a|\mathbf{X})h^{2}}{p(a|\mathbf{X})}-\frac{p(a|\mathbf{X})\partial_{aa}^{2}\mathbb{E}_{\mathcal{P}|\mathbb{P}(\mathbf{X})}[Z|A=a,\mathbf{X}]h^{2}}{2p(a|\mathbf{X})}\bigg\}+O_{P}(h^{3})
\end{align*}
}\noindent
and
{\small
\begin{align*}
&\|\mathbb{E}_{\mathbb{P}(A),\mathcal{P}|\mathbb{P}(\mathbf{X})}\big[F(A,\mathbf{X},Z)|\mathbf{X}\big]\|\\
&\lesssim\|\partial_{a}\mathbb{E}_{\mathcal{P}|\mathbb{P}(\mathbf{X})}[Z|A=a,\mathbf{X}]\||\partial_{a}p(a|\mathbf{X})|h^{2}+|p(a|\mathbf{X})|\|\partial_{aa}^{2}\mathbb{E}_{\mathcal{P}|\mathbb{P}(\mathbf{X})}[Z|A=a,\mathbf{X}]\|h^{2}+O_{P}(h^{3}).
\end{align*}
}\noindent
We bound $\mathbb{E}_{\mathbb{P}(\mathbf{X})}[\|\mathbb{E}_{\mathbb{P}(A),\mathcal{P}|\mathbb{P}(\mathbf{X})}\big[G(A,\mathbf{X},Z)|\mathbf{X}\big]\|^{2}]$ and $\mathbb{E}_{\mathbb{P}(\mathbf{X})}[\|\mathbb{E}_{\mathbb{P}(A),\mathcal{P}|\mathbb{P}(\mathbf{X})}\big[F(A,\mathbf{X},Z)|\mathbf{X}\big]\|^{2}]$. Note that
\begin{align*}
&\mathbb{E}_{\mathbb{P}(\mathbf{X})}[\|\mathbb{E}_{\mathbb{P}(A),\mathcal{P}|\mathbb{P}(\mathbf{X})}\big[G(A,\mathbf{X},Z)|\mathbf{X}\big]\|^{2}] \\ 
&\lesssim\big(\mathbb{E}_{\mathbb{P}(\mathbf{X})}[\|(m(a;\mathbf{X})-\tilde{m}^{k}(a;\mathbf{X}))\|^{4}]\big)^{\frac{1}{2}}\times\big(\mathbb{E}_{\mathbb{P}(\mathbf{X})}[|p(a|\mathbf{X})-\hat{p}^{k}(a|\mathbf{X})|^{4}]\big)^{\frac{1}{2}}+O(h^{4})
\end{align*}
and
{\small
\begin{align*}
\mathbb{E}_{\mathbb{P}(\mathbf{X})}[\|\mathbb{E}_{\mathbb{P}(A),\mathcal{P}|\mathbb{P}(\mathbf{X})}\big[F(A,\mathbf{X},Z)|\mathbf{X}\big]\|^{2}]\lesssim O(h^{4}).
\end{align*}
}\noindent 
Hence, we conclude that $\text{I}_{2}=O(h\rho_{p}^{2}\rho_{m}^{2}+h^{5})$. Combining all the results, we can conclude that $\sqrt{N}\underset{k=1,2}{\sum}\frac{N_{k}}{N}\text{I}=o_{P}(1)$.
\noindent Consider $\left\|\text{III}\right\|$. We have
{\small
\begin{align*}
\left\|\text{III}\right\|&\leq\left\|\mathbb{E}_{N_{k}}\bigg[\frac{\{\tilde{m}_{a}^{k}(\mathbf{X})-m(a;\mathbf{X})\}}{\hat{p}^{k}(a|\mathbf{X})}\times\right.\left.\mathbb{E}_{N_{k}}[\{\hat{p}^{k}(a|\mathbf{X})-K_{h}(A-a)\}|\mathbf{X}]\bigg]\right\|\\
&+\left\|\sqrt{h}\mathbb{E}_{N_{k}}\bigg[K_{h}(A-a)(Z-m(a;\mathbf{X}))\times\right.\left.\frac{(p(a|\mathbf{X})-\hat{p}^{k}(a|\mathbf{X}))}{\hat{p}^{k}(a|\mathbf{X})p(a|\mathbf{X})}\bigg]\right\|\\
&\lesssim\sqrt{h}\left\|\mathbb{E}_{N_{k}}\bigg[\{\tilde{m}^{k}(a;\mathbf{X})-m(a;\mathbf{X})\}\{\hat{p}^{k}(a|\mathbf{X})-p(a|\mathbf{X})-\frac{h^{2}}{2}\partial_{aa}^{2}p(a|\mathbf{X})\int u^{2}K(u)du + O_{P}(h^{3})\}\bigg]\right\|+h^{\frac{5}{2}}\rho_{p}+O(h^{\frac{7}{2}})\\
&\lesssim\sqrt{h}\mathbb{E}_{N_{k}}\bigg[\left\|\{\tilde{m}^{k}(a;\mathbf{X})-m(a;\mathbf{X})\}\right\|\times|\{\hat{p}^{k}(a|\mathbf{X})-p(a|\mathbf{X})\}|\bigg]\\
&\quad +\sqrt{h}\mathbb{E}_{N_{k}}\bigg[\frac{h^{2}}{2}\left\|\{\tilde{m}^{k}(a;\mathbf{X})-m(a;\mathbf{X})\}\times\partial_{aa}^{2}p(a|\mathbf{X})\int u^{2}K(u)du\right\|\bigg]+ O(h^{\frac{7}{2}})+h^{\frac{5}{2}}\rho_{p}\\
&\lesssim\sqrt{h}\biggl(\mathbb{E}_{N_{k}}\big[\left\|\tilde{m}^{k}(a;\mathbf{X})-m(a;\mathbf{X})\right\|^{2}\big]\biggl)^{\frac{1}{2}}\times\biggl(\mathbb{E}_{N_{k}}\big[\mid\hat{p}^{k}(a|\mathbf{X})-p(a|\mathbf{X})\mid^{2}\big]\biggl)^{\frac{1}{2}}\\
&\;+\sqrt{h}\frac{h^{2}\bigg(\int u^{2}K(u)du\bigg)}{2}\times\bigg(\mathbb{E}_{N_{k}}\big[\left\|\tilde{m}^{k}(a;\mathbf{X})-m(a;\mathbf{X})\right\|^{2}\big]\bigg)^{\frac{1}{2}}\times\bigg(\mathbb{E}_{N_{k}}\big[|\partial_{aa}^{2}p(a|\mathbf{X})|^{2}\big]\bigg)^{\frac{1}{2}} + O(h^{\frac{7}{2}})+h^{\frac{5}{2}}\rho_{p}.
\end{align*}
}\noindent
We therefore conclude that 
\begin{equation*}
\begin{aligned}
\text{III}=O(h^{\frac{5}{2}}\rho_{p}+h^{\frac{1}{2}}\rho_{p}\rho_{m}+h^{\frac{5}{2}}\rho_{m}+h^{\frac{7}{2}}),
\end{aligned}
\end{equation*}
and hence $\sqrt{N}\underset{k=1,2}{\sum}\frac{N_{k}}{N}\text{III}=o_{P}(1)$.
					
\noindent Consider the term IV. Note that
\begin{align*}
\left\|\text{IV}\right\|^{2}&=\text{IV}_{1}+\text{IV}_{2},
\end{align*}
where
{\small
\begin{align*}
\text{IV}_{1}&=\frac{1}{N_{k}^{2}}\underset{i\in\mathcal{D}_{k}}{\sum}\left\|\big\{1-\frac{K_{h}(A_{i}-a)}{\hat{p}^{k}(a|\mathbf{X}_{i})}\big\}\{D_{a}^{k}(\mathbf{X}_{i})\}\right\|^{2}\\
\text{IV}_{2}&=\frac{1}{N_{k}^{2}}\underset{\substack{i,j\in\mathcal{D}_{k}\\i\neq j}}{\sum}\langle\big\{1-\frac{K_{h}(A_{i}-a)}{\hat{p}^{k}(a|\mathbf{X}_{i})}\big\}\{D_{a}^{k}(\mathbf{X}_{i})\},\big\{1-\frac{K_{h}(A_{j}-a)}{\hat{p}^{k}(a|\mathbf{X}_{j})}\big\}\{D_{a}^{k}(\mathbf{X}_{j})\}\rangle.\\
\end{align*}
}\noindent
It can be shown that $\text{IV}_{1}\lesssim \frac{1}{N_{k}}\underset{i\in\mathcal{D}_{k}}{\sum}\left\|D_{a}^{k}(\mathbf{X}_{i})\right\|^{2}$. Besides, we can show that 
{\small
\begin{align*}
\vvvert \hat{m}^{k}(a;\cdot)-\tilde{m}^{k}(a;\cdot)\vvvert^{2}=\frac{1}{N_{k}}\mathbb{E}_{\mathbb{P}(\mathbf{X})}\bigg[\underset{i\in\mathcal{D}_{k}}{\sum}\big\|D_{a}^{k}(\mathbf{X}_{i})\big\|^{2}\bigg].
\end{align*}
}\noindent
Now, for any $\xi>0$, using Markov inequality gives
{\small
\begin{align*}
&\mathbb{P}\biggl\{\frac{1}{N_{k}}\underset{i\in\mathcal{D}_{k}}{\sum}\left\|D_{a}^{k}(\mathbf{X}_{i})\right\|^{2}\geq \xi^{-1}\vvvert \hat{m}^{k}(a;\cdot)-\tilde{m}^{k}(a;\cdot)\vvvert^{2}\biggl\} \\
&\leq\xi\frac{\frac{1}{N_{k}}\mathbb{E}_{\mathbb{P}(\mathbf{X})}\big[\underset{i\in\mathcal{D}_{k}}{\sum}\left\|D_{a}^{k}(\mathbf{X}_{i})\right\|^{2}\big]}{\vvvert \hat{m}^{k}(a;\cdot)-\tilde{m}^{k}(a;\cdot)\vvvert^{2}}=\xi.
\end{align*}		
}\noindent
Under the Convergence Assumptions, we have
\begin{align*}
\text{IV}_{1}&=O_{P}(\vvvert \hat{m}^{k}(a;\cdot)-\tilde{m}^{k}(a;\cdot)\vvvert^{2})\\
&=O_{P}(N^{-2}+N^{-1}\nu_{N}^{2}+N^{-1}\alpha_{N}^{2}).
\end{align*}
\noindent For the quantity $\text{IV}_{2}$, we notice that 
{\small
\begin{align*}
&\text{IV}_{2}\leq \frac{1}{N_{k}^{2}}\underset{\substack{i,j\in\mathcal{D}_{k}\\i\neq j}}{\sum}\biggl\|\biggl\{1-\frac{K_{h}(A_{i}-a)}{\hat{p}^{k}(a|\mathbf{X}_{i})}\biggl\}\{D_{a}^{k}(\mathbf{X}_{i})\}\biggl\|\times\biggl\|\biggl\{1-\frac{K_{h}(A_{j}-a)}{\hat{p}^{k}(a|\mathbf{X}_{j})}\biggl\}\{D_{a}^{k}(\mathbf{X}_{j})\}\biggl\|\\
&\leq \frac{N_{k}-1}{N_{k}}\frac{1}{N_{k}}\underset{i\in\mathcal{D}_{k}}{\sum}\biggl\|\biggl\{1-\frac{K_{h}(A_{i}-a)}{\hat{p}^{k}(a|\mathbf{X}_{i})}\biggl\}\{D_{a}^{k}(\mathbf{X}_{i})\}\biggl\|^{2}\\
&\leq \frac{1}{N_{k}}\underset{i\in\mathcal{D}_{k}}{\sum}\biggl\|\biggl\{1-\frac{K_{h}(A_{i}-a)}{\hat{p}^{k}(a|\mathbf{X}_{i})}\biggl\}\{D_{a}^{k}(\mathbf{X}_{i})\}\biggl\|^{2}.
\end{align*}
}\noindent
Similarly, we can show that $\text{IV}_{2}=O_{P}(N^{-2}+N^{-1}\nu_{N}^{2}+N^{-1}\alpha_{N}^{2})$. Hence, $\text{IV}=O_{P}(N^{-1}+N^{-\frac{1}{2}}\nu_{N}+N^{-\frac{1}{2}}\alpha_{N})$ which implies that $\sqrt{N}\underset{k=1,2}{\sum}\frac{N_{k}}{N}\text{IV}=o_{P}(1)$.
					
\noindent Consider the term V. Note that
{\small
\begin{align*}
\mathbb{P}_{N_{k}}\bigg[\frac{K_{h}(A-a)R}{\hat{p}^{k}(a|\mathbf{X})}\bigg]
=\mathbb{P}_{N_{k}}\bigg[\frac{K_{h}(A-a)R}{p(a|\mathbf{X})}\bigg]+\mathbb{P}_{N_{k}}\bigg[\frac{K_{h}(A-a)R}{\hat{p}^{k}(a|\mathbf{X})}-\frac{K_{h}(A-a)R}{p(a|\mathbf{X})}\bigg].
\end{align*}
}\noindent
The second term is dominated by the first term since the second term involves the difference between the estimated density function $\hat{p}^{k}(a|\mathbf{X})$ and the true density function $p(a|\mathbf{X})$. Now, we consider the first term and we have
{\small
\begin{align*}
&\mathbb{E}_{(\mathbb{P}(A),\mathbb{P}(\mathbf{X}),\mathcal{P})}\bigg[\frac{1}{N_{k}}\underset{i=1}{\overset{N_{k}}{\sum}}\left\|\frac{K_{h}(A_{i}-a)R_{i}}{p(a|\mathbf{X}_{i})}\right\|\bigg]\\
\leq&\frac{c}{N_{k}}\underset{i=1}{\overset{N_{k}}{\sum}}\mathbb{E}_{\mathbb{P}(\mathbf{X}),\mathcal{P}}[\mathbb{E}_{\mathbb{P}(A)|\mathbb{P}(\mathbf{X})}[K_{h}(A_{i}-a)|\mathbf{X}_{i}]\left\|R_{i}\right\|]\\
=&O(h^{3})+c\bigg\{\frac{1}{N_{k}}\underset{i=1}{\overset{N_{k}}{\sum}}\mathbb{E}_{\mathbb{P}(\mathbf{X}),\mathcal{P}}[p(a|\mathbf{X}_{i})\left\|R_{i}\right\|]+\frac{h^{2}\int u^{2}K(u)du}{2}\frac{1}{N_{k}}\underset{i=1}{\overset{N_{k}}{\sum}}\mathbb{E}_{\mathbb{P}(\mathbf{X}),\mathcal{P}}[\partial_{aa}^{2}p(a|\mathbf{X}_{i})\left\|R_{i}\right\|]\bigg\}\\
\lesssim&(1+h^{2})\bigg(\mathbb{E}_{\mathcal{P}}\bigg[\frac{1}{N_{k}}\underset{i=1}{\overset{N_{k}}{\sum}}\left\|R_{i}\right\|^{2}\bigg]\bigg)^{\frac{1}{2}}+O(h^{3}).
\end{align*}
}\noindent

Using Lemma \ref{lemma:2Lemma} and assumptions on $\alpha_{N}$ and $\nu_{N}$, we have $\text{V}=O_{P}((1+h^{2})(\alpha_{N}^2+\nu_{N}^2)\times\sqrt{h}+h^{3}\times\sqrt{h})$ which implies that $\sqrt{N}\underset{k=1,2}{\sum}\frac{N_{k}}{N}\text{V}=o_{P}(1)$. As a result, we have
{\small
\begin{align*}
&\sqrt{Nh}\big(\hat{\Theta}^{DML}(a)-\Theta(a)\big)\\
=&\sqrt{Nh}\Biggl\{(\mathbb{P}_{N}-\mathbb{E}_{N})\{\varphi(A,\mathbf{X},\dutchcal{Y})\}+\mathbb{E}_{N}\bigg[\frac{K_{h}(A-a)(\dutchbcal{L}\dutchcal{Y}-m(a;\mathbf{X}))}{p(a|\mathbf{X})}\bigg]\Biggl\}+o_{P}(1).
\end{align*}
}\noindent
Thus, we can rewrite the above equality as follows:
{\small
\begin{align*}
&\sqrt{Nh}\biggl\{\hat{\Theta}^{DML}(a)-\Theta(a)-\mathbb{E}_{(\mathbb{P}(A),\mathbb{P}(\mathbf{X}),\mathcal{P})}\bigg[\frac{K_{h}(A-a)(\dutchcal{Y}^{-1}-m(a;\mathbf{X}))}{p(a|\mathbf{X})}\bigg]\biggl\}\\
&=\sqrt{Nh}\bigg[(\mathbb{P}_{N}-\mathbb{E}_{(\mathbb{P}(A),\mathbb{P}(\mathbf{X}),\mathcal{P})})\varphi(A,\mathbf{X},\dutchcal{Y})\bigg]+o_{P}(1).
\end{align*}
}\noindent
Now, note that
{\small
\begin{align}
\begin{split}\label{eqt:causal quantity appendix}
&\mathbb{E}_{(\mathbb{P}(A),\mathbb{P}(\mathbf{X}),\mathcal{P})}\bigg[\frac{K_{h}(A-a)(\dutchcal{Y}^{-1}-m(a;\mathbf{X}))}{p(a|\mathbf{X})}\bigg]\\
&=\mathbb{E}_{\mathbb{P}(\mathbf{X})}\bigg[\frac{1}{p(a|\mathbf{X})}\times\mathbb{E}_{\mathbb{P}(A),\mathcal{P}|\mathbb{P}(\mathbf{X})}[K_{h}(A-a)(\dutchcal{Y}^{-1}-m(a;\mathbf{X}))|\mathbf{X}]\bigg].
\end{split}
\end{align}
}\noindent
Detailed derivations show that Eqn. \eqref{eqt:causal quantity appendix} equals the following quantity:
{\small
\begin{align*}
&h^{2}\bigg(\int u^{2}K(u)du\bigg)\times\\
&\biggl\{\mathbb{E}_{\mathbb{P}(\mathbf{X})}\biggl[\partial_{a}\mathbb{E}_{\mathcal{P}|\mathbb{P}(\mathbf{X})}\big[\dutchcal{Y}^{-1}|\mathbf{X},A=a\big]\frac{\partial_{a}p(a|\mathbf{X})}{p(a|\mathbf{X})}\bigg]+\mathbb{E}_{\mathbb{P}(\mathbf{X})}\biggl[\frac{\partial_{aa}^{2}\mathbb{E}_{\mathcal{P}|\mathbb{P}(\mathbf{X})}\big[\dutchcal{Y}^{-1}|\mathbf{X},A=a\big]}{2}\biggl]\bigg\}+O(h^{3})\\
=&h^{2}\bigg(\int u^{2}K(u)du\bigg)\times\biggl\{\mathbb{E}_{\mathbb{P}(\mathbf{X})}\biggl[\partial_{a}m(a;\mathbf{X})\frac{\partial_{a}p(a|\mathbf{X})}{p(a|\mathbf{X})}\bigg]+\mathbb{E}_{\mathbb{P}(\mathbf{X})}\biggl[\frac{\partial_{aa}^{2}m(a;\mathbf{X})}{2}\biggl]\bigg\}+O(h^{3}).
\end{align*}
}\noindent
Finally, by the Central Limit Theorem, we conclude that $\sqrt{Nh}\big[(\mathbb{P}_{N}-\mathbb{E}_{(\mathbb{P}(A),\mathbb{P}(\mathbf{X}),\mathcal{P})})\{\varphi(A,\mathbf{X},\dutchcal{Y})\}\big]$ converges weakly to a Gaussian process. 

\noindent For the case when $w=IPW$, following the above derivations by setting $m(a;\mathbf{X})=0$, $\hat{m}^{k}(a;\mathbf{X})=0$, and $D_{a}^{k}=0$, we obtain
{\small
\begin{align*}
&\sqrt{Nh}\biggl\{\hat{\Theta}^{IPW}(a)-\Theta(a)+\mathbb{E}_{(\mathbb{P}(A),\mathbb{P}(\mathbf{X}),\mathcal{P})}\bigg[\frac{K_{h}(A-a)\dutchcal{Y}^{-1}}{p(a|\mathbf{X})}\bigg]\biggl\}\\
=&\sqrt{Nh}\bigg[(\mathbb{P}_{N}-\mathbb{E}_{(\mathbb{P}(A),\mathbb{P}(\mathbf{X}),\mathcal{P})})\{\varphi(A,\mathbf{X},\dutchcal{Y})\}\bigg]+o_{P}(1).
\end{align*}
}\noindent
After undergoing detailed derivations, we show that $\mathbb{E}_{(\mathbb{P}(A),\mathbb{P}(\mathbf{X}),\mathcal{P})}\bigg[\frac{K_{h}(A-a)\dutchcal{Y}^{-1}}{p(a|\mathbf{X})}\bigg]$ equals the following quantity:
{\small
\begin{align*}
&h^{2}\bigg(\int u^{2}K(u)du\bigg)\times\biggl\{\mathbb{E}_{\mathbb{P}(\mathbf{X})}\biggl[\frac{m(a;\mathbf{X})\partial_{a}^{2}p(a|\mathbf{X})}{2p(a|\mathbf{X})}\bigg]+\mathbb{E}_{\mathbb{P}(\mathbf{X})}\biggl[\frac{\partial_{a}m(a;\mathbf{X})\partial_{a}p(a|\mathbf{X})}{p(a|\mathbf{X})}\biggl]+\mathbb{E}_{\mathbb{P}(\mathbf{X})}\biggl[\frac{\partial_{aa}^{2}m(a;\mathbf{X})}{2}\biggl]\bigg\}+O(h^{3}).
\end{align*}
}\noindent
The proof is completed.
\end{proof}
\subsection{Proofs of Proposition \ref{thm:ffjord conditional density}}\label{appendix:proofs of Theorem ffjord}
\begin{proof} \
We first find an equation which relates $\log p(z(\tau_{0}),\mathbf{X})$ and $\log p(z(\tau_{1}),\mathbf{X})$. Suppose that $G(\cdot)$ is a bijective function and differentiable. The proof requires the change of variables in probability density theorem, i.e., given the variables $(Z,\mathbf{X})$ and the corresponding density function $p(z,\mathbf{x})$, the density function of $(G(Z),\mathbf{X})$ is $p(G(z),\mathbf{x})$ such that
{\small
\begin{align*}
p(G(z),\mathbf{x})&=p(z,\mathbf{x})\biggl|\det\begin{bmatrix}
&\frac{\partial G(z)}{\partial z} & \mathbf{0}_{1\times D}\\
&\mathbf{0}_{D\times 1} & \mathbb{I}_{D}
\end{bmatrix}\biggl|^{-1}\\
\Rightarrow \log \frac{p(G(z),\mathbf{x})}{p(z,\mathbf{x})}&=-\log \biggl|\det\begin{bmatrix}
&\frac{\partial G(z)}{\partial z} & \mathbf{0}_{1\times D}\\
&\mathbf{0}_{D\times 1} & \mathbb{I}_{D}
\end{bmatrix}\biggl|.
\end{align*}
}\noindent
Write $\mathbf{z}(\tau)=[z(\tau),\mathbf{X}]^{\top}$. From the integral equation $\begin{bmatrix}
						z(\tau_{0})\\
						\mathbf{X}(\tau_{0})
					\end{bmatrix}=\begin{bmatrix}
						a\\
						\mathbf{X}
					\end{bmatrix}+\int_{\tau_{1}}^{\tau_{0}}\begin{bmatrix}
						g(z(\tau),\mathbf{X}, \tau; \theta)\\
						0
					\end{bmatrix}d\tau$, the corresponding differential equation is
					{\small
						\begin{align*}
							\frac{\partial \mathbf{z}(\tau)}{\partial \tau}=\begin{bmatrix}
								\frac{\partial z(\tau)}{\partial \tau}\\
								\frac{\partial \mathbf{X}(\tau)}{\partial \tau}
							\end{bmatrix}=\begin{bmatrix}
								g(z(\tau),\mathbf{X}, \tau; \theta)\\
								\mathbf{0}
							\end{bmatrix},\;\text{\normalsize where}\; \tau_{0}\leq \tau\leq \tau_{1}.
						\end{align*}
					}\noindent
Consider $\frac{\partial \log p(\mathbf{z}(\tau))}{\partial \tau}=\frac{\partial \log p(z(\tau), \mathbf{X})}{\partial \tau}$. Write  $\mathbf{z}(\tau+\epsilon)=[z(\tau+\epsilon),\mathbf{X}]^{\top}=[T_{\epsilon}(\tau),\mathbf{X}]^{\top}$. From the first principle of derivatives, we have
{\small
\begin{align*}
&\frac{\partial \log p(\mathbf{z}(\tau))}{\partial \tau}=\frac{\partial \log p(z(\tau), \mathbf{X}(\tau))}{\partial \tau}\\
&=\underset{\epsilon\rightarrow 0^{+}}{\lim}\frac{\log p(\mathbf{z}(\tau+\epsilon))-\log p(\mathbf{z}(\tau))}{\epsilon}\\
&=\underset{\epsilon\rightarrow 0^{+}}{\lim}\frac{\log p(T_{\epsilon}(z(\tau)),\mathbf{X})-\log p(z(\tau),\mathbf{X})}{\epsilon}\\
&=\underset{\epsilon\rightarrow 0^{+}}{\lim}\frac{-\log \biggl|\det\begin{bmatrix}
&\frac{\partial T_{\epsilon}(z(\tau))}{\partial z(\tau)} & \mathbf{0}_{1\times D}\\
&\mathbf{0}_{D\times 1} & \mathbb{I}_{D}
\end{bmatrix}\biggl|}{\epsilon}\\
&\overset{\text{L'H\^{o}pital}}{=}-\underset{\epsilon\rightarrow 0^{+}}{\lim}\frac{\partial}{\partial \epsilon}\log \biggl|\det\begin{bmatrix}
&\frac{\partial T_{\epsilon}(z(\tau))}{\partial z(\tau)} & \mathbf{0}_{1\times D}\\
&\mathbf{0}_{D\times 1} & \mathbb{I}_{D}
\end{bmatrix}\biggl|\\
&=-\underset{\epsilon\rightarrow 0^{+}}{\lim}\frac{\frac{\partial}{\partial \epsilon}\biggl|\det\begin{bmatrix}
&\frac{\partial T_{\epsilon}(z(\tau))}{\partial z(\tau)} & \mathbf{0}_{1\times D}\\
&\mathbf{0}_{D\times 1} & \mathbb{I}_{D}
\end{bmatrix}\biggl|}{\biggl|\det\begin{bmatrix}
&\frac{\partial T_{\epsilon}(z(\tau))}{\partial z(\tau)} & \mathbf{0}_{1\times D}\\
&\mathbf{0}_{D\times 1} & \mathbb{I}_{D}
\end{bmatrix}\biggl|}\\
&=\underbrace{\frac{\underbrace{-\underset{\epsilon\rightarrow 0^{+}}{\lim}\frac{\partial}{\partial \epsilon}\biggl|\det\begin{bmatrix}
&\frac{\partial T_{\epsilon}(z(\tau))}{\partial z(\tau)} & \mathbf{0}_{1\times D}\\
&\mathbf{0}_{D\times 1} & \mathbb{I}_{D}
\end{bmatrix}\biggl|}_{\text{bounded}}}{\underset{\epsilon\rightarrow 0^{+}}{\lim}\biggl|\det\begin{bmatrix}
&\frac{\partial T_{\epsilon}(z(\tau))}{\partial z(\tau)} & \mathbf{0}_{1\times D}\\
&\mathbf{0}_{D\times 1} & \mathbb{I}_{D}
\end{bmatrix}\biggl|}}_{1}\\
&=-\underset{\epsilon\rightarrow 0^{+}}{\lim}\frac{\partial}{\partial \epsilon}\biggl|\det\begin{bmatrix}
&\frac{\partial T_{\epsilon}(z(\tau))}{\partial z(\tau)} & \mathbf{0}_{1\times D}\\
&\mathbf{0}_{D\times 1} & \mathbb{I}_{D}
\end{bmatrix}\biggl|.
\end{align*}
}\noindent

Applying the Jacobi’s formula, we have
{\small
\begin{align*}
&\frac{\partial \log p(\mathbf{z}(\tau))}{\partial \tau}\\
&=-\underset{\epsilon\rightarrow 0^{+}}{\lim}\operatorname{Tr}\biggl(\operatorname{adj}\bigg(\begin{bmatrix}
&\frac{\partial T_{\epsilon}(z(\tau))}{\partial z(\tau)} & \mathbf{0}_{1\times D}\\
&\mathbf{0}_{D\times 1} & \mathbb{I}_{D}
\end{bmatrix}\bigg)\times\frac{\partial}{\partial \epsilon}\begin{bmatrix}
&\frac{\partial T_{\epsilon}(z(\tau))}{\partial z(\tau)} & \mathbf{0}_{1\times D}\\
&\mathbf{0}_{D\times 1} & \mathbb{I}_{D}
\end{bmatrix}\biggl)\\
&=-\operatorname{Tr}\biggl(\underbrace{\underset{\epsilon\rightarrow 0^{+}}{\lim}\operatorname{adj}\bigg(\begin{bmatrix}
&\frac{\partial T_{\epsilon}(z(\tau))}{\partial z(\tau)} & \mathbf{0}_{1\times D}\\
&\mathbf{0}_{D\times 1} & \mathbb{I}_{D}
\end{bmatrix}\bigg)}_{\mathbb{I}_{D}}\times\underset{\epsilon\rightarrow 0^{+}}{\lim}\frac{\partial}{\partial \epsilon}\begin{bmatrix}
&\frac{\partial T_{\epsilon}(z(\tau))}{\partial z(\tau)} & \mathbf{0}_{1\times D}\\
&\mathbf{0}_{D\times 1} & \mathbb{I}_{D}
\end{bmatrix}\biggl)\\
&=-\underset{\epsilon\rightarrow 0^{+}}{\lim}\biggl(\frac{\partial}{\partial \epsilon}\frac{\partial T_{\epsilon}(z(t))}{\partial z(t)}\biggl).
\end{align*}
}\noindent
Applying Taylor series expansion on $T_{\epsilon}(z(\tau))$ w.r.t. $\epsilon$ and taking the limit, we have
{\small
\begin{align*}
&\frac{\partial \log p(\mathbf{z}(\tau))}{\partial \tau}=-\underset{\epsilon\rightarrow 0^{+}}{\lim}\biggl(\frac{\partial}{\partial \epsilon}\frac{\partial T_{\epsilon}(z(\tau))}{\partial z(\tau)}\biggl)\\
&=-\underset{\epsilon\rightarrow 0^{+}}{\lim}\biggl(\frac{\partial}{\partial \epsilon}\frac{\partial}{\partial z(\tau)}(z(\tau)+\frac{\partial z(\tau)}{\partial \tau}\epsilon+O(\epsilon^{2}))\biggl)\\
&=-\underset{\epsilon\rightarrow 0^{+}}{\lim}\biggl(\frac{\partial}{\partial \epsilon}(1+\frac{\partial g(z(\tau),\mathbf{X},\tau;\theta)}{\partial z(\tau)}\epsilon+O(\epsilon^{2}))\biggl)\\
&=-\frac{\partial g(z(\tau),\mathbf{X},\tau;\theta)}{\partial z(\tau)}.
\end{align*}
}\noindent
As such, we have
\begin{align*}
\int_{\tau_{0}}^{\tau_{1}} \frac{\partial \log p(\mathbf{z}(\tau))}{\partial \tau}d\tau &= \int_{\tau_{0}}^{\tau_{1}} -\frac{\partial g(z(\tau),\mathbf{X},\tau;\theta)}{\partial z(\tau)} d\tau\\
\Rightarrow \log \frac{p(\mathbf{z}(\tau_{1}))}{p(\mathbf{z}(\tau_{0}))} &= \int_{\tau_{1}}^{\tau_{0}} \frac{\partial g(z(\tau),\mathbf{X},\tau;\theta)}{\partial z(\tau)} d\tau\\
\Rightarrow \log \frac{p(z(\tau_{1}),\mathbf{X})}{p(z(\tau_{0}),\mathbf{X})} &= \int_{\tau_{1}}^{\tau_{0}} \frac{\partial g(z(\tau),\mathbf{X},\tau;\theta)}{\partial z(\tau)} d\tau.
\end{align*}
\end{proof}

\section{Bandwidth selection} \label{appendix:bandwidth selection}
Since we estimate $m(a; \mathbf{X})$ at $9$ quantiles, we have $h^{*}=\underset{h}{\arg\min}\underset{s\in\{0.1,\cdots,0.9\}}{\overset{}{\sum}}[h^{4}[\hat{B}_{a}(s)]^{2}+\frac{\hat{C}(s,s)}{Nh}]$. In fact, $h^{*}=\bigg(\frac{\underset{s\in\{0.1,\cdots,0.9\}}{\overset{}{\sum}}C(s,s)}{4N\underset{s\in\{0.1,\cdots,0.9\}}{\overset{}{\sum}}(B_{a}(s))^{2}}\bigg)^{\frac{1}{5}}$, where
\begin{subequations}
{\small
\begin{align}
B_{a}(s)&=\frac{{\hat{\Theta}^{b}(a)(s)}-{\hat{\Theta}^{\epsilon b}(a)(s)}}{b^{2}(1-\epsilon^{2})},\label{eqt:bandwidth cond1}\\
C(s,\bar{s})&=\frac{h}{N}\underset{i=1}{\overset{N}{\sum}}(\hat{V}_{i}(s)-\bar{V}(s))(\hat{V}_{i}(\bar{s})-\bar{V}(\bar{s})),\label{eqt:bandwidth cond2}\\
\hat{V}_{i}&=\frac{K_{h}(A_{i}-a)(\hat{\dutchcal{Y}}_{i}^{-1}-\hat{m}(a;\mathbf{X}_{i}))}{\hat{f}(a|\mathbf{X}_{i})}+\hat{m}(a;\mathbf{X}_{i}),\label{eqt:bandwidth cond23}\\
\bar{V}&=\frac{1}{N}\underset{i=1}{\overset{N}{\sum}}\hat{V}_{i},\label{eqt:bandwidth cond3}\\
\hat{\bigtriangleup}_{a}^{b}&=\frac{K_{b}(A_{i}-a)(\hat{\dutchcal{Y}}_{i}^{-1}-\hat{m}(a;\mathbf{X}_{i}))}{\hat{f}(a|\mathbf{X}_{i})}+\hat{m}(a;\mathbf{X}_{i}),\label{eqt:bandwidth cond4}\\
\hat{\bigtriangleup}_{a}^{\epsilon b}&=\frac{K_{\epsilon b}(A_{i}-a)(\hat{\dutchcal{Y}}_{i}^{-1}-\hat{m}(a;\mathbf{X}_{i}))}{\hat{f}(a|\mathbf{X}_{i})}+\hat{m}(a;\mathbf{X}_{i}).\label{eqt:bandwidth cond5}
\end{align}
}\noindent
\end{subequations}
Here, we choose $\epsilon=0.5\in[0,1]$, $b=2h$, $h=c\sigma_{A}N^{-0.2}$, and $\sigma_{A}$ is the standard deviation of treatment $A$. In the settings $\epsilon=0.5\in[0,1]$, $b=2h$, $h=c\sigma_{A}N^{-0.2}$, we calculate $B_{a}(s)$ and $C(s,s)$. After that, we compute $h^{*}$ by $\bigg(\frac{\underset{s\in\{0.1,\cdots,0.9\}}{\overset{}{\sum}}C(s,s)}{4N\underset{s\in\{0.1,\cdots,0.9\}}{\overset{}{\sum}}(B_{a}(s))^{2}}\bigg)^{\frac{1}{5}}$.

\section{Simulation of Gaussian Process} \label{appendix:Gaussian Process Simulation}
In this section, we present how to simulate a centered Gaussian process with a specified covariance function \(C(s,t)\). The simulation process can be decomposed into several steps:
\begin{description}
\item[Step 1] Randomly draw \(\mathfrak{M}\) points from \([0,1]\) where the \(\mathfrak{M}\) points are uniformly distributed.
\item[Step 2] Denote the drawn sample by \(t_{1},\cdots,t_{\mathfrak{M}}\). Compute \(\hat{\mathbf{C}}=[\hat{C}_{ij}]_{\mathfrak{M}\times\mathfrak{M}}\) where \(\hat{C}_{ij}=C(t_{i},t_{j})\).
\item[Step 3] Perform the Cholesky decomposition (or eigenvalue decomposition if \(\hat{\mathbf{C}}\) is not positive definite) on \(\hat{\mathbf{C}}\) such that \(\hat{\mathbf{C}}=\mathbf{L}\mathbf{L}^{\top}\).
\item[Step 4] Generate \(\mathbf{Z}\) from \(\mathcal{N}(\mathbf{0},\mathbf{I}_{\mathfrak{M}})\) such that the size of \(\mathbf{Z}\) is \(\mathfrak{M}\times\mathfrak{N}\).
\item[Step 5] Compute \(\mathbf{Y}=\mathbf{L}\mathbf{Z}\). Each column of \(\mathbf{Y}_{\mathfrak{M}\times\mathfrak{N}}\) represents a simulated centered Gaussian process with the specified covariance function.
\end{description}
\noindent







\end{document}